\documentclass[pdflatex,JoC/sn-nature]{sn-jnl}% Style for submissions to Nature Portfolio journals
\usepackage{pgfplots}
\usepackage{natbib}
\usepackage[dvipsnames,table]{xcolor}

\usepackage{graphicx}%
\usepackage{multirow}%
\usepackage{amsmath,amssymb,amsfonts}%
\usepackage{amsthm}%
\usepackage{mathrsfs}%
\usepackage[title]{appendix}%
\usepackage{xcolor}%
\usepackage{textcomp}%
\usepackage{manyfoot}%
\usepackage{booktabs}%
\usepackage{algorithm}%
\usepackage{algorithmicx}%
\usepackage{algpseudocode}%
\usepackage{listings}%

\newcommand{\up}{\textcolor{ForestGreen}{$^{\pmb{\uparrow}}$}}
\newcommand{\down}{\textcolor{BrickRed}{$^{\pmb{\downarrow}}$}}
\theoremstyle{thmstyleone}%
\theoremstyle{thmstyletwo}%

\theoremstyle{thmstylethree}%

\begin{document}

\title[Molecular Embedding–Based Algorithm Selection in Protein–Ligand Docking]{Molecular Embedding–Based Algorithm Selection in Protein–Ligand Docking}


\author[1]{\fnm{Jiabao Brad} \sur{Wang}}\email{jb.wang@dukekunshan.edu.cn}

\author[1]{\fnm{Siyuan} \sur{Cao}}\email{siyuan.cao@dukekunshan.edu.cn}

\author[1]{\fnm{Hongxuan} \sur{Wu}}\email{hongxuan.wu@dukekunshan.edu.cn}

\author[2]{\fnm{Yiliang} \sur{Yuan}}\email{yiliang.yuan@mbzuai.ac.ae}
  
\author*[1]{\fnm{Mustafa} \sur{M\i s\i r}}\email{mustafa.misir@dukekunshan.edu.cn}

\affil*[1]{\orgdiv{Division of Natural and Applied Sciences}, \orgname{Duke Kunshan University}, \orgaddress{\street{ 8 Duke Av.}, \city{Suzhou}, \postcode{215316}, \state{Jiangsu}, \country{China}}}

\affil[2]{\orgdiv{Machine Learning Department}, \orgname{Mohamed bin Zayed University of Artificial Intelligence}, \orgaddress{\street{Building 1B}, \city{Masdar City}, \state{Abu Dhabi}, \country{UAE}}}


\abstract{
Selecting an effective docking algorithm is highly context-dependent, and no single method performs reliably across structural, chemical, and protocol regimes. MolAS is a lightweight algorithm-selection model that predicts per-algorithm performance from pretrained protein and ligand embeddings using attentional pooling and a shallow residual decoder. With hundreds to a few thousand labelled complexes, MolAS achieves up to a 15 percentage-point absolute improvement over the single-best solver (SBS) and closes 17--66\% of the Virtual Best Solver (VBS)--SBS gap across five docking benchmarks. Analyses of selection frequencies, margin-conditioned reliability, and benchmark-level oracle structure indicate that MolAS is most effective when the workflow-defined oracle landscape has low winner entropy and a reasonably separable top-solver region, but degrades under protocol mismatch that shifts solver rankings and changes the induced labels. These results suggest that, in the evaluated regime, robustness is limited less by representational capacity than by workflow- and protocol-induced instability in solver hierarchies, positioning MolAS as an in-domain selector for fixed pipelines and as a diagnostic tool for assessing when docking algorithm selection is well-posed.

\textbf{Scientific Contribution.} MolAS introduces a controlled, embedding-based selector that reduces dependence on heavy graph encoders, enabling a cleaner separation between representational choices and workflow-defined label structure. A cross-benchmark and cross-protocol analysis links selection success and failure to oracle entropy, near-ties among top solvers, and protocol-induced ranking shifts, providing an evidence-backed diagnostic account of when docking algorithm selection is likely to yield gains. The findings differentiate this work from prior docking AS studies that report in-domain improvements under a single fixed workflow by explicitly characterising protocol dependence and motivating protocol-aware modelling as a route to stronger generalisation.

\textbf{Graphical abstract.} A visual demo of our work is shown in Fig.~\ref{fig:pipeline}.
\begin{figure*}
    \centering
    \includegraphics[width=\linewidth]{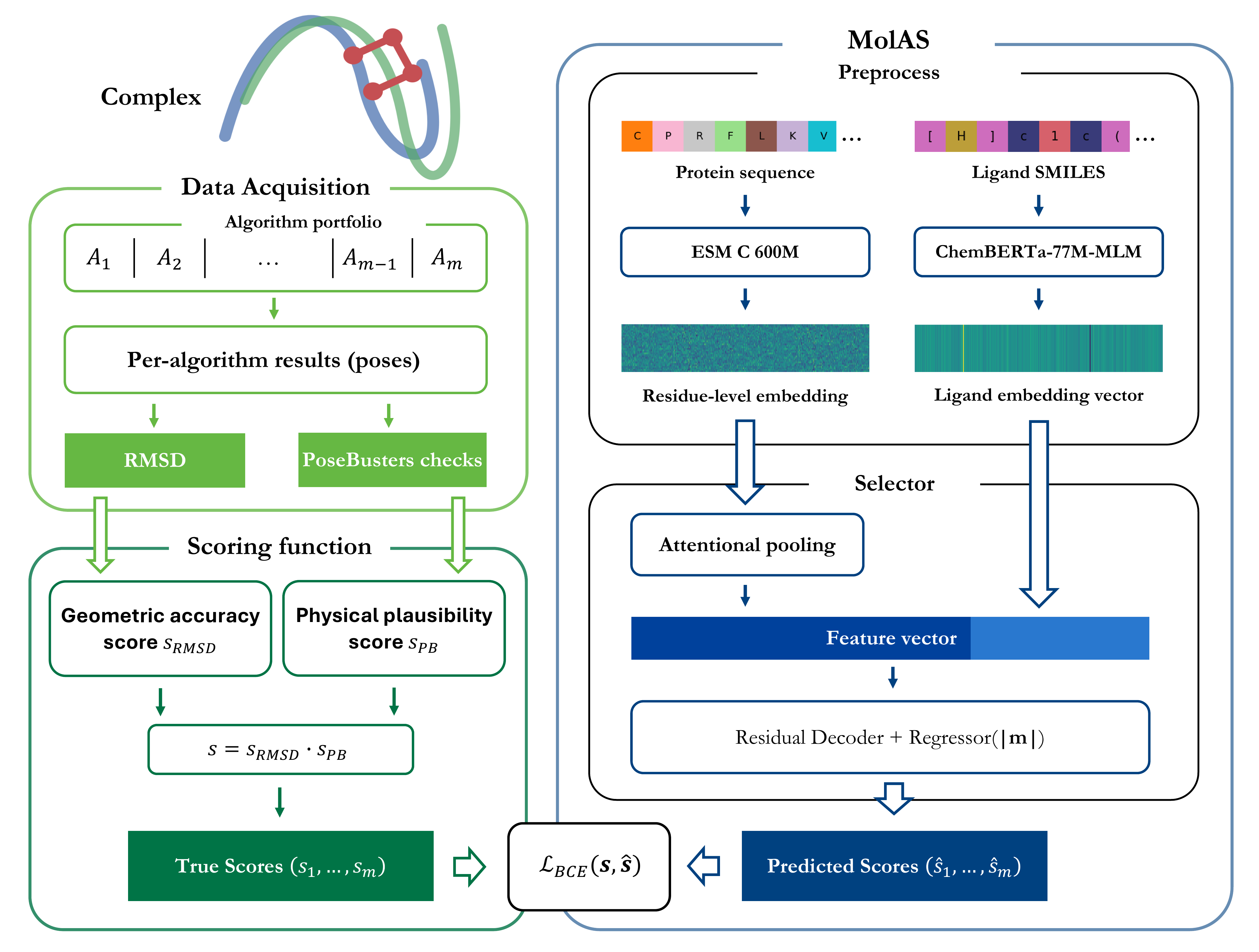}
    \caption{Overview of MolAS for workflow-specific molecular docking algorithm selection. (Left) Label data acquisition and scoring. (Right) MolAS model.}
    \label{fig:pipeline}
\end{figure*}
}




\maketitle

\section{Introduction}
\label{sec:intro}

Molecular docking is the computational process that predicts the binding configuration of a ligand to its target protein at atomic resolution and is central to structure-based drug discovery~\citep{fan2019progress}. Traditional docking pipelines rely on empirical scoring functions coupled with heuristic search~\citep{trott2010autodock,koes2013lessons,hassan2017protein}, while classical scoring functions have historically spanned physics-based, empirical, and knowledge-based formulations~\citep{ballester2010machine,cheng2010knowledge,meng2011molecular}. These methods balance speed and physical interpretability, but their docking scores often show limited correlation with experimentally measured binding affinities or binding constants~\citep{su2018comparative}. Recent machine learning (ML)–based models such as DiffDock~\citep{corso2022diffdock} and SurfDock~\citep{cao2025surfdock} aim to improve pose generation or scoring by learning protein–ligand interaction patterns directly from data. Despite decades of development, no single method consistently excels across all docking scenarios, reflecting both the \textit{No Free Lunch Theorem}~\citep{wolpert2002no} and the profound context dependence of protein–ligand recognition~\citep{yuan2024algorithm}.

Algorithm Selection (AS)~\citep{rice1976algorithm} offers a principled route to adaptivity in such heterogeneous regimes. 
Rather than committing to a single solver or fixed configuration, AS systems learn to recommend the algorithm predicted to perform best for each instance~\citep{kerschke2019automated}. 
Beyond docking, AS has been operationalised at scale in AutoML systems such as Auto-WEKA~\citep{thornton2013auto} and Auto-sklearn~\citep{feurer2015efficient}, and has shown consistent benefits in molecular modelling tasks such as QSAR regression via meta-learning~\citep{olier2018meta}.

In docking, initial AS studies focused on \emph{within-engine} choice. 
\citet{chen2023algorithm} proposed the first docking-specific AS framework using the \textit{ALORS} recommender~\citep{misir2017alors} to select among AutoDock configurations from tabular descriptors, and subsequent work explored related regression-based parameter tuning~\citep{zayed2025optimizing}. 
More recent efforts extend AS to \emph{cross-algorithm} selection across multiple docking engines, using molecular descriptors or learned representations to predict per-solver performance~\citep{yuan2024algorithm,yuan2024enhancing,yuan2024gnnas,cao2025mc}. 
These approaches demonstrate that instance-wise solver choice can outperform a single best solver (SBS) in-domain, but they are typically evaluated under a fixed benchmark and protocol, and often rely on relatively heavy graph-based encoders.

This combination leaves an important ambiguity unresolved: observed gains can reflect genuine improvements in molecular representation, but can also arise from benchmark- and workflow-specific effects, such as protocol-induced shifts in solver rankings and instability of the empirical oracle used to define training labels. 
When both the representation and the workflow vary implicitly, it is difficult to separate representational benefit from distributional or procedural alignment.
Table~\ref{tab:prior_as_compare} summarises representative docking AS formulations and highlights which factors are controlled (representation, objective, and protocol/workflow consistency).

\begin{table}[t]
\centering
\small
\caption{High-level comparison of MolAS with representative docking algorithm-/protocol-selection approaches, focusing on what is controlled in the formulation and evaluation.}
\label{tab:prior_as_compare}
\hspace*{-5em}
\begin{tabular}{|p{3.2cm}|p{2.0cm}|p{3cm}|p{3cm}|p{2.5cm}|}
\hline
\textbf{Method} & \textbf{Selection scope} & \textbf{Input representation} & \textbf{Learning objectives}& \textbf{Protocol variation evaluated}\\
\hline
ALORS-based config selection~\citep{chen2023algorithm}
& within-engine
& Tabular ligand features (RDKit descriptors + PubChem substructure fingerprints)
& Recommender-style selection/ranking of configurations (ALORS)& Fixed benchmark/protocol with multiple within-engine configurations\\
\hline
Parameter/protocol tuning~\citep{zayed2025optimizing}
& within-engine
& Tabular ligand features (molecular descriptors + substructure fingerprints)
& predicts docking scores  with MAE/MSE/RMSE& Fixed benchmark/protocol with multiple within-engine configurations\\
\hline
Cross-algorithm selection (descriptors)~\citep{yuan2024algorithm}
& cross-engine
& Tabular protein statistics + ligand RDKit descriptors
& Predicts RMSD (squared loss)& Fixed benchmark/protocol and portfolio\\
\hline
GNNAS-Dock~\citep{yuan2024gnnas}
& cross-engine
& Residue-level protein graph + atom-level ligand graph (GNN features; stacking/meta-model)
& Accuracy model predicts RMSD prediction; Efficiency model predicts binary success (RMSD $<2$\,\AA) + runtime& Fixed benchmark/protocol and portfolio\\
\hline
MC-GNNAS-Dock~\citep{cao2025mc}
& cross-engine
& Residue-level protein graph (pretrained node embedding) + atom-level ligand graph (physicochemical node embedding)
& Predicts PoseBuster-validity-gated RMSD score with BCE loss; ranking loss tested& Fixed benchmark/protocol and portfolio\\
\hline
\textbf{MolAS (this work)}
& cross-engine
& Residue-level pretrained protein node embedding + pretrained ligand embedding
& Predicts PoseBuster-validity-gated RMSD score with BCE loss; ranking loss tested& multiple benchmarks \& post-processing\\
\hline
\end{tabular}
\end{table}

The present work introduces \textbf{MolAS} (Molecular Embedding--Based Algorithm Selector), a lightweight selector designed to probe this ambiguity. 
MolAS replaces graph encoders with pretrained molecular language model embeddings for proteins (ESM C~\citep{esm2024cambrian}) and ligands (ChemBERTa~\citep{ahmad2022chemberta}), coupled with a minimal attentional pooler and shallow residual decoder. 
This design aims to isolate the role of molecular \emph{representation} from that of \emph{workflow-defined} oracle landscapes, enabling a more direct assessment of whether docking AS is primarily constrained by representational capacity or by data and protocol variability.

We evaluate MolAS on a curated BindingMOAD dataset and across three modern docking benchmarks spanning diverse structural regimes and pose-generation protocols. In most in-domain settings, MolAS outperforms the single best solver (SBS) and closes a substantial portion of the virtual best solver (VBS)--SBS gap with hundreds to a few thousand labelled complexes, depending on the benchmark. However, performance deteriorates sharply when training and test protocols differ, indicating that MolAS inherits the oracle hierarchy of the workflow on which it is trained. This highlights a practical constraint of docking AS: changes in dataset curation, docking engines, or post-processing can effectively redefine the selection problem.

Ablation studies show that neither deeper encoders nor alternative objective functions provide systematic improvements, reinforcing that solver-hierarchy instability, benchmark-specific oracle entropy, and protocol-induced shifts in ranking dominate performance. Overall, MolAS is positioned as a workflow-adaptive selector under fixed docking protocols and as a diagnostic tool that clarifies when docking AS is feasible (non-trivial VBS--SBS gap and separable solver regimes) versus when an SBS-like policy is a reasonable default. These findings suggest that robust cross-protocol AS will require explicit modelling of workflow changes rather than further architectural scaling.

In summary, the contributions are threefold. 
Methodologically, MolAS provides a deliberately lightweight, embedding-based selector that uses pretrained protein and ligand language model representations with a minimal pooling and decoding head, so that selection performance can be studied without conflating it with increasingly complex graph encoders. 
Empirically, MolAS is evaluated across a curated BindingMOAD split and multiple modern docking benchmarks spanning distinct protocols and post-processing settings, including systematic ablations that test whether heavier encoders or alternative objectives yield consistent gains. 
Conceptually, the study frames docking algorithm selection as a \emph{workflow-defined} problem: solver hierarchies and training labels are shaped by dataset curation and pipeline choices, and cross-protocol shifts and oracle instability can dominate apparent progress, motivating MolAS both as a workflow-adaptive selector under fixed protocols and as a diagnostic tool for assessing when selection is well-posed.

% \begin{figure*}[ht!]
% \centering
% \begin{tikzpicture}[
%   node distance=4cm,
%   every node/.style={font=\small, align=center},
%   >=stealth
% ]

% % Nodes
% \node (x) [draw, rounded corners, minimum width=2.6cm, minimum height=0.8cm] {Problem instance \\ $\mathbf x \in \mathcal{X}$};
% \node (f) [draw, rounded corners, right of=x] {Features \\ $f(x) \in \mathbb{R}^d$};
% \node (phi) [draw, rounded corners, right of=f] {Predicted scores \\ $\hat{\Phi}(\mathbf x) = g(f(\mathbf x))\in \mathbb{R}^m$};
% \node (s) [draw, rounded corners, right of=phi] {Selected alg. \\ $S(x) \in \mathcal{A}$};

% % Arrows
% \draw[->] (x) -- (f) node[midway, above] {$f$};
% \draw[->] (f) -- (phi) node[midway, above] {$g$};
% \draw[->] (phi) -- (s) node[midway, above] {$\arg\max$};

% \end{tikzpicture}
% \caption{Schematic of an algorithm selection system. 
% % Each instance $x$ is mapped to a feature vector $g(x)$, then to predicted performance $\hat{\Phi}(x)$ across algorithms in $\mathcal{A}$. The best-performing algorithm is selected via $\arg\min$.
% }
% \label{fig:algorithm-selection}
% \end{figure*}

%%%%%%%%%%%%%%%%%%%%%%%%%%%%%%%%%%%%%%%%%%%%%%%%
%%%%%%%%%%%%%%%%%%%%%%%%%%%%%%%%%%%%%%%%%%%%%%%%
\section{Materials and Methods}
\label{sec:materialsMethods}
%%%%%%%%%%%%%%%%%%%%%%%%%%%%%%%%%%%%%%%%%%%%%%%%
%%%%%%%%%%%%%%%%%%%%%%%%%%%%%%%%%%%%%%%%%%%%%%%%

\subsection{Pipeline overview}
MolAS is an algorithm selection (AS) system. Let $\mathcal{X}$ denote a class of problem instances and $\mathcal{A} = \{A_1, \dots, A_m\}$ a finite set of $m$ candidate algorithms. An AS system is a mapping
\begin{equation}
    S : \mathcal{X} \to \mathcal{A}
\end{equation}
that selects, for each $\mathbf x \in \mathcal{X}$, an algorithm expected to perform well under a task-specific performance measure $\phi : \mathcal{A} \times \mathcal{X} \to \mathbb{R}_{\geq 0}$~\citep{rice1976algorithm}. In practice, $\phi$ is unknown and approximated by a learned model $\hat{\phi} : \mathcal{A} \times \mathcal{X} \to \mathbb{R}_{\geq 0}$. The system then defines
\begin{equation}
    S(x) := \arg\max_{A \in \mathcal{A}} \hat{\phi}(A, \mathbf x),
\end{equation}
thereby inducing selection via performance prediction. Often, it is modelled $(\hat{\phi}(A_1, \mathbf x), ..., \hat{\phi}(A_m, \mathbf x)) = \hat{\boldsymbol \phi}(\mathbf x) = g(f(\mathbf x))$ for a feature extractor (encoder) $f : \mathcal{X} \to \mathbb{R}^d$ and performance predictor (decoder) $g:\mathbb R^d \rightarrow \mathbb R^m$ learned to reconstruct the performance/score labels  $\mathbf s\in\mathcal S$, yielding the composition
\begin{equation}
    S = \arg\max \circ g \circ f.
\end{equation}
Such a system, as shown in Fig.~\ref{fig:algorithm-selection}, induces a total ranking over $\mathcal{A}$ for each problem instance by its characteristics, with selection yielding the top-ranked algorithm. In our work, each $\mathbf x = (\mathbf x_P, \mathbf x_L)$ represents the input pair constructed from a protein-ligand pair (\S\ref{subsec:preprocess}), $f, g$ are learned by an attentional pooler and a residual MLP decoder respectively (\S\ref{subsec:model_architecture}), and the target performance metrics are formulated with a physics-inspired composite scoring function (\S\ref{subsec:score}).

\begin{figure*}[ht!]
\centering
\hspace*{-3em}
\begin{tikzpicture}[
  node distance=4cm,
  every node/.style={font=\small, align=center},
  >=stealth
]

% Nodes
\node (x) [draw, rounded corners, minimum width=2.7cm, minimum height=0.9cm] 
  {Problem instance \\ $\mathbf{x} \in \mathcal{X}$};

\node (f) [draw, rounded corners, right of=x, node distance=3.2cm] 
  {Representation \\ $f(\mathbf{x}) \in \mathbb{R}^d$};

\node (phi) [draw, rounded corners, right of=f, text width=2.9cm, node distance=3.4cm] 
  {Predicted per--algorithm\\ performance \\ 
   $\hat{\mathbf s} = \hat{\boldsymbol \phi}(\mathbf{x}) \in \mathbb{R}^{m}$};

\node (s) [draw, rounded corners, right of=phi, node distance=5.1cm] 
  {Selector \\ $S(\mathbf{x}) = 
  \arg\max_{a \in \mathcal{A}} \hat{s}_a(\mathbf{x})$};

% Arrows
\draw[->] (x) -- (f) node[midway, above] {$f$};
\draw[->] (f) -- (phi) node[midway, above] {$g$};
\draw[->] (phi) -- (s) node[midway, above] {$\arg\max$};

\end{tikzpicture}

\caption{Schematic of the algorithm selection mapping in the sense of~\cite{rice1976algorithm}. 
Each instance $\mathbf{x}$ is mapped to a feature vector $f(\mathbf{x})$, from which a predictor $g$ produces a vector 
$\hat{\mathbf s}$ of estimated performances for all algorithms in the portfolio $\mathcal{A}$. 
The selector chooses the algorithm with maximal predicted performance.}
\label{fig:algorithm-selection}
\end{figure*}

% \begin{figure*}[ht!]
% %%\vspace{-1em}
% \centering
% \begin{tikzpicture}[
%   node distance=3.4cm,
%   every node/.style={font=\small, align=center},
%   >=stealth
% ]

% % Nodes
% \node (x) [draw, rounded corners, minimum width=2.6cm, minimum height=0.8cm] {Problem instance \\ $\mathbf x \in \mathcal{X}$};
% \node (g) [draw, rounded corners, right of=x] {Features \\ $g(\mathbf x) \in \mathbb{R}^d$};
% \node (phi) [draw, rounded corners, right of=g] {Predicted perf. \\ $\hat{\Phi}(\mathbf x) \in \mathbb{R}^m$};
% \node (s) [draw, rounded corners, right of=phi] {Selected alg. \\ $S(\mathbf x) \in \mathcal{A}$};

% % Arrows
% \draw[->] (x) -- (g) node[midway, above] {$g$};
% \draw[->] (g) -- (phi) node[midway, above] {$f$};
% \draw[->] (phi) -- (s) node[midway, above] {$\arg\min$};

% \end{tikzpicture}
% \caption{Schematic of an algorithm selection system. 
% % Each instance $x$ is mapped to a feature vector $g(x)$, then to predicted performance $\hat{\Phi}(x)$ across algorithms in $\mathcal{A}$. The best-performing algorithm is selected via $\arg\min$.
% }
% \label{fig:algorithm-selection}
% %%\vspace{-1em}
% \end{figure*}

\begin{table*}[ht!]
\centering
\caption{Candidate docking algorithms for algorithm selection}
\hspace*{-7em}
\begin{tabular}{l|l|l}
\toprule
\textbf{Method} & \textbf{Type} & \textbf{Central mechanism}\\
\midrule
Smina \citep{koes2013lessons} & Classical & Vina with empirical scoring and local search\\
\hline
Qvina-W \citep{hassan2017protein} & Classical & Vina with fast stochastic global optimisation\\
\hline
DiffDock \citep{corso2022diffdock} & ML-based& Diffusion model for direct pose generation\\
\hline
DiffDockL \citep{corso2024deep} & ML-based& DiffDock with flexible ligand modeling\\
\hline
SurfDock \citep{cao2025surfdock} & ML-based& $SE(3)$-equivariant diffusion GNN\\
\hline
Gnina \citep{mcnutt2021gnina} & ML-based& $3$D CNN-based rescoring of docking poses\\
\hline
Uni-Mol Docking V2 \citep{alcaide2025uni} & ML-based& Transformer on atomic coordinates with pretraining\\
\hline
KarmaDock \citep{zhang2023efficient} & ML-based& $SE(3)$-equivariant attention\\
\bottomrule
\end{tabular}
\label{tab:docking_tools}
\end{table*}

\subsection{Datasets and preprocessing}

\subsubsection{Dataset}
A proper AS dataset involves problem instances (protein--ligand pairs) $\mathcal{X} = \{\mathbf{x}_1,\ldots,\mathbf{x}_n\}$ and score targets $\mathcal{S}\in\mathbb{R}^{n\times m}$ that contain per-instance, per-algorithm performance data. 
Acquiring such data is inherently challenging: candidate docking pipelines are heterogeneous, their training data may overlap with common benchmarks, and exhaustive evaluation over large instance sets is costly. 
This necessitates a strategic selection of both representative molecular instances and a portfolio that is (i) methodologically diverse and (ii) available under consistent, documented protocols.

Accordingly, the main AS experiments use a curated portfolio of eight docking algorithms, chosen to span strong baselines and learning-based approaches while remaining tractable for systematic analysis across workflows (Table~\ref{tab:docking_tools}). 
Portfolio completeness is not claimed; rather, this set aims to analyse algorithm selection under a representative and protocol-consistent set of solvers.

Since most of these algorithms have parsed \textit{PDBBind}~\citep{liu2017forging} as the training set, we evaluate them on \textit{BindingMOAD}~\citep{hu2005binding}, a comprehensive dataset of experimentally determined protein--ligand complexes. 
To reduce potential overlap with common training corpora and to obtain a chemically standardised evaluation split, a \textit{MOAD-curated} subset is constructed from BindingMOAD using a fixed, two-stage filtering procedure that does not consult docking scores, solver rankings, RMSD, or pose validity at any point.

First, starting from BindingMOAD, a working pool of approximately 6700 complexes is formed by removing complexes whose PDB IDs appear in the PDBBind v2020 collection and by excluding overlaps with the DockGen protein lists, followed by basic dataset housekeeping (e.g., resolving duplicates and incomplete entries). 
Second, the working pool is filtered using fixed protein and ligand quality criteria implemented in a standalone script: on the protein side, only structures composed of the 20 canonical amino acids are retained; on the ligand side, RDKit-based checks enforce standard Lipinski-style constraints (molecular weight $\leq 500~\mathrm{Da}$, $\log P \leq 5$, $\leq 5$ hydrogen bond donors, and $\leq 10$ hydrogen bond acceptors). 
Complexes failing any criterion are removed. 
The surviving 3179 complexes constitute the final \textit{MOAD-curated} dataset used in the main experiments.

Solver performance is analysed post hoc on this fixed dataset to characterise solver dominance and the resulting VBS--SBS gap. Each algorithm demonstrates strong performance in certain docking scenarios, ensuring the presence of a non-trivial selection problem.
In addition, complementary analyses use extra benchmarks that provide released per-complex outputs for a substantially broader catalogue of docking and co-folding pipelines (including recent methods such as Interformer) under documented settings, which we describe next.

%%%%%%%%%%%%%%%%%%%%%%%%%%%%%%%%%%
\subsubsection{Extra benchmarks}
%%%%%%%%%%%%%%%%%%%%%%%%%%%%%%%%%%
In addition to MOAD-curated, two external benchmarks, \textit{PoseX}~\citep{jiang2025posex} and \textit{PoseBusters}~\citep{buttenschoen2024posebusters} are included to assess the model's performance under distinct data distributions. 

\paragraph{PoseX}
\textit{PoseX} is a docking benchmark including \textbf{718 self-docking} cases (\textit{PoseX-SD}) and \textbf{1312 cross-docking} cases (\textit{PoseX-CD}). 
Self-docking refers to docking a ligand back into the same protein conformation from which it was crystallised and is the appropriate analogue of standard \emph{redocking}-based docking-power evaluation.
Cross-docking docks each ligand into non-cognate conformations of the same protein, introducing receptor mismatch (backbone and side-chain differences) that better reflects scenarios where the exact receptor structure is unavailable; it is therefore treated here as a robustness regime rather than a definition of docking power.
These cases, alongside another benchmark dataset of \textbf{85} complexes, the \textit{Astex Diverse Set}~\citep{hartshorn2007diverse}, are evaluated across \textbf{24} algorithms.
These methods span physics-based docking like AutoDock Vina, machine learning–based docking models such as DiffDock and SurfDock, and AI co-folding frameworks that can implicitly orient ligands by joint protein–ligand structure prediction, e.g. AlphaFold~3.
Each algorithm is treated as an end-to-end docking pipeline executed under its recommended/default settings, including any method-internal preparation, repair, or refinement steps used to produce the final pose.
These default settings include method-specific search and sampling budgets (e.g., number of generated poses/samples and method-internal ranking procedures), which can materially affect docking quality and therefore shape the induced solver-performance landscape on which algorithm selection is defined. PoseX evaluates the \emph{top-1 ranked} pose returned by each pipeline, and documents per-method inference/sampling settings; accordingly, sampling budget is treated here as part of the solver protocol definition rather than a controlled variable. Standardising budgets across methods could change solver rankings and is naturally subsumed under protocol-aware selection, rather than the present out-of-the-box pipeline selection setting.

PoseX additionally provides an optional post-processing step termed \textbf{relaxation}, intended to correct steric clashes and strained local geometries by locally adjusting atom positions~\citep{jiang2025posex}.
To separate the effect of this universal wrapper from solver-specific processing, results are reported both on the native pipeline outputs (\emph{no post-processing}) and after applying relaxation once to each pipeline’s output pose; the associated changes in docking RMSD and PoseBusters-validity are summarised in Appendix~\ref{app:data_supplements}.
Overall, PoseX yields a benchmark that captures both near-native and cross-conformation docking scenarios while retaining comparability across diverse algorithm families.

\paragraph{PoseBusters}
The \textit{PoseBusters} (V2) benchmark provides a comprehensive validation framework for ligand poses. Its focus extends beyond RMSD accuracy, prioritizing the physical plausibility of docking outputs.
It contains \textbf{428} complexes and provides predictions from \textbf{7} modern docking models. Alongside with the standard output poses, a unified \textbf{molecular-mechanics energy minimisation} post-processing step is applied to generate a mirror group of refined poses. During this process, the ligand is optimised with OpenMM using the AMBER ff14SB (protein) and SMIRNOFF/Sage (ligand) force fields while the receptor is held fixed, producing an MM-minimised (MM-min) pose for standardised comparison~\citep{buttenschoen2024posebusters}. 

We note that several widely used datasets in molecular modelling, such as CASF~\citep{su2018comparative} and CrossDocked 2020~\citep{francoeur2020three}, are unsuitable for algorithm selection. 
These benchmarks provide either single-model submissions, aggregated scoring-function results, or affinity labels, but do not release per-algorithm pose predictions across a solver portfolio. 
Since AS requires instance-level performance profiles for multiple algorithms to define SBS, VBS, and supervised training labels, such datasets cannot be used without recomputing all predictions under a controlled protocol. 
For this reason, our evaluation focuses on \textit{PoseX} and \textit{PoseBusters}.

%%%%%%%%%%%%%%%%%%%%%%%%%%%%%%%%%%
\subsubsection{Preprocessing}\label{subsec:preprocess}
%%%%%%%%%%%%%%%%%%%%%%%%%%%%%%%%%%
Protein inputs are represented at the residue level. For each protein, we encode per-residue embeddings using \textbf{ESM C 600M}~\citep{esm2024cambrian}, yielding a sequence of vectors $\mathbf{x}_P \in \mathbb{R}^{|\text{residues}| \times 1152}$. 
Ligand inputs are encoded using \textbf{ChemBERTa-77M-MLM}~\citep{ahmad2022chemberta} to a single pooled embedding $\mathbf{x}_L \in \mathbb{R}^{384}$. 
Both models supply pretrained molecular language representations.
ESM C provides residue-level vectors that capture evolutionary and structural regularities, offering a high signal-to-noise representation without requiring supervised pocket annotations. ChemBERTa supplies a chemically informed ligand embedding that captures functional groups and local reactivity patterns more reliably than handcrafted descriptors. Using these pretrained models allows MolAS to operate in the small-data regime by leveraging representations already optimised on large biomolecular corpora, reducing the need for task-specific feature engineering.
For each complex, the final input therefore consists of a pair of a protein residue embedding $\mathbf{x}_P$ and a ligand embedding $\mathbf{x}_L$. 

\subsection{Model architecture}\label{subsec:model_architecture}
%%%%%%%%%%%%%%%%%%%%%%%%%%%%%%%%%%
The \textbf{encoder} $f$ consists of an attentional pooler applied to the protein embeddings, followed by a ligand–protein fusion module. The attention layer compresses the residue-level representations into a single protein vector, which is then concatenated with the ligand embedding and linearly projected into a lower-dimensional joint space.

Inspired by the demonstrated stability and variation preservation of ResNet~\citep{he2020resnet}, the fused representation is passed to a residual multilayer perceptron (MLP) that outputs per-algorithm performance scores. The \textbf{decoder} $g$ comprises three consecutive residual blocks, each containing a linear layer, batch normalisation, and ReLU activation with a skip connection. The outputs of all residual blocks are concatenated and processed by a final linear projection head to produce the predicted score vector $\hat{\mathbf s}\in\mathbb R^m$ over the algorithm portfolio.

\subsection{Scoring function}\label{subsec:score}
%%%%%%%%%%%%%%%%%%%%%%%%%%%%%%%%%%%%%%%%
% The algorithms are evaluated and ranked based on a scoring function (for each predicted pose). Our composite score takes account for unlinearly favourable geometric accuracy and physical plausibility by PoseBusters. 

\subsubsection{Geometric accuracy}

\begin{figure}[ht!]
    \centering
    \includegraphics[width=0.65\linewidth]{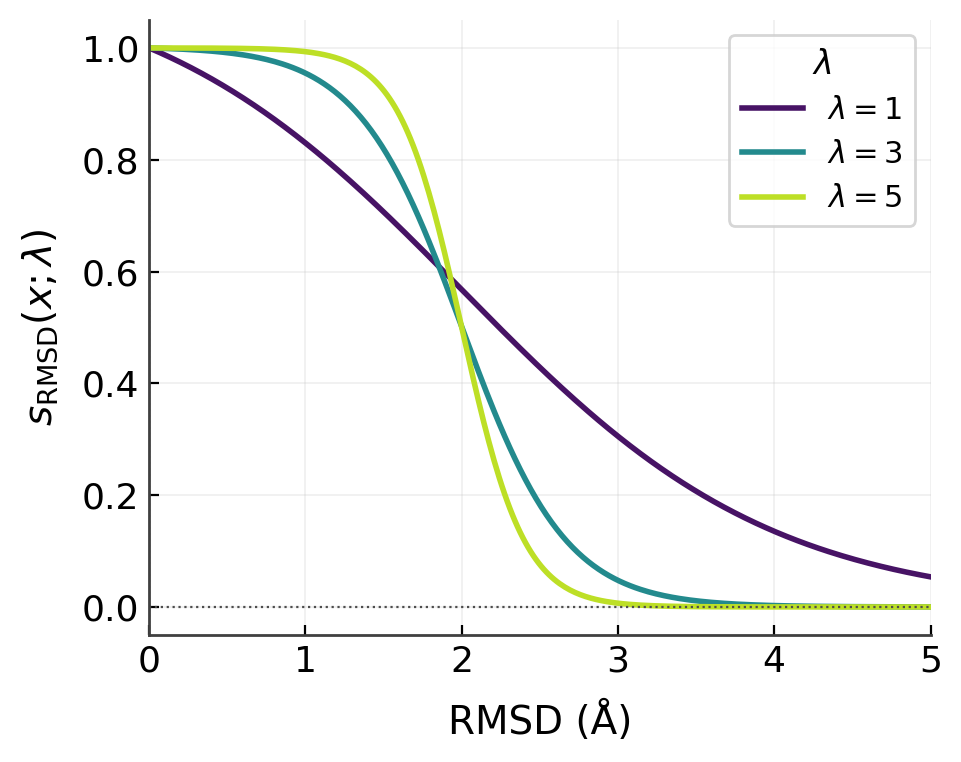}
    \caption{$s_{RMSD}(x;\lambda)$ curves when $\lambda\in\{1, 3, 5\}$.}
    \label{fig:s_rmsd}
\end{figure}

RMSD measures the spatial distance from a predicted pose to the crystal reference in Angstrom (\AA). A pose with RMSD $\leq2$~\AA~ is typically considered geometrically valid. The exponential scoring function used by MC-GNNAS-Dock~\citep{cao2025mc} captures this diminishing returns in RMSD improvements but assigns zero value to all poses with RMSD above a hard threshold ($M = \ln 11 \approx 2.39$), thereby neglecting informative differences between moderately poor poses.
To address this, we propose a smooth sigmoid-based score function:
\begin{equation}
    s_{\mathrm{RMSD}}(x;\lambda) \;=\; \frac{1 + e^{-2\lambda}}{1 + e^{\lambda(x - 2)}}, \qquad \lambda > 0,
\end{equation}
which is centred at the critical threshold of 2~\AA, where the maximum gradient occurs, ensuring the model rewards borderline improvements most strongly, while gains or losses away from this boundary yield diminishing marginal effect (Fig.~\ref{fig:s_rmsd}). The parameter $\lambda$ controls the sensitivity of the score to changes in RMSD, and is set to $\lambda=3$ with an ablation conducted in \S\ref{subsec:ablation}.

\subsubsection{Physicochemical plausibility.}
The PB-validity score (PoseBusters-validity) of a pose is evaluated by a binary accept–reject gate that mirrors current practice in crystallographic validation:
\begin{equation}\label{eq:pb_score}
   s_{\text{PB}}
  \ =\
  \begin{cases}
    1 \quad & \text{passes all $18$ PoseBusters checks,} \\
    0 & \text{otherwise,}
  \end{cases} 
\end{equation}
which ensures that no geometrically plausible yet chemically impossible pose can receive a decent score. Our final score is then the RMSD score gated by PB-validity
\begin{equation}\label{eq:score}
s = s_{\text{RMSD}} \cdot s_{PB}.
\end{equation}

\begin{algorithm}[t]
  \caption{MolAS}
  \label{alg:molAS}
  \begin{algorithmic}[1]
    \Require Raw data of protein-ligand pairs $\{(P_i, L_i)\}_{i=1}^n$; per-algorithm targets
      $\{\mathbf{s}_i \in \mathbb{R}^m\}_{i=1}^n$;
      docking portfolio $\mathcal{A} = \{a_1,\dots,a_m\}$;
      fusion encoder $f_\theta$ and decoder head $g_\theta$ with parameters $\theta$.
    % \Ensure Trained selector and selection rule.

    \Statex
    \Statex \textbf{Preprocessing}
    \For{each $(P_i, L_i)$}
      \State $\mathbf{x}_{P_i} \gets \text{ESM}(P_i)$
      \State $\mathbf{x}_{L_i} \gets \text{ChemBERTa}(L_i)$
    \EndFor
    \State $\mathcal D \gets \{((\mathbf x_{P_i}, \mathbf x_{L_i}), \mathbf{s}_i)\}_{i=1}^n$

    \Statex
    \Statex \textbf{Training}
    \State Initialize $\theta$.
      \For{mini-batch $\mathcal{B} \subset \mathcal D$}
        \For{$((\mathbf x_P, \mathbf x_L), \mathbf{s}) \in \mathcal{B}$}
          \State $\mathbf{z} \gets f_\theta(\mathbf{x}_P, \mathbf{x}_L)$
          \State $\hat{\mathbf{s}} \gets g_\theta(\mathbf{z}) \in \mathbb{R}^m$
        \EndFor
        \State Compute loss $\mathcal{L}_{\text{BCE}}(\theta; \mathcal{B})$
               between $\{\hat{\mathbf{s}}\}$ and $\{\mathbf{s}\}$.
        \State $\theta \gets \theta - \eta \nabla_\theta \mathcal{L}_{\text{BCE}}$
      \EndFor

    \Statex
    \Statex \textbf{Inference (selection)}
    \Function{S}{$\mathbf{x}_P,\mathbf{x}_L$}
      \State $\hat{\mathbf{s}} \gets g_\theta(f_\theta(\mathbf{x}_P, \mathbf{x}_L)) \in \mathbb{R}^m$
      \State $j^\star \gets \arg\max_{j \in \{1,\dots,m\}} \hat{s}_j$
      \State \Return $a_{j^\star}$
    \EndFunction

  \end{algorithmic}
\end{algorithm}

\subsection{Training objective}

MolAS is trained to predict per-algorithm gated score $s_a = s_{\mathrm{RMSD},a}\cdot s_{\mathrm{PB},a}$. Since multiple algorithms may simultaneously achieve high scores on the same complex, outputs are modelled independently per algorithm using sigmoid activations rather than a softmax normalisation across algorithms.

The loss uses binary cross-entropy with logits as a calibration objective for these bounded, multi-label targets:
\[
\mathcal{L}_{\text{BCE}}(\hat{\mathbf s}, \mathbf s)
= - \sum_{a \in \mathcal{A}}
\Big[\, s_a \log\sigma(\hat{s}_a)
      + (1 - s_a)\log\bigl(1 - \sigma(\hat{s}_a)\bigr)
\Big],
\]
where $\hat{s}_a$ is the logit output and $\sigma$ is the logistic sigmoid.
BCE-with-logits is a proper scoring rule for Bernoulli targets and naturally supports soft labels $s_a\in[0,1]$, yielding stable optimisation for bounded success-type scores.
Multi-class cross-entropy is not appropriate because it would enforce competition between algorithms via normalisation, even when several solvers achieve similarly high scores on the same instance.
Mean-squared error regression was also tested as a direct regression objective for $\mathbf{s}$ but did not yield systematic improvements in this setting; ranking-aware alternatives (pairwise logistic and listwise NDCG@3) were explored as auxiliary objectives (Table~\ref{tab:ablation_architecture}) and likewise did not provide consistent gains.
Training minimises the mean loss over mini-batches.

The procedures of MolAS are summarised in Alg.~\ref{alg:molAS}.

\subsection{Experimental Setup}

\subsubsection{Implementation details}
All models are implemented in \texttt{Python} using \texttt{PyTorch}. Docking outputs for PoseX and Astex are obtained from the official PoseX benchmark release~\citep{jiang2025posex}, while those for PoseBusters are taken from the official PoseBusters release~\citep{buttenschoen2024posebusters}. The code and pretrained weights of the ESM-C 600M model are sourced from~\cite{evolutionaryscale_2024}. Models are trained using the Adam optimiser with a cosine learning-rate schedule and early stopping based on validation accuracy. All experiments are conducted on a dual-socket server equipped with two AMD EPYC 7763 CPUs and eight NVIDIA RTX 3090 GPUs, running \texttt{Ubuntu 22.04.5 LTS}.

\subsubsection{Reporting protocol}
We evaluate MolAS under two standard AS baselines. The \textbf{Single Best Solver (SBS)} is the single docking method achieving the highest mean performance on the tested benchmark. The \textbf{Virtual Best Solver (VBS)} is an oracle that chooses the highest-scoring method for each complex, i.e., the theoretical maximum or limit. In addition to absolute accuracy, we report the fraction of the \textbf{VBS-SBS gap closed}, denoted as $\%\frac{\text{AS}-\text{SBS}}{\text{VBS}-\text{SBS}}$, which quantifies the actionable improvement achieved by MolAS relative to the upper bound imposed by the portfolio.

For interpretability, some results involve \textbf{Selected@K} and \textbf{VBS@K}, which summarise the behaviour of the selector beyond the top choice. Here, Selected@K denotes the \emph{algorithm} ranked $K^{th}$ by MolAS selection frequency over the tested benchmark, e.g., the second most frequently selected algorithm when $K=2$. VBS@K denotes a virtual \emph{selector} that selects the $K^{th}$ best algorithm for each instance.

Two evaluation regimes are used. For in-domain learning, each benchmark is evaluated using a 5-fold split with an 8–2 train–test ratio per fold; results report the mean over all folds. For cross-benchmark learning, the entire source benchmark is used for training and the entire target benchmark for testing, with no overlap between complexes. All metrics are reported for success rates of both strict (\text{RMSD} $\leq 1$~\AA~ \& PB-valid) and relaxed (\text{RMSD} $\leq 2$~\AA~ \& PB-valid) success criteria.

\section{Results}
\label{sec:results}

We report MolAS' in-domain performance (\S\ref{subsec:in_domain}), analyse its failure cases and their underlying causes (\S\ref{subsec:why_fail}), and evaluate cross-benchmark generalisation (\S\ref{subsec:cross_bench}). 
We further compare MolAS with the prior GNN-based MC-GNNAS-Dock framework (\S\ref{subsec:comparison_mc_gnnas_dock}) and provide architectural and data-driven ablations (\S\ref{subsec:ablation}).

\subsection{Benchmark Performance}

% Detailed performance are displayed in Fig.~\ref{fig:representative_result}, and Tables~\ref{tab:in_domain_results,tab:cross_benchmark_results}.

\begin{table*}[ht!]
\centering
\caption{
Averaged 5-fold MolAS performance v.s. SBS across benchmarks.
Columns list PoseBusters-validated pose rates within 1~\AA\ and 2~\AA\ RMSD, and the VBS–SBS gap closed.
Bold marks improvement over SBS, and an improvement with * indicates $p<0.05$ (paired test between MolAS and SBS).
}
\setlength{\tabcolsep}{6pt}
\hspace*{-8em}
\begin{tabular}{l|c|c|ccc|ccc}
\toprule
\textbf{Dataset}& \textbf{Post-process}&\textbf{SBS}& 
\multicolumn{3}{c}{$(\%) \text{RMSD}\leq 1\,\text{\AA}\ \&\ \text{PB-valid}$} & \multicolumn{3}{c}{$(\%) \text{RMSD}\leq 2\,\text{\AA}\ \&\ \text{PB-valid}$ } \\
\cmidrule(lr){4-6}\cmidrule(lr){7-9}& && SBS & MolAS& $\%\frac{AS-SBS}{VBS-SBS}$ & SBS & MolAS& $\%\frac{AS-SBS}{VBS-SBS}$  \\
 \midrule \multicolumn{9}{l}{\textbf{With-and-without relaxation trained together}} \\ \midrule
 MOAD-curated&Mixed&Uni-Mol& 43.41& \textbf{50.01}*& \textbf{17.42}& 68.17& \textbf{74.68}*&\textbf{22.40} \\
 PoseX + Astex&Mixed&SurfDock (relax)& 50.24& \textbf{58.61}*& \textbf{21.42}& 74.08& \textbf{79.8}*&\textbf{24.24} \\
 PoseX-SD&Mixed&SurfDock (relax)& 50.08& \textbf{51.62}& \textbf{3.66}& 74.21& \textbf{74.63}&\textbf{1.67} \\
 PoseX-CD&Mixed&SurfDock (relax)& 49.08& \textbf{66.39}*&  \textbf{45.50}&  73.02& \textbf{87.50}*&\textbf{61.28} \\
 PoseBusters&Mixed&AutoDock& 34.34& \textbf{36.69}& \textbf{8.90}& 51.17& \textbf{54.91}&\textbf{11.87} \\
 \midrule \multicolumn{9}{l}{\textbf{With-and-without relaxation trained separately}} \\ \midrule
 \multirow{2}{*}{PoseX + Astex}&No&Boltz1x& 42.48& \textbf{57.05}*& \textbf{32.12}& 61.02& \textbf{76.58}*&\textbf{42.67} \\
 & Relaxation& SurfDock (relax)& 50.24& \textbf{58.94}*& \textbf{23.32}& 74.08& \textbf{80.84}*&\textbf{28.99} \\ \hline
 \multirow{2}{*}{PoseX-SD}&No&Uni-Mol& 40.73& \textbf{43.51}& \textbf{5.56}& 60.53& \textbf{62.62}&\textbf{5.41} \\
 & Relaxation& SurfDock (relax)& 50.08& \textbf{50.22}& \textbf{0.37}& 74.21& 73.93&-1.10 \\\hline
 \multirow{2}{*}{PoseX-CD}&No&Boltz1x& 44.82& \textbf{61.66}*&  \textbf{41.48}&  65.16& \textbf{83.38}*&\textbf{58.29} \\
 & Relaxation& SurfDock (relax)& 49.08& \textbf{66.99}*& \textbf{49.27}& 73.02& \textbf{88.41}*&\textbf{66.43} \\\hline
 \multirow{2}{*}{PoseBusters}&No&AutoDock& 34.34& \textbf{36.44}& \textbf{11.67}& 51.17& \textbf{53.97}&\textbf{9.83} \\
 & MM-min& Gold (MM-min)& 27.58& 26.87& -3.24& 45.10& \textbf{45.81}&\textbf{2.34} \\
\bottomrule
\end{tabular}
\label{tab:in_domain_results}
\end{table*}

\subsubsection{In-domain learning}\label{subsec:in_domain}

Table~\ref{tab:in_domain_results} reports within-distribution performance across benchmarks. On the composite \textit{MOAD-curated} set, MolAS attains success rates of \textbf{50.01\%} under the strict (RMSD $\leq 1$~\AA\ \& PB-valid) criterion and \textbf{74.68\%} under the relaxed (RMSD $\leq 2$~\AA\ \& PB-valid) criterion, improving over the SBS (Uni-Mol Docking V2) by \textbf{6.6\%} and \textbf{6.51\%} respectively. These gains close roughly \textbf{17--23\% }of the VBS--SBS gap. Improvements are more pronounced on \textit{PoseX-CD}, where MolAS closes approximately \textbf{45\%} and \textbf{61\%} of the gap under the strict and relaxed criteria, consistent with cross-docking acting as a stronger receptor-mismatch stress test that induces larger shifts in solver rankings. On the hybrid \textit{PoseX + Astex} dataset, absolute gains of \textbf{8.38\%} (strict) and \textbf{5.72\%} (relaxed) correspond to \textbf{21--24\%} gap closure, indicating robust behaviour under substantial structural and chemical diversity when trained on comparably heterogeneous data. All these improvements above SBS are statistically significant at $p < 0.05$ (paired test).

The improvements on \textit{PoseBusters} from (\textbf{34.3\%}, \textbf{51.2\%}) to (\textbf{36.7\%}, \textbf{54.9\%}) are relatively marginal. A similar pattern is also seen on \textit{PoseX-SD}, where MolAS performs closely to the SBS -- (\textbf{50.1\%}, \textbf{74.2\%}) v.s. (\textbf{51.6\%}, \textbf{74.6\%}). 
In these regimes, the VBS--SBS gap is comparatively smaller and solver hierarchies are more stable, leaving less headroom for selection-driven gains; correspondingly, MolAS tends to default to the dominant solver more often, which is analysed in Sec.~\ref{subsec:why_fail}.

To further quantify the effect of relaxation, we compared models trained on mixed versus post-process-separated datasets using the VBS--SBS gap closed as the metric of interest, as SBS performance varies between regimes. The post-process flag indicates similar trends.
For the hybrid \textit{PoseX~+~Astex} benchmark, separating the relaxation regimes yielded higher performance in both subsets, suggesting that independent modelling of (relax) and un(relax) outputs slightly improves consistency. 
For \textit{PoseX-SD}, non-uniform trends are observed: the non-(relax) regime yields an incremental improvement relative to the mixed baseline, while the (relax) regime is marginally worse; given the small magnitude, these are rather considered as non-systematic fluctuations.
\textit{PoseX-CD} maintains strong performance across all conditions, again with only minor fluctuations that are not systematic across folds. 
In contrast, on \textit{PoseBusters}, MolAS performs comparably to mixed data in the non-(MM-min) subset but shows a marked drop under the (MM-min) regime, likely reflecting the different energy-minimization procedure used in that dataset.

Overall, MolAS outperforms the SBS over most tested scenarios and remains robust across post-processing regimes, while exhibiting clear protocol-specific failure patterns (e.g., on \textit{PoseX-SD}), which are investigated subsequently.

\begin{figure*}[ht!]
\centering
\vspace*{-2em}
\hspace*{-7em}
\begin{tikzpicture}
% ===============================
% COLOR SCHEME
% ===============================
\definecolor{vir1}{HTML}{de4968}
\definecolor{vir2}{HTML}{fa7d5e}
\definecolor{vir3}{HTML}{fecf92}

% ===============================
% RIGHT SIDE (Hyper DMRs)
% ===============================
\begin{axis}[
    name=like,
    scale only axis,
    xbar,
    width=3.7cm, height=14cm,
    bar width=5pt,
    xmin=0, xmax=100,
    xlabel={(\%) $\text{RMSD}\leq 2\,\text{\AA}\ \&\ \text{PB-valid}$},
    xlabel style={font=\scriptsize},
    % symbolic y coords={
    %     MOADcur,PXAstex,PXSD,PXCD,PB,
    %     PXAstexNoRel,PXAstexRel,
    %     PXSDNoRel,PXSDRel,
    %     PXCDNoRel,PXCDRel,
    %     PBNoRel,PBRel},
    symbolic y coords={
        PXCDRel, 
        PXCD,
        PXCDNoRel,
        PXAstexNoRel,
        PXAstexRel,
        PXAstex,
        MOADcur,
        PB,
        PBNoRel,
        PXSDNoRel,
        PBRel,
        PXSD,
        PXSDRel},
    yticklabels={},
    ytick=data,
    y dir=reverse,
    axis y line=none,
    axis x line=bottom,
    xmajorgrids=true,
    major grid style={dashed, thin, black!30},
    enlarge y limits=0.05,
    clip=false,
    nodes near coords,
    point meta=x,
    nodes near coords align={horizontal},
    every node near coord/.append style={
        anchor=west, font=\tiny, color=black
    }
]

\addplot+[fill=vir3!80, draw=vir3!95] coordinates {
    (42.87,MOADcur)
    (57.428,PXAstex) 
    (61.788,PXSD)
    (62.5,PXCD) 
    (35.982,PB)
    (57.90,PXAstexNoRel) 
    (74.078,PXAstexRel)
    (60.53,PXSDNoRel) 
    (61.788,PXSDRel)
    (54.648,PXCDNoRel) 
    (73.016,PXCDRel)
    (14.026,PBNoRel) 
    (35.982,PBRel)};
\addplot+[fill=vir2!70, draw=vir2!90] coordinates {
    (53.66,MOADcur) 
    (61.02,PXAstex) 
    (52.3,PXSD)
    (60.67,PXCD) 
    (48.364,PB)
    (61.022,PXAstexNoRel) 
    (59.462,PXAstexRel)
    (59.558,PXSDNoRel) 
    (51.88,PXSDRel)
    (65.162,PXCDNoRel) 
    (60.67,PXCDRel)
    (48.364,PBNoRel) 
    (44.15,PBRel)};
\addplot+[fill=vir1!70, draw=vir1!90] coordinates {
    (68.16,MOADcur) 
    (74.078,PXAstex) 
    (74.21,PXSD)
    (58.53,PXCD) 
    (51.17,PB)
    (54.778,PXAstexNoRel) 
    (57.428,PXAstexRel)
    (52.3,PXSDNoRel) 
    (74.21,PXSDRel)
    (58.534,PXCDNoRel) 
    (62.5,PXCDRel)
    (51.166,PBNoRel) 
    (45.10,PBRel)};

\node[xshift=-11.7cm,align=center,font=\scriptsize] at (axis cs:0,MOADcur) {\textit{MOAD-curated}}; 
\node[xshift=-11.7cm,align=center,font=\scriptsize] at (axis cs:0,PXAstex) {\textit{PoseX + Astex}\\ \textit{(mixed)}};
\node[xshift=-11.7cm,align=center,font=\scriptsize] at (axis cs:0,PXSD) {\textit{PoseX-SD}\\ \textit{(mixed)}};
\node[xshift=-11.7cm,align=center,font=\scriptsize] at (axis cs:0,PXCD) {\textit{PoseX-CD}\\ \textit{(mixed)}}; 
\node[xshift=-11.7cm,align=center,font=\scriptsize] at (axis cs:0,PB) {\textit{PoseBusters}\\ \textit{(mixed)}};
\node[xshift=-11.7cm,align=center,font=\scriptsize] at (axis cs:0,PXAstexNoRel) {\textit{PoseX + Astex}\\ \textit{(no relax)}};
\node[xshift=-11.7cm,align=center,font=\scriptsize] at (axis cs:0,PXAstexRel) {\textit{PoseX + Astex}\\ \textit{(relax)}}; 
\node[xshift=-11.7cm,align=center,font=\scriptsize] at (axis cs:0,PXSDNoRel) {\textit{PoseX-SD}\\ \textit{(no relax)}};
\node[xshift=-11.7cm,align=center,font=\scriptsize] at (axis cs:0,PXSDRel) {\textit{PoseX-SD}\\ \textit{(relax)}};
\node[xshift=-11.7cm,align=center,font=\scriptsize] at (axis cs:0,PXCDNoRel) {\textit{PoseX-CD}\\ \textit{(no relax)}};
\node[xshift=-11.7cm,align=center,font=\scriptsize] at (axis cs:0,PXCDRel,) {\textit{PoseX-CD}\\ \textit{(relax)}}; 
\node[xshift=-11.7cm,align=center,font=\scriptsize] at (axis cs:0,PBNoRel) {\textit{PoseBusters}\\ \textit{(no MM-min)}};
\node[xshift=-11.7cm,align=center,font=\scriptsize] at (axis cs:0,PBRel) {\textit{PoseBusters}\\ \textit{(MM-min)}};

\node[anchor=east,xshift=-8.4cm,align=right,font=\tiny] at (axis cs:0,MOADcur)
{\textbf{Uni-Mol} \\ SurfDock \\ GNINA};
\node[anchor=east,xshift=-8.4cm,align=right,font=\tiny] at (axis cs:0,PXAstex)
{\textbf{SurfDock (relax)} \\ Boltz1x \\ Chai (relax)};
\node[anchor=east,xshift=-8.4cm,align=right,font=\tiny] at (axis cs:0,PXSD)
{\textbf{SurfDock (relax)} \\ Boltz1x \\ GNINA (relax)};
\node[anchor=east,xshift=-8.4cm,align=right,font=\tiny] at (axis cs:0,PXCD)
{AlphaFold 3 \\ Chai (relax) \\ AlphaFold 3 (relax)};
\node[anchor=east,xshift=-8.4cm,align=right,font=\tiny] at (axis cs:0,PB)
{\textbf{AutoDock} \\ Gold \\ DiffDock (MM-min)};
\node[anchor=east,xshift=-8.4cm,align=right,font=\tiny] at (axis cs:0,PXAstexNoRel)
{AlphaFold 3 \\ \textbf{Boltz1x} \\ Uni-Mol};
\node[anchor=east,xshift=-8.4cm,align=right,font=\tiny] at (axis cs:0,PXAstexRel)
{Chai (relax) \\ AlphaFold 3 (relax) \\ \textbf{SurfDock (relax)}};
\node[anchor=east,xshift=-8.4cm,align=right,font=\tiny] at (axis cs:0,PXSDNoRel)
{Boltz1x \\ GNINA \\ \textbf{Uni-Mol}};
\node[anchor=east,xshift=-8.4cm,align=right,font=\tiny] at (axis cs:0,PXSDRel)
{\textbf{SurfDock (relax)} \\ Boltz1x (relax) \\ GNINA (relax)};
\node[anchor=east,xshift=-8.4cm,align=right,font=\tiny] at (axis cs:0,PXCDNoRel)
{AlphaFold 3 \\ \textbf{Boltz1x} \\ Uni-Mol};
\node[anchor=east,xshift=-8.4cm,align=right,font=\tiny] at (axis cs:0,PXCDRel)
{AlphaFold 3 (relax) \\ Chai (relax) \\ \textbf{SurfDock (relax)}};
\node[anchor=east,xshift=-8.4cm,align=right,font=\tiny] at (axis cs:0,PBNoRel)
{\textbf{AutoDock} \\ Gold \\ DiffDock};
\node[anchor=east,xshift=-8.4cm,align=right,font=\tiny] at (axis cs:0,PBRel)
{\textbf{Gold (MM-min)} \\ AutoDock (MM-min) \\ DiffDock (MM-min)};

\end{axis}

% ===============================
% Middle SIDE (Hyper DMRs)
% ===============================
\begin{axis}[
    name=like,
    at={(like.north west)},anchor=north east,xshift=-0.5cm,
    scale only axis,
    xbar,
    width=3.7cm, height=14cm,
    bar width=5pt,
    xmin=0, xmax=100,
    xlabel={(\%) VBS pick proportion},
    xlabel style={font=\scriptsize},
    % symbolic y coords={
    %     MOADcur,PXAstex,PXSD,PXCD,PB,
    %     PXAstexNoRel,PXAstexRel,
    %     PXSDNoRel,PXSDRel,
    %     PXCDNoRel,PXCDRel,
    %     PBNoRel,PBRel},
    symbolic y coords={
        PXCDRel, 
        PXCD,
        PXCDNoRel,
        PXAstexNoRel,
        PXAstexRel,
        PXAstex,
        MOADcur,
        PB,
        PBNoRel,
        PXSDNoRel,
        PBRel,
        PXSD,
        PXSDRel},
    yticklabels={},
    ytick=data,
    y dir=reverse,
    axis y line=none,
    axis x line=bottom,
    xmajorgrids=true,
    major grid style={dashed, thin, black!30},
    enlarge y limits=0.05,
    clip=false,
    nodes near coords,
    point meta=x,
    nodes near coords align={horizontal},
    every node near coord/.append style={
        anchor=west, font=\tiny, color=black
    }
]

\addplot+[fill=vir3!80, draw=vir3!95] coordinates {
    (15.854,MOADcur)
    (3.9262,PXAstex) 
    (2.6499,PXSD)
    (4.9542,PXCD) 
    (15.4206,PB)
    (5.818,PXAstexNoRel) 
    (18.3538,PXAstexRel)
    (6.555,PXSDNoRel) 
    (4.60251,PXSDRel)
    (5.41,PXCDNoRel) 
    (18.902,PXCDRel)
    (8.64,PBNoRel) 
    (19.63,PBRel)};
\addplot+[fill=vir2!70, draw=vir2!90] coordinates {
    (36.0805,MOADcur) 
    (2.7909,PXAstex) 
    (2.092,PXSD)
    (5.1067,PXCD) 
    (16.355,PB)
    (6.575,PXAstexNoRel) 
    (12.725,PXAstexRel)
    (5.4393,PXSDNoRel) 
    (3.06834,PXSDRel)
    (7.6981,PXCDNoRel) 
    (8.918,PXCDRel)
    (34.81,PBNoRel) 
    (20.33,PBRel)};
\addplot+[fill=vir1!70, draw=vir1!90] coordinates {
    (30.2296,MOADcur) 
    (11.8732,PXAstex) 
    (10.3208,PXSD)
    (12.4238,PXCD) 
    (17.523,PB)
    (19.7729,PXAstexNoRel) 
    (7.569,PXAstexRel)
    (5.29986,PXSDNoRel) 
    (18.8285,PXSDRel)
    (22.637,PXCDNoRel) 
    (14.7866,PXCDRel)
    (37.62,PBNoRel) 
    (25.70,PBRel)};

\end{axis}

% ===============================
% LEFT SIDE (Hypo DMRs)
% ===============================
\begin{axis}[
    at={(like.north west)},anchor=north east,xshift=-0.5cm,
    scale only axis,
    xbar,
    % x dir=reverse,
    width=3.7cm, height=14cm,
    bar width=5pt,
    xmin=0, xmax=100,
    xlabel={(\%) MolAS pick proportion},
    xlabel style={font=\scriptsize},
    ytick=data,
    y dir=reverse,
    % symbolic y coords={
    %     MOADcur,PXAstex,PXSD,PXCD,PB,
    %     PXAstexNoRel,PXAstexRel,
    %     PXSDNoRel,PXSDRel,
    %     PXCDNoRel,PXCDRel,
    %     PBNoRel,PBRel},
    % yticklabels={
    %     {MOAD-curated},
    %     {PoseX + Astex},
    %     {PoseX-SD},
    %     {PoseX-CD},
    %     {PoseBusters},
    %     {PoseX + Astex\\ (no relaxation)},
    %     {PoseX + Astex\\ (relax)},
    %     {PoseX-SD\\ (no relaxation)},
    %     {PoseX-SD\\ (relax)},
    %     {PoseX-CD\\ (no relaxation)},
    %     {PoseX-CD\\ (relax)},
    %     {PoseBusters\\ (no relaxation)},
    %     {PoseBusters\\ (relax)}
    % },
    symbolic y coords={
        PXCDRel, 
        PXCD,
        PXCDNoRel,
        PXAstexNoRel,
        PXAstexRel,
        PXAstex,
        MOADcur,
        PB,
        PBNoRel,
        PXSDNoRel,
        PBRel,
        PXSD,
        PXSDRel},
    yticklabels={
        {PoseX-CD\\ (relax)},
        {PoseX-CD},
        {PoseX-CD\\ (no relaxation)},
        {PoseX + Astex\\ (no relaxation)},
        {PoseX + Astex\\ (relax)},
        {PoseX + Astex},
        {MOAD-curated},
        {PoseBusters},
        {PoseBusters\\ (no relaxation)},
        {PoseX-SD\\ (no relaxation)},
        {PoseBusters\\ (relax)},
        {PoseX-SD},
        {PoseX-SD\\ (relax)}
    },
    yticklabel style={font=\small, align=center, text width=3cm},
    axis y line=none,
    axis x line=bottom,
    xmajorgrids=true,
    major grid style={dashed, thin, black!30},
    enlarge y limits=0.05,
    legend style={
        at={(1.25,-0.08)},
        anchor=north,
        legend columns=4,
        /tikz/every even column/.append style={column sep=0.5cm},
        font=\small
    },
    legend reversed=true,
    nodes near coords,
    point meta=x,
    nodes near coords align={horizontal},
    every node near coord/.append style={
        anchor=west, font=\tiny, color=black
    }
]

\addplot+[
    fill=vir3!80,
    draw=vir3!95,
    yshift=10.5pt
] coordinates {
    (11.64,MOADcur) % GNINA
    (7.710,PXAstex) % Chai (relax)
    (4.1841,PXSD) % GNINA (relax)
    (9.7561,PXCD) % AlphaFold 3 (relax)
    (20.5607,PB) % DiffDock (relax)
    (13.3869,PXAstexNoRel)  % Uni-Mol
    (12.2044,PXAstexRel) % SurfDock (relax)
    (17.8522,PXSDNoRel)  % Uni-Mol
    (5.5788,PXSDRel) % GNINA (relax)
    (12.3476,PXCDNoRel)  % Uni-Mol
    (10.2896,PXCDRel) % SurfDock (relax)
    (3.03738,PBNoRel)  % DiffDock
    (25.47,PBRel)}; % DiffDock (relax)
\addplot+[
    fill=vir2!70,
    draw=vir2!90,
    yshift=10.5pt
] coordinates {
    (19.63,MOADcur) % Uni-Mol
    (13.01,PXAstex) % Boltz1x
    (14.5049,PXSD) % Boltz1x
    (10.8232,PXCD)   % Chai (relax)
    (24.0654,PB)  % Gold
    (15.8467,PXAstexNoRel)   % Boltz1x
    (12.5355,PXAstexRel)  % AlphaFold 3 (relax)
    (21.06,PXSDNoRel)   % GNINA
    (7.67085,PXSDRel)  % Boltz1x (relax)
    (15.0152,PXCDNoRel)   % Boltz1x
    (16.0823,PXCDRel)  % Chai (relax)
    (35.2804,PBNoRel)   % Gold
    (29.67,PBRel)}; % AutoDock (relax)
\addplot+[
    fill=vir1!70,
    draw=vir1!90,
    yshift=10.5pt
] coordinates {
    (63.13,MOADcur) % SurfDock
    (14.71,PXAstex)  % SurfDock (relax)
    (65.69,PXSD) % SurfDock (relax)
    (12.04,PXCD)  % AlphaFold 3
    (37.8505,PB) % AutoDock
    (21.48,PXAstexNoRel)  % AlphaFold 3
    (13.5289,PXAstexRel) % Chai (relax)
    (25.523 ,PXSDNoRel)  % Boltz1x
    (76.2901,PXSDRel) % SurfDock (relax)
    (27.0579,PXCDNoRel)  % AlphaFold 3
    (16.7683,PXCDRel) % AlphaFold 3 (relax)
    (61.215,PBNoRel)  % AutoDock
    (41.59,PBRel)}; % Gold (relax)

\addplot+[fill=vir3!150, draw=vir3!150, yshift=-10.5pt, nodes near coords={}] coordinates {
    (4.215,MOADcur) % GNINA
    (0.993,PXAstex) % Chai (relax)
    (0.2789,PXSD) % GNINA (relax)
    (0.9146,PXCD) % AlphaFold 3 (relax)
    (5.3738,PB) % DiffDock (relax)
    (2.507,PXAstexNoRel)  % Uni-Mol
    (3.4059,PXAstexRel) % SurfDock (relax)
    (1.674,PXSDNoRel)  % Uni-Mol
    (0.55788,PXSDRel) % GNINA (relax)
    (2.591,PXCDNoRel)  % Uni-Mol
    (3.811,PXCDRel) % SurfDock (relax)
    (1.168,PBNoRel)  % DiffDock
    (8.88,PBRel)}; % DiffDock (relax)

\addplot+[fill=vir2!150, draw=vir2!150, yshift=-10.5pt, nodes near coords={}] coordinates {
    (9.8459,MOADcur) % Uni-Mol
    (0.709,PXAstex) % Boltz1x
    (0.4184,PXSD) % Boltz1x
    (1.143,PXCD)   % Chai (relax)
    (4.9065,PB)  % Gold
    (2.318,PXAstexNoRel)   % Boltz1x
    (4.068,PXAstexRel)  % AlphaFold 3 (relax)
    (1.534,PXSDNoRel)   % GNINA
    (0.13947,PXSDRel)  % Boltz1x (relax)
    (1.98171,PXCDNoRel)   % Boltz1x
    (2.591,PXCDRel)  % Chai (relax)
    (14.252,PBNoRel)   % Gold
    (6.54,PBRel)}; % AutoDock (relax)

\addplot+[fill=vir1!150, draw=vir1!150, yshift=-10.5pt, nodes near coords={}
] coordinates {
    (24.4416,MOADcur) % SurfDock
    (2.507,PXAstex)  % SurfDock (relax)
    (8.2287,PXSD) % SurfDock (relax)
    (2.21,PXCD)  % AlphaFold 3
    (9.112,PB) % AutoDock
    (8.751,PXAstexNoRel)  % AlphaFold 3
    (2.129,PXAstexRel) % Chai (relax)
    (1.813 ,PXSDNoRel)  % Boltz1x
    (15.2022,PXSDRel) % SurfDock (relax)
    (12.0427,PXCDNoRel)  % AlphaFold 3
    (5.79,PXCDRel) % AlphaFold 3 (relax)
    (25.467,PBNoRel)  % AutoDock
    (10.98,PBRel)}; % Gold (relax)
    
% Reset cycle index before custom legends
\pgfplotsset{cycle list shift=-6}

% % ---- precise selection: triple bar swatch ----
% \matrix[
%     matrix of nodes,
%     row sep=1pt,
%     column sep=6pt,
% ] at (axis cs:0) {

%     % Three small rectangles in one legend cell
%     \node[inner sep=0pt,outer sep=0pt,anchor=base] {
%         \tikz{\draw[fill=vir1!150,draw=vir1!150] (0,0) rectangle (6pt,4pt);}
%         \tikz{\draw[fill=vir2!150,draw=vir2!150] (0,0) rectangle (6pt,4pt);}
%         \tikz{\draw[fill=vir3!150,draw=vir3!150] (0,0) rectangle (6pt,4pt);}
%     }; &
%     \node[anchor=base west,font=\small] {Precise selection}; \\

% };

% \addlegendimage{area legend,style={},
%   legend image code/.code={
%     \path[fill=vir1!150] (0pt,0pt) rectangle (3pt,6pt);
%     \path[fill=vir2!150] (3pt,0pt) rectangle (6pt,6pt);
%     \path[fill=vir3!150] (6pt,0pt) rectangle (9pt,6pt);
%   }
% }
\addlegendimage{
  legend image code/.code={
    \path[fill=vir1!150] (0pt,-3.2pt) rectangle (6pt,3.2pt);
    \path[fill=vir2!150] (6pt,-3.2pt) rectangle (12pt,3.2pt);
    \path[fill=vir3!150] (12pt,-3.2pt) rectangle (18pt,3.2pt);
  }
}
\addlegendentry{Correct selection}

% ---- Selected@3 ----
\addlegendimage{area legend,style={},fill=vir3!80,draw=vir3!95}
\addlegendentry{Selected@3}

% ---- Selected@2 ----
\addlegendimage{area legend,style={},fill=vir2!70,draw=vir2!90}
\addlegendentry{Selected@2}

% ---- Selected@1 ----
\addlegendimage{area legend,style={},fill=vir1!70,draw=vir1!90}
\addlegendentry{Selected@1}

\end{axis}

\end{tikzpicture}
% \caption{The selection frequencies by MolAS (left bar chart) and by the VBS (middle bar chart) and their success rates under the relaxed ($\text{RMSD}\leq x~\text{\AA}\ \&\ \text{PB-valid}$) criterion (right bar chart) of the \textbf{top-3 selected algorithms by MolAS} in the concatenated 5-fold results over benchmarks. The benchmarks are ordered descending based on the VBS-SBS gap closed by MolAS under the relaxed criterion. The correct picks by MolAS (those that accords with the VBS) are coloured in deeper colours and the SBS method for each benchmark is in \textbf{bold}.}
\caption{Selection distribution diagnostics across benchmarks. For each benchmark (rows), the left panel reports MolAS selection frequencies for its top-3 most frequently chosen solvers; the middle panel reports the VBS frequencies for the same solvers; and the right panel reports their realised success rates under the $\mathrm{RMSD}\le 2~\text{\AA}\ \&\ \mathrm{PB}$-valid criterion. Darker segments mark cases where MolAS’ top-1 choice agrees with the VBS, and the SBS for each benchmark is shown in \textbf{bold}. The figure highlights two regimes: (i) \emph{selection collapse}, where MolAS concentrates on a single solver substantially more than the VBS and gains over SBS are negligible, and (ii) \emph{distributed selection}, where MolAS allocates mass across multiple competitive solvers and closes a larger fraction of the VBS--SBS gap even when exact oracle matching is below 50\%.}
\label{fig:in_domain_top_3}
\end{figure*}

\begin{figure*}[h!]

\centering

% ---------- Row 1 ----------
\begin{minipage}[t]{0.48\textwidth}
    \centering
    \textbf{Reliability curves}
    \hspace*{-4em}
    \includegraphics[width=1.2\textwidth]{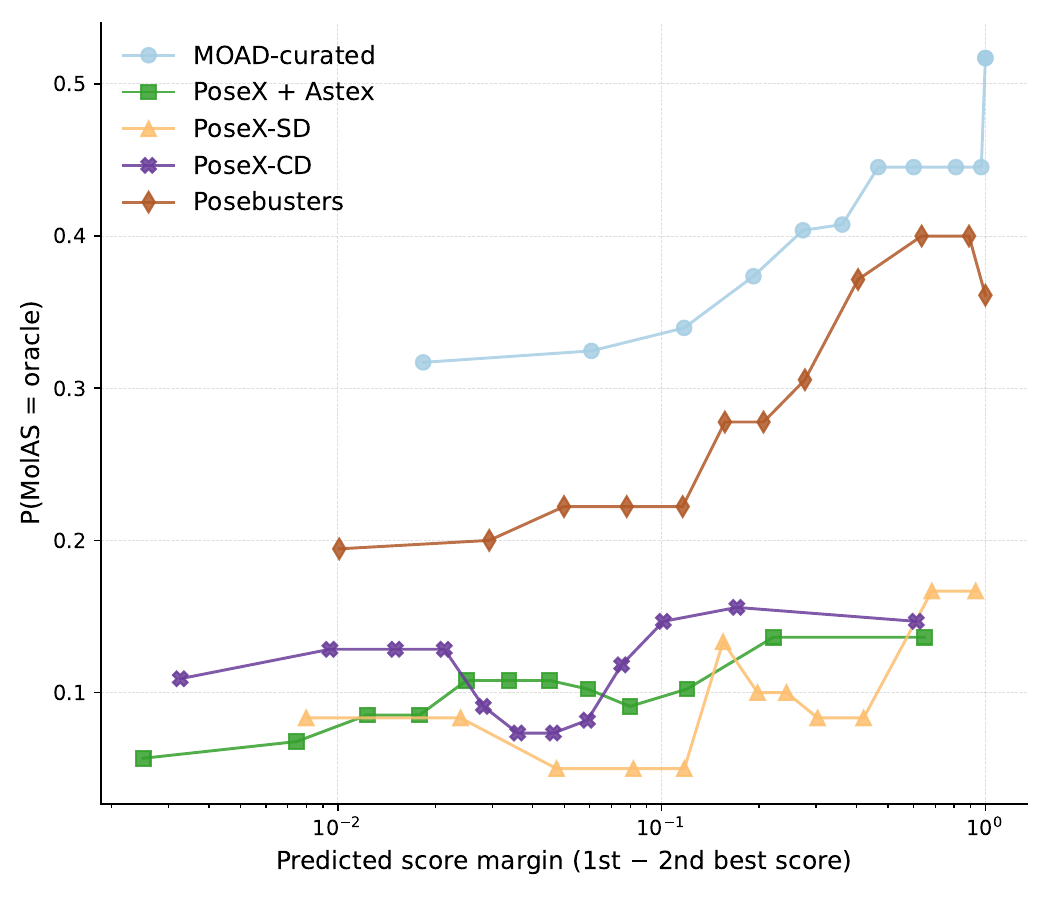}
\end{minipage}
\hfill
\begin{minipage}[t]{0.48\textwidth}
    \centering
    \textbf{Advantage-over-SBS curves}
    \includegraphics[width=1.2\textwidth]{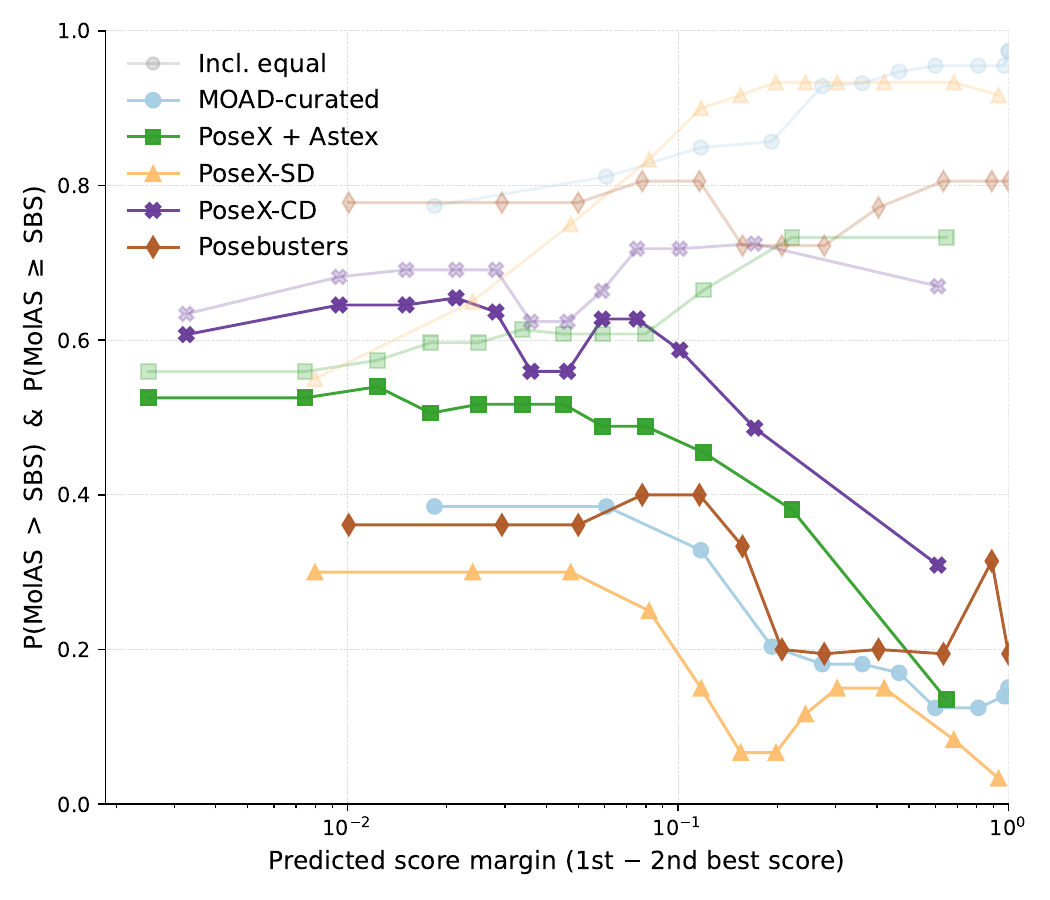}
\end{minipage}
% \caption{(Left) reliability curves showing the probability that MolAS matches the oracle as a function of predicted score margin (12 equal-mass bins). (Right) Advantageness curves obtained by replacing reliability (MolAS accords with VBS) with advantageness over SBS. Solid lines correspond to MolAS outperforming SBS ($s_{\text{MolAS}} > s_{\text{SBS}}$) and faded lines for MolAS performing as well as SBS ($s_{\text{MolAS}} \geq s_{\text{SBS}}$). Thereby, the y-axis indicates the probability that MolAS achieves higher true score than the SBS within each margin bin.}
\caption{Confidence diagnostics via MolAS score margins (12 equal-mass bins). Left: reliability, $\mathbb{P}(\text{MolAS}=\text{VBS})$, as a function of the predicted margin between the top-1 and top-2 solvers. Right: advantageness over SBS, $\mathbb{P}(s_{\text{MolAS}} > s_{\text{SBS}})$ (solid) and $\mathbb{P}(s_{\text{MolAS}} \ge s_{\text{SBS}})$ (faded), over the same bins. Across benchmarks, larger margins tend to increase oracle agreement but often coincide with re-selection of the SBS, so margin is a benchmark-dependent confidence proxy rather than a universally calibrated indicator of improvement over SBS.}
\label{fig:reliability}
\end{figure*}

\subsubsection{Operational boundaries of MolAS}\label{subsec:why_fail}
The behaviour of MolAS deteriorates on PoseX-SD and PoseBusters under the post-processing regimes and in several cross-benchmark settings. We therefore examined these regimes in more detail, (i) focusing on how the selector distributes its choices across the portfolio, (ii) whether those choices are conditionally reliable, and (iii) how the embedding space and algorithm landscape affect the performance. 

Fig.~\ref{fig:in_domain_top_3} characterises a key failure mode of MolAS: \emph{selection collapse}. In difficult regimes such as PoseX-SD (relax) and PoseBusters (MM-min), the selector concentrates most probability mass on a single solver far more aggressively than the VBS does, effectively behaving like a near-constant policy. This collapse is consistent with regimes where (i) the SBS dominates the oracle landscape or (ii) multiple solvers are near-tied, so that the embedding-derived signal is insufficient to support stable instance-wise differentiation; in both cases, little headroom remains for selection gains beyond SBS.

In contrast, when MolAS closes a substantial fraction of the VBS--SBS gap, its selections remain distributed across several competitive solvers and better reflect the oracle’s multi-modality. Notably, exact oracle matching can remain below 50\% while still yielding sizable gains, because the relevant success region is often \emph{flat near the top}: selecting a near-best solver is frequently sufficient when multiple solvers achieve similar composite success. This is consistent with the observation that VBS@2 and VBS@3 already exceed SBS performance on these benchmarks (Appendix~\ref{app:verbose}).

Fig.~\ref{fig:reliability} then links these outcomes to the predicted score margin between the top two solvers. Reliability $\mathbb{P}(\text{MolAS}=\text{VBS})$ generally increases with margin, but the probability of strictly outperforming SBS often decreases at large margins. This indicates that high margins frequently correspond to confident re-selection of the SBS (or another dominant solver) rather than confident identification of a better alternative. Consequently, margin behaves as a partially informative, benchmark-dependent confidence signal: it can reflect oracle agreement on low-entropy regimes, but it is not a reliable monotone indicator of improvement over SBS under high-entropy or near-tie regimes.

We additionally examined whether MolAS failures correlate with gross changes in the embedding space. 
The PCA spectra, t-SNE projections, centroid-distance heatmaps, and quantitative cluster/oracle statistics are reported in Appendix~\ref{app:failure_supplements} (Fig.~\ref{fig:tsne}, Fig.~\ref{fig:inter-algo-distance}, Table~\ref{tab:embed_stats}). 
Briefly, the PCA spectra are broadly similar across benchmarks, indicating no obvious low-rank collapse. 
However, embedding-based separability remains weak overall, and in the most difficult regimes (e.g., PoseX-SD), high oracle diversity coincides with poor separability, consistent with the observed tendency of MolAS to collapse towards an SBS-like strategy.

In summary, MolAS is most effective when multiple algorithms are competitive and occupy at least partially distinct embedding regions. When embeddings fail to separate algorithms, the selector becomes overconfident and defaults to a near-SBS policy. This defines a clear operational boundary for learned docking algorithm selection.

\begin{table*}[ht!]
\centering
\caption{Cross-benchmark results of MolAS performance v.s. SBS (of the test set) across benchmark pairs.
Columns list PoseBusters-validated pose rates within 1~\AA\ and 2~\AA\ RMSD, and the VBS–SBS gap closed.
Bold marks improvement over SBS, and an improvement with $*$ indicates $p<0.05$ (paired test between MolAS and SBS).}
\setlength{\tabcolsep}{6pt}
\hspace*{-9em}
\begin{tabular}{l|c|c|c|ccc|ccc}
\toprule
\textbf{Train} & \textbf{Test}  & \textbf{Post-Process}&\textbf{SBS}& 
\multicolumn{3}{c}{$\%\leq 1\,\text{\AA}\ \&\ \text{PB-valid}$} & \multicolumn{3}{c}{$\%\leq 2\,\text{\AA}\ \&\ \text{PB-valid}$} \\
\cmidrule(lr){5-7}\cmidrule(lr){8-10}
 & & && SBS & MolAS& $\%\frac{AS-SBS}{VBS-SBS}$ & SBS & MolAS& $\%\frac{AS-SBS}{VBS-SBS}$ \\
 \midrule \multicolumn{10}{l}{\textbf{With-and-without relaxation trained together}} \\ \midrule
 PoseX-SD &PoseX-CD &   Mixed&SurfDock (relax)& 49.09& \textbf{50.53}& \textbf{3.79}& 73.02& \textbf{74.01}&\textbf{4.19}\\
 PoseX-CD &PoseX-SD &  Mixed&SurfDock (relax)& 50.07& 44.63& -12.96& 74.2& 62.2&-47.81\\
 PoseX-SD & Astex &  Mixed&SurfDock (relax)& 69.41& 65.88& -11.54& 89.41& 87.06&-22.19\\
 PoseX-CD & Astex &  Mixed&SurfDock (relax)& 69.41& 65.88& -11.54& 89.41& 82.53&-66.67\\
 \midrule \multicolumn{10}{l}{\textbf{With-and-without relaxation trained separately}} \\ \midrule
 \multirow{2}{*}{PoseX-SD} &  \multirow{2}{*}{PoseX-CD} &  No&Boltz1x& 44.82& 40.7& -10.14& 65.17& 59.98&-16.61\\
 & & Relaxation& SurfDock (relax)& 49.09& \textbf{50.0}& \textbf{2.50}& 73.02& 72.94&-0.35\\ \hline
 \multirow{2}{*}{PoseX-CD} &  \multirow{2}{*}{PoseX-SD} &  No&Uni-Mol& 40.73& 38.91& -3.64& 60.53& 57.04&-9.03\\
 & & Relaxation& SurfDock (relax)& 50.07& 42.68& -18.53& 74.2& 59.55&-58.37\\ \hline
 \multirow{2}{*}{PoseX-SD} & \multirow{2}{*}{Astex} &  No&Uni-Mol& 77.65& 58.82& -84.25& 85.88& 75.29&-75.00\\
 & & Relaxation& SurfDock (relax)& 69.41& 67.06& -7.68& 69.41& 87.06&-22.19\\ \hline
 \multirow{2}{*}{PoseX-CD} & \multirow{2}{*}{Astex} &  No&Uni-Mol& 77.65& 56.47& -94.77& 85.88& 70.59&-108.29\\
 & & Relaxation& SurfDock (relax)& 69.41& 67.06& -7.68& 69.41& 81.18&-77.71\\ \hline
\end{tabular}
\label{tab:cross_benchmark_results}
\end{table*}

\begin{table*}[t]
\centering
\caption{Protocol-pair solver-ranking stability. For each protocol pair, Spearman's $\rho$ and Kendall's $\tau_b$ are computed between solver rankings induced by mean per-solver scores. $J_1$ is the top-1 Jaccard overlap, and $\overline{J}=\frac{1}{m}\sum_{k=1}^{m} J_k$ is the mean top-$k$ Jaccard overlap over $k\in\{1,\dots,m\}$.}
\label{tab:protocol_pair_stability}
\begin{tabular}{ll|c|cccc}
\toprule
\textbf{Post-Process} & \textbf{Benchmark pair} & $m$ & Spearman $\rho$ & Kendall $\tau_b$ & $J_1$ & $\overline{J}$ \\
\midrule
% \multicolumn{7}{l}{\textit{Mixed post-processing}} \\
\multirow{3}{*}{Mixed}& (PoseX-SD, PoseX-CD)& 48 & 0.841 & 0.679 & 1.000 & 0.660 \\
& (PoseX-SD, Astex)& 48 & 0.858 & 0.702 & 1.000 & 0.715 \\
& (PoseX-CD, Astex)& 48 & 0.880 & 0.729 & 1.000 & 0.716 \\
\midrule
% \multicolumn{7}{l}{\textit{No post-processing}} \\
\multirow{3}{*}{No}& (PoseX-SD, PoseX-CD)& 24 & 0.843 & 0.710 & 0.000 & 0.647 \\
 & (PoseX-SD, Astex)& 24 & 0.804 & 0.652 & 1.000 & 0.720 \\
 & (PoseX-CD, Astex)& 24 & 0.849 & 0.681 & 0.000 & 0.686 \\
\midrule
% \multicolumn{7}{l}{\textit{Relaxation}} \\
\multirow{3}{*}{Relaxation} & (PoseX-SD, PoseX-CD)& 24 & 0.817 & 0.645 & 1.000 & 0.669 \\
 & (PoseX-SD, Astex)& 24 & 0.857 & 0.710 & 1.000 & 0.731 \\
 & (PoseX-CD, Astex)& 24 & 0.856 & 0.717 & 1.000 & 0.715 \\
\bottomrule
\end{tabular}
\end{table*}

\begin{figure}
    \centering
    \includegraphics[width=0.7\linewidth]{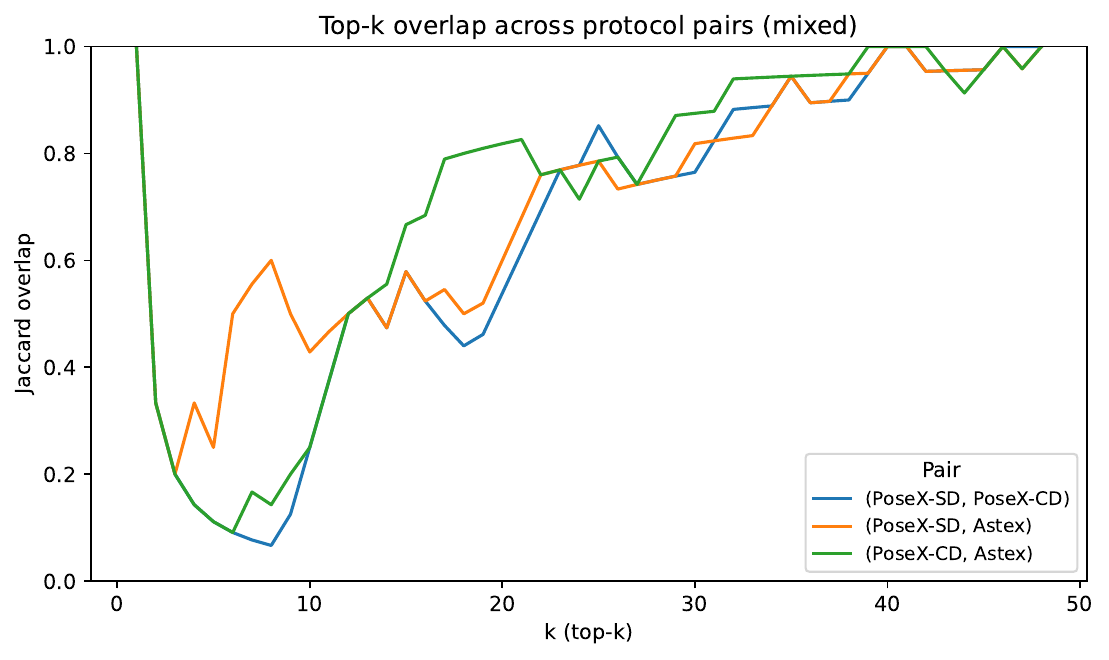}
    \caption{Top-$k$ overlap of solver rankings across docking protocols. Each curve reports the Jaccard similarity between the sets of the top-$k$ solvers under a benchmark pair $(p,q)$, over $k=1,\dots,m$, where $m=|\mathcal{A}|$ is the number of solvers. Higher overlap indicates a more stable solver hierarchy across protocols, while low overlap at small $k$ highlights instability among the highest-ranked solvers.}
    \label{fig:topk_jaccard_mixed}
\end{figure}

\subsubsection{Cross-benchmark generalisation}\label{subsec:cross_bench}
The cross-benchmark results in Table~\ref{tab:cross_benchmark_results} indicate that MolAS does not achieve meaningful generalisation across docking distributions. Although a few \textit{PoseX SD$\rightarrow$CD} transfers produce small positive gap closures (3.79\% and 4.19\% in the mixed regime, 2.50\% in the relax-only regime), none of these improvements reach statistical significance, and all are numerically marginal. Without statistical support, these fluctuations are better interpreted as noise rather than evidence of generalisable behaviour.

In contrast, the majority of cross-distribution transfers show consistent and large degradations, including around $13\%$ to $48\%$ \emph{VBS--SBS gap widening} for \textit{PoseX CD$\rightarrow$SD}, $7\%$ to $84\%$ for \textit{PoseX-SD$\rightarrow$Astex}, and up to $108\%$ gap widening for \textit{PoseX-CD}$\rightarrow$\textit{Astex}. These failures span both the 1~\AA\ and 2~\AA\ thresholds and persist regardless of whether relaxation and non-relaxation data are trained jointly or separately. The overall pattern is that when the target benchmark differs materially in curation and protocol, including the distribution of targets/ligands and the evaluation pipeline, transfer degrades towards the SBS baseline or worse. Consistently, in poorly transferring regimes, the selector's outputs contract onto a small subset of methods (often a single dominant solver), diverging from the oracle distributions of the target benchmarks.

To quantify the protocol dependence underlying these observations, Table~\ref{tab:protocol_pair_stability} reports explicit solver-hierarchy stability diagnostics across benchmark pairs, separately for the mixed, no-post-processing, and relaxation settings. Across protocol pairs, rank correlations of the solver ordering induced by mean scores are moderate-to-high ($\rho\in[0.80,0.88]$, $\tau_b\in[0.64,0.73]$), indicating that the global ranking is not arbitrarily permuted. However, these global correlations do not imply stability of the upper tail that determines oracle labels for algorithm selection. The top-$k$ overlap diagnostics reveal materially weaker agreement among the highest-ranked solvers: the mean top-$k$ overlap $\overline{J}$ ranges from 0.65 to 0.73, and the top-1 overlap $J_1$ is zero for some no-post-processing benchmark pairs, implying that even the identity of the best solver can change across protocols. 

The top-$k$ overlap curves (Figure~\ref{fig:topk_jaccard_mixed}) provide a complementary view of where ranking disagreement concentrates.
Across benchmark pairs, overlap is typically high at $k=1$ (shared best solver in the mixed setting; Table~\ref{tab:protocol_pair_stability}) but drops substantially for small-to-moderate $k$ (e.g., top-5/top-10), before increasing steadily as $k$ grows.
This indicates that protocols agree more on the single best method than on the set of near-best contenders, i.e., protocol shifts primarily reorder the upper tail rather than arbitrarily permuting the full portfolio.
Such near-top reordering is particularly relevant for algorithm selection, because oracle labels are mostly determined by differences among a small set of high-performing solvers; perturbations within this set can therefore reduce transferability even when global rank correlations remain high.
The no-post-processing and relaxation settings exhibit the same qualitative behaviour; full top-$k$ curves for these settings are reported in Appendix~\ref{app:topk-overlap}.

Taken together, Table~\ref{tab:cross_benchmark_results} and Table~\ref{tab:protocol_pair_stability} support a conservative interpretation of the transfer failures: MolAS does not learn an invariant mapping from molecular geometry to the optimal solver under a fixed portfolio; instead, it fits decision boundaries aligned to the training protocol's score landscape and solver hierarchy. When the target protocol induces a different upper-tail hierarchy (even if the global ordering remains broadly correlated), the learned decision surface becomes misaligned with the target oracle, yielding systematic negative transfer. This limitation is intrinsic to portfolio-based docking AS: supervision is defined by protocol-specific performance profiles, and cross-protocol robustness cannot be assumed without additional mechanisms (e.g., explicit domain adaptation, calibration on the target protocol, or selectors designed to be stable under hierarchy perturbations).

Finally, it is useful to distinguish cross-protocol instability from within-protocol oracle diversity. The VBS entropy in Table~\ref{tab:embed_stats} summarises the within-protocol dispersion of oracle winners and indicates that, within each benchmark, multiple solvers can be competitive. The protocol-pair stability diagnostics above instead quantify how the induced hierarchy shifts across protocols; both aspects are relevant, but they capture different failure modes for transfer.

\begin{figure}
\centering
\scalebox{1}{
\begin{tikzpicture}
% --- Viridis color picks ---
\definecolor{vir1}{HTML}{f8765c} % blue-violet
\definecolor{vir2}{HTML}{440f76} % teal
\definecolor{vir3}{HTML}{2a788e} % green
\definecolor{vir4}{HTML}{BDBDBD} % neutral (100% reference)

\begin{axis}[
    xbar,
    y dir=reverse,
    bar width=5pt,
    width=8cm,
    height=17cm,
    xlabel={(\%) $\text{RMSD}\leq 2\,\text{\AA}\ \&\ \text{PB-valid}$},
    xmin=0, xmax=100,
    symbolic y coords={
        MOADcur,
        PXAstex,
        PXSD,
        PXCD,
        PB,
        PXAstexNoRel,
        PXAstexRel,
        PXSDNoRel,
        PXSDRel,
        PXCDNoRel,
        PXCDRel,
        PBNoRel,
        PBRel
    },
    yticklabels={
        {MOAD-curated},
        {PoseX + Astex},
        {PoseX-SD},
        {PoseX-CD},
        {PoseBusters},
        {PoseX + Astex\\ (no relaxation)},
        {PoseX + Astex\\ (relax)},
        {PoseX-SD\\ (no relaxation)},
        {PoseX-SD\\ (relax)},
        {PoseX-CD\\ (no relaxation)},
        {PoseX-CD\\ (relax)},
        {PoseBusters\\ (no relaxation)},
        {PoseBusters\\ (relax)}
    },
    ytick=data,
    yticklabel style={font=\scriptsize, align=center},
    enlarge y limits=0.05,
    nodes near coords,
    point meta=x,
    nodes near coords align={horizontal},
    every node near coord/.append style={
        anchor=east, xshift=1pt, font=\tiny
    },
    legend reversed=true,
    legend style={
        at={(0.4,-0.1)},
        anchor=north,
        legend columns=4,
        /tikz/every even column/.append style={column sep=0.5cm},
        font=\small
    },
    xmajorgrids=true,
    major grid style={dashed, thin, black!30},
]

% --- Reference 100% (vir4) ---
\addplot+[
    fill=vir4!90, draw=vir4!70,
    every node near coord/.append style={text=black, anchor=east, xshift=-1pt}
] coordinates {
    (97.23,MOADcur)
    (97.68,PXAstex)
    (99.36,PXSD)
    (96.65,PXCD)
    (82.68,PB)
    (97.49,PXAstexNoRel)
    (97.40,PXAstexRel)
    (99.16,PXSDNoRel)
    (99.66,PXSDRel)
    (96.42,PXCDNoRel)
    (96.19,PXCDRel)
    (79.65,PBNoRel)
    (75.44,PBRel)
};

% --- MolAS (vir3) ---
\addplot+[
    fill=vir3!80, draw=vir3!95,
    every node near coord/.append style={text=white}
] coordinates {
    (74.68,MOADcur)
    (79.80,PXAstex)
    (74.63,PXSD)
    (87.50,PXCD)
    (54.91,PB)
    (76.58,PXAstexNoRel)
    (80.84,PXAstexRel)
    (62.62,PXSDNoRel)
    (73.93,PXSDRel)
    (83.38,PXCDNoRel)
    (88.41,PXCDRel)
    (53.97,PBNoRel)
    (45.81,PBRel)
};

% --- MC-GNNAS-Dock (vir2) ---
\addplot+[
    fill=vir2!70, draw=vir2!90,
    every node near coord/.append style={text=white}
] coordinates {
    (68.98,MOADcur)
    (76.34,PXAstex)
    (67.23,PXSD)
    (80.11,PXCD)
    (49.30,PB)
    (70.10,PXAstexNoRel)
    (75.32,PXAstexRel)
    (56.76,PXSDNoRel)
    (68.35,PXSDRel)
    (76.45,PXCDNoRel)
    (80.41,PXCDRel)
    (47.18,PBNoRel)
    (40.90,PBRel)
};

% --- SBS (vir1) ---
\addplot+[
    fill=vir1!70, draw=vir1!90,
    every node near coord/.append style={text=white}
] coordinates {
    (68.17,MOADcur)
    (74.08,PXAstex)
    (74.21,PXSD)
    (73.02,PXCD)
    (51.17,PB)
    (61.02,PXAstexNoRel)
    (74.08,PXAstexRel)
    (60.53,PXSDNoRel)
    (74.21,PXSDRel)
    (65.16,PXCDNoRel)
    (73.02,PXCDRel)
    (51.17,PBNoRel)
    (45.10,PBRel)
};

% Legend (custom)
\addlegendimage{area legend,style={},fill=vir4!90,draw=vir4!70}
\addlegendentry{VBS}

\addlegendimage{area legend,style={},fill=vir3!80,draw=vir3!95}
\addlegendentry{MolAS}

\addlegendimage{area legend,style={},fill=vir2!70,draw=vir2!90}
\addlegendentry{MC-GNNAS-Dock}

\addlegendimage{area legend,style={},fill=vir1!70,draw=vir1!90}
\addlegendentry{SBS}

\end{axis}
\end{tikzpicture}
}
\caption{Comparison among SBS, MC-GNNAS-Dock, MolAS, and VBS in averaged 5-fold results in in PoseBusters-validated pose rate within 2~\AA\ RMSD  across benchmarks.}
\label{fig:comparison_mc_gnnas}
\end{figure}

%%%%%%%%%%%%%%%%%%%%%%%%%%%%%%%%%%%%%%%%%%%%%
\subsection{Comparison to MC-GNNAS-Dock}\label{subsec:comparison_mc_gnnas_dock}
%%%%%%%%%%%%%%%%%%%%%%%%%%%%%%%%%%%%%%%%%%%%%
We benchmarked MC-GNNAS-Dock~\citep{cao2025mc} across the same distributions to assess whether MolAS offers a substantive improvement over a heavier, GNN-based selector. As summarised in Fig.~\ref{fig:comparison_mc_gnnas}, MolAS consistently outperforms MC-GNNAS-Dock across all evaluated benchmarks. The gains range from approximately $3.5–8\%$ in the $2~\text{\AA}$ \& PB-validity metric. Notably, these improvements are achieved with a substantially smaller model: $\sim 638\text{k}$ parameters for MolAS versus $\sim 3.2\text{M}$ for MC-GNNAS-Dock. This suggests that in the current data regime, lightweight embedding-level architectures exploit available signal more efficiently than deeper graph-based models, aligning with the earlier ablation results that point to a data-limited rather than architecture-limited setting.

\begin{table}
\centering
\caption{Component-wise ablation of architectural modules and optimisation objectives based on the MolAS baseline, with the default choices indicated in the brackets. \up denotes improvements over MolAS and \down denotes declinations.}
\label{tab:ablation_architecture}
\begin{tabular}{lccll}
\toprule
    \textbf{Ablated items}& \multicolumn{4}{c}{$\%\leq$x~\AA\ \& PB-valid}\\
\cmidrule(lr){2-5}
 & \multicolumn{2}{c}{MOAD-curated}& \multicolumn{2}{c}{PoseX + Astex} \\
\cmidrule(lr){2-3}\cmidrule(lr){4-5}
    & $x=1$ & $x=2$   & $x=1$ &$x=2$    \\
\midrule
    \textbf{SBS}& 43.41& 68.17& 50.24&74.08 \\
    \textbf{MolAS} & 50.01& 74.68& 58.61&79.80\\
\midrule
    \multicolumn{5}{l}{\textbf{GNN encoder (MolAS: None)}} \\
    GCN-GAT-GINE& 48.38\down& 71.34\down& 56.81\down& 78.81\down\\
    EGNN-GAT-GINE & 47.37\down& 71.81\down& 57.19\down& 78.67\down\\
 % Graph Transformer& 47.97\down& 71.12\down& 56.34\down&77.67\down\\
\midrule
    \multicolumn{5}{l}{\textbf{Protein attention heads (MolAS: 1)}} \\
    2 & 48.22\down& 72.70\down& 58.37\down& 79.90\up\\
    4 & 49.01\down& 72.92\down& 58.70\up& 79.85\up\\
\midrule
    \multicolumn{5}{l}{\textbf{Decoder (MolAS: 4 ResBlock(128, 256))}} \\
    6 ResBlock(128, 256) & 48.19\down& 71.75\down& 59.93\up& 80.27\up\\
    2 ResBlock(128, 256) & 47.81\down& 71.56\down& 59.13\up& 80.37\up\\
    4 ResBlock(256, 512) & 48.79\down& 72.29\down& 58.23\down& 79.28\down\\
    4 ResBlock(64, 128) & 48.57\down& 71.94\down& 57.71\down& 80.42\up\\
    Single linear layer & 48.88\down& 72.57\down& 59.84\up& 80.98\up\\
\midrule
    \multicolumn{5}{l}{\textbf{Scoring function (MolAS: $\lambda=3$)}} \\
    $\lambda = 1$& 48.75\down& 70.75\down& 60.12\up&78.01\down \\
    $\lambda = 5$& 48.48\down& 72.85\down& 57.52\down&80.80\up \\
\midrule
    \multicolumn{5}{l}{\textbf{Ranking-aware loss (MolAS: BCE)}} \\
    BCE + PL& 48.50\down& 71.47\down& 59.65\up&77.15\down \\
    BCE + NDCG@3& 48.47\down& 72.32\down& 59.70\up&80.51\up\\
    BCE + PL + NDCG@3& 48.22\down& 71.25\down& 60.12\up&78.01\down \\
\midrule
    \multicolumn{5}{l}{\textbf{Modality dropout (MolAS: both embeddings)}} \\
    Protein-only (embeddings) & 45.58\down& 68.83\down& 53.83\down&74.93\down\\
    Ligand-only (embeddings)& 48.28\down& 73.10\down& 58.66\up&80.61\up\\
    Protein-only (graph+embeddings) with GCN-GAT-GINE & 46.71\down& 69.55\down& 52.93\down&73.93\down\\
    Ligand-only (graph+embeddings) with GCN-GAT-GINE & 48.38\down& 73.10\down& 58.66\up&80.61\up\\
\bottomrule
\end{tabular}
\end{table}

%%%%%%%%%%%%%%%%%%%%%%%%%%%%%%%%%%%%%%%%%%%%%
\subsection{Ablation}\label{subsec:ablation}
%%%%%%%%%%%%%%%%%%%%%%%%%%%%%%%%%%%%%%%%%%%%%
MolAS uses separate protein and ligand embeddings and does not explicitly encode protein--ligand 3D contact graphs. As discussed in prior analyses of bias and shortcut learning in structure-based ML benchmarks~\citep{sieg2019need,gorantla2023proteins}, this design choice motivates explicit controls for whether performance is driven by a single modality or by protocol-specific artefacts. Table~\ref{tab:ablation_architecture} therefore includes modality-dropout variants (protein-only, ligand-only, and full input), reported under the same evaluation protocols as the main model.

Across the two major benchmark sets, \textit{MOAD-curated} and \textit{PoseX + Astex}, parallel GNN encoder stacks (GCN--GAT--GINE, inspired by graphLambda~\citep{mqawass2024graphlambda}, and an equivariant EGNN--GAT--GINE variant) fail to surpass the embedding-only model, indicating that pretrained residue-level ESM-C and ligand ChemBERTa representations are already sufficient to support in-domain selection in this data regime. Increasing the number of attention heads produces only sub-percentage fluctuations with no consistent pattern. Decoder capacity exhibits a similarly limited effect, with deeper or wider residual stacks yielding only marginal gains on one benchmark at the expense of performance on the other. A single linear decoder achieves peak performance on the \textit{PoseX + Astex} set under the $2~\text{\AA}$ \& PB-validity criterion; however, its performance on the \textit{MOAD-curated} benchmark decreases by approximately $2\%$, suggesting limited benefit from architectural scaling at the current data scale.

The modality-dropout controls further clarify how signal is distributed across inputs. In the embedding-only setting, removing either the protein or ligand embedding leads to a moderate degradation relative to the full model, indicating that neither modality alone trivially explains the selection labels. In the graph+embedding setting, removing protein information causes a substantially larger drop than removing ligand information, while removing ligand information has a smaller effect. Taken together, these ablations do not support a degenerate single-modality shortcut explanation on the presented benchmarks, while leaving open the possibility that explicit protein--ligand contact features could improve robustness under stronger distribution shifts (Sec.~\ref{sec:discussion}).

% Sensitivity to the scoring-function hyperparameter is modest: altering the RMSD--validity trade-off from $\lambda=3$ to $\lambda=1$ or $5$ shifts scores by at most $2\%$, and the directions of change differ across benchmarks, implying that MolAS is relatively insensitive to reasonable weighting choices. Ranking-aware objectives, including pairwise logistic (PL), Normalized Discounted Cumulative Gain (NDCG@3), and their combinations with BCE, produce only small variations with no consistent gain (Table~\ref{tab:ablation_architecture}). 
% A plausible reason is that many instances exhibit near-ties among top solvers and multiple solvers can simultaneously achieve high composite scores, making strict ranking supervision comparatively noisy under the available sample sizes. 

% Taken together, these results support the interpretation that MolAS is \emph{data-limited} rather than \emph{architecture-limited}. Additional architectural complexity fails to yield reliable improvements. The baseline configuration therefore represents an appropriate balance of expressiveness and stability for the scale and heterogeneity of the available data.

Sensitivity to the scoring-function hyperparameter is modest: altering the RMSD--validity trade-off from $\lambda=3$ to $\lambda=1$ or $5$ shifts scores by at most $\sim2\%$, and the directions of change differ across benchmarks, implying that MolAS is relatively insensitive to reasonable weighting choices in the current regime. Ranking-aware objectives, including pairwise logistic (PL), Normalized Discounted Cumulative Gain (NDCG@3), and their combinations with BCE, likewise produce only small variations with no consistent gain (Table~\ref{tab:ablation_architecture}). 

These negative results are informative when viewed through the oracle-landscape diagnostics. On several benchmarks the oracle is high-entropy and shallow near the top: multiple solvers frequently achieve similar composite success scores, and small changes in post-processing can reorder winners. In such regimes, the supervision signal exhibits (i) \emph{near-ties} among top solvers, (ii) \emph{non-smooth ranking transitions} where small score perturbations flip the top-1 label, and (iii) \emph{label noise induced by protocol differences} when the same complex is evaluated under different workflows. Together, these properties make strict ranking supervision brittle: pairwise/listwise losses treat many near-ties as hard constraints, amplifying noise without adding stable information beyond a calibrated success tendency. Consistently, increasing model capacity (deeper decoders or GNN encoders) does not yield systematic gains, suggesting that the limiting factor is not expressiveness but the stability and separability of the workflow-defined labels.

Taken together, the ablations support an evidence-backed hypothesis that MolAS is \emph{data- and label-regime limited} rather than \emph{architecture-limited}: when oracle entropy is high and top-rank boundaries are non-smooth, additional capacity or more aggressive ranking losses provide limited benefit and can degrade transfer across benchmarks. The baseline configuration therefore represents an appropriate balance of expressiveness and stability for the scale and heterogeneity of the available data.

%%%%%%%%%%%%%%%%%%%%%%%%%%%%%%%%%%%%%%%
\section{Discussion}
\label{sec:discussion}
%%%%%%%%%%%%%%%%%%%%%%%%%%%%%%%%%%%%%%%

\subsection{When is MolAS useful in practice?}
\label{subsec:practical_utility}

MolAS is best interpreted as a \emph{workflow-adaptive selector} rather than a universal docking policy. Practical use is most straightforward when a project fixes, for a period, a target domain (protein family and ligand chemistry), a solver portfolio, and an evaluation/post-processing pipeline. Under such a fixed workflow, a modest in-domain calibration set can be labelled by running the portfolio on that subset, after which MolAS can be trained to reduce expected regret relative to a static solver, e.g. SBS, by exploiting systematic instance-dependent differences among solvers.

Empirically, the strongest gains appear in regimes where the induced selection problem is non-trivial: the VBS--SBS gap is appreciable and oracle winners are not concentrated in a single method. In these settings, performance tends to saturate with hundreds to a few thousand labelled complexes, consistent with the observed gap closure on benchmarks such as \textit{MOAD-curated} and \textit{PoseX-CD}. Conversely, when the oracle landscape is effectively dominated by a single solver or when multiple solvers are near-indistinguishable under the evaluation protocol, headroom for selection is limited and MolAS tends to converge to an SBS-like policy, which is locally reasonable but yields negligible improvement.

Workflow changes define the main operational boundary. Cross-benchmark experiments indicate that changes in docking settings, dataset curation, or post-processing can shift solver scores and reorder solver rankings, thereby changing the induced supervision signal. In practice, this implies that selectors trained without explicit workflow descriptors should not be assumed to transfer reliably across pipelines; when the portfolio or protocol changes, re-labelling and retraining on the new workflow may generally be required. When oracle diversity is low or the VBS--SBS gap is small, an SBS fallback remains an appropriate default.

\subsection{What does MolAS reveal about docking algorithm selection?}
\label{subsec:diagnostic_insights}

MolAS was designed to separate representational considerations from workflow-defined effects. Across the reported ablations, increasing architectural capacity (including graph-based encoders) and changing optimisation objectives (including ranking-aware losses) did not yield systematic gains. Within the evaluated regime, this pattern suggests that simply scaling model complexity is not the dominant lever for improving docking AS, at least when strong pretrained protein and ligand embeddings are available and supervision is defined by a workflow-specific oracle.

Instead, performance is strongly associated with properties of the \emph{workflow-defined oracle landscape}. Several regimes exhibit high oracle diversity and shallow separation among top solvers, so that small perturbations of scores can flip the top-1 label. Combined with protocol-induced ranking shifts under post-processing or benchmark changes, this creates a supervision signal that is effectively non-smooth and partially noisy with respect to molecular inputs alone. This diagnostic picture provides an evidence-backed hypothesis for the observed ablation failures: ranking-aware losses can over-emphasise unstable near-ties, and additional capacity can overfit workflow-specific characteristics without improving robustness, yielding limited or inconsistent benefit beyond calibrated success prediction.

Importantly, the observed protocol dependence is not an impossibility result. Stronger generalisation across workflows may be achievable if protocol variables are modelled explicitly, e.g. by conditioning the selector on workflow descriptors, by learning representations that are invariant across a family of protocols, or by applying explicit domain adaptation when transferring between pipelines. Related difficulties with out-of-distribution generalisation have been reported in algorithm selection for other domains when solver rankings shift under distributional change, motivating adaptation mechanisms rather than assuming invariance~\citep{cenikj2025landscape}.

A further limitation is that MolAS does not represent explicit protein--ligand contact geometry. While modality-dropout controls indicate that both protein and ligand embeddings contribute on the evaluated benchmarks, explicit cross-molecular interaction features may be required for stronger robustness under protocol and distribution shifts, particularly when oracle-defined solver regimes depend on fine-grained contact patterns not captured reliably by unimodal embeddings.

\subsection{Future directions}
\label{subsec:future}

Several directions follow naturally from these findings. First, protocol-aware selection appears necessary for reliable transfer: incorporating workflow descriptors, modelling protocol families jointly, or using explicit adaptation across pipelines offers a principled route to improving cross-protocol performance. Second, more robust confidence estimation could mitigate winner-take-all collapses by enabling abstention or controlled fallback to SBS-like policies when the selector’s margins are not informative under high-entropy regimes. Third, incorporating explicit interaction geometry (e.g., contact features or cross-molecular graphs) remains a plausible path to improving robustness, provided that evaluation protocols and label definitions are controlled carefully enough to separate representational gains from workflow-induced effects. Together, these considerations delimit a realistic scope for docking AS: substantial gains are achievable under fixed workflows with non-trivial oracle diversity, while generalisation across workflows is a modelling problem rather than a property that can be assumed.

\backmatter

\subsubsection*{Abbreviations}

\begin{tabular}{@{}p{2.2cm}p{12cm}@{}}
AS  & Algorithm selection \\
BCE & Binary cross-entropy \\
CD   & Cross docking \\
EGNN & Equivariant graph neural network \\
GAT  & Graph attention network \\
GINE & Graph isomorphism network \\
GNN  & Graph neural network \\
ML   & Machine learning \\
MM-min & Molecular-mechanics energy minimisation \\
MolAS & Molecular embedding-based algorithm selector \\
NDCG  & Normalized Discounted Cumulative Gain \\
PB   & PoseBusters \\
PL   & Pairwise logistic \\
RMSD & Root-mean-square deviation \\
SBS  & Single best solver \\
SD   & Self docking \\
VBS  & Virtual best solver \\
\end{tabular}

% \bmhead{Supplementary information}

% If your article has accompanying supplementary file/s please state so here. 

% Authors reporting data from electrophoretic gels and blots should supply the full unprocessed scans for key as part of their Supplementary information. This may be requested by the editorial team/s if it is missing.

% Please refer to Journal-level guidance for any specific requirements.

\section*{Declarations}

\bmhead{Availability of data and materials}
The dataset supporting the conclusions of this article including processed complexes, per-algorithm performances, and in-domain checkpoints is available on Zenodo at \href{https://zenodo.org/records/17760688}{https://zenodo.org/records/17760688}. 
The codes for generating embeddings, training and testing MolAS are available in the
MolAS GitHub repository at \href{https://github.com/BradWangW/MolAS}{https://github.com/BradWangW/MolAS}.

\bmhead{Competing interests}
No competing interest are declared.

\bmhead{Funding}
Not applicable.

\bmhead{Authors' contributions}
Y.Y. developed the base GNNAS-Dock framework. S.C. and H.W. designed and implemented the MC-GNNAS-Dock extensions with input from all authors. J.B.W. performed the MolAS experiments and analyses and drafted the manuscript. M.M. conceptualised the research direction, supervised all experiments and analyses, and edited the manuscript. All authors reviewed and approved the final version.  

\bmhead{Acknowledgements}
The research results of this article are sponsored by the Wang-Cai Biochemistry Lab and Synear Food Molecular Biology Lab.

% Some journals require declarations to be submitted in a standardised format. Please check the Instructions for Authors of the journal to which you are submitting to see if you need to complete this section. If yes, your manuscript must contain the following sections under the heading `Declarations':

% \begin{itemize}
% \item Conflict of interest/Competing interests (check journal-specific guidelines for which heading to use)
% \item Ethics approval and consent to participate
% \item Consent for publication
% \item Materials availability
% \end{itemize}

% \noindent
% If any of the sections are not relevant to your manuscript, please include the heading and write `Not applicable' for that section. 

\bigskip
\begin{flushleft}%
Editorial Policies for:

\bigskip\noindent
Springer journals and proceedings: \url{https://www.springer.com/gp/editorial-policies}

\bigskip\noindent
Nature Portfolio journals: \url{https://www.nature.com/nature-research/editorial-policies}

\bigskip\noindent
\textit{Scientific Reports}: \url{https://www.nature.com/srep/journal-policies/editorial-policies}

\bigskip\noindent
BMC journals: \url{https://www.biomedcentral.com/getpublished/editorial-policies}
\end{flushleft}

\begin{appendices}

\section{Appendix}

\subsection{Supplements on dataset} \label{app:data_supplements}

\begin{figure}[t]
    \centering
    \includegraphics[width=0.67\linewidth]{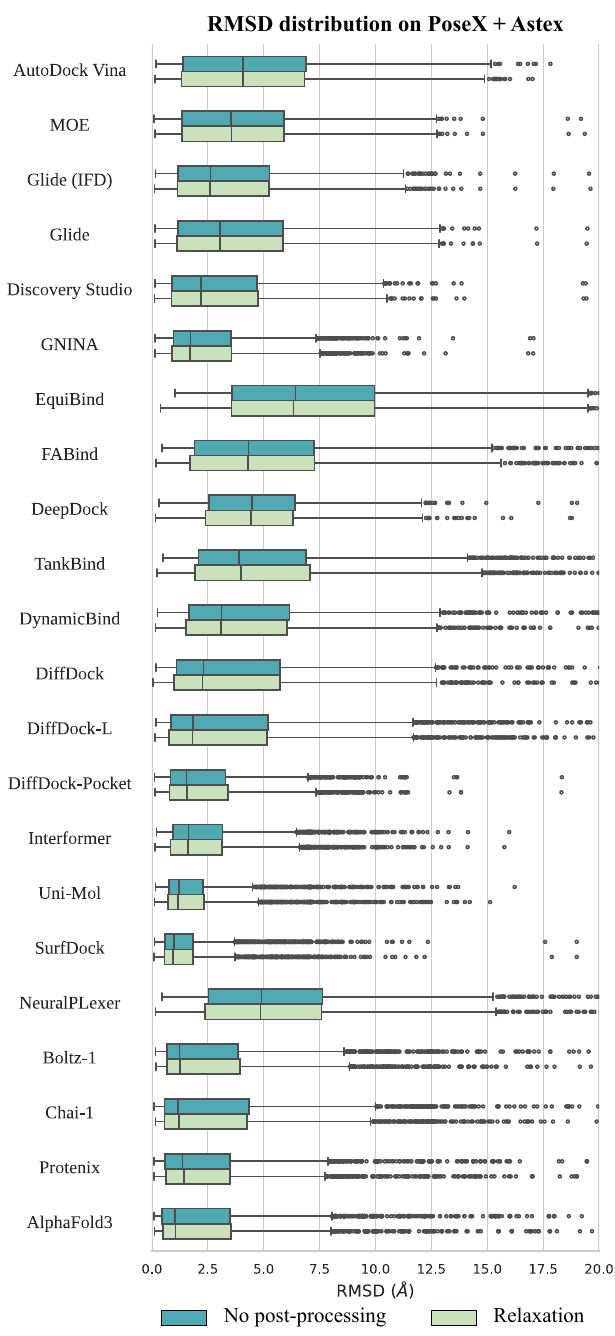}
    \caption{Distribution of ligand RMSD-to-reference for each method before (raw) and after PoseX relaxation, evaluated on PoseX + Astex. Boxes show the interquartile range with median; whiskers or outliers follow the standard boxplot convention.}
    \label{fig:rmsd_before_after_relaxtion}
\end{figure}

\paragraph{Effect of relaxation on RMSD and PoseBusters validity.}
Figure~\ref{fig:rmsd_before_after_relaxtion} compares RMSD distributions before and after applying the PoseX relaxation step, and Table~\ref{tab:paired_stats_relaxation} summarises per-complex changes over $N{=}2115$ protein--ligand complexes (PoseX + Astex).
Across all methods, relaxation rarely converts a PoseBusters-valid pose into an invalid one (0--36 complexes; $\leq 1.7\%$), indicating that geometry correction is typically not detrimental to validity.
In contrast, the number of validity improvements varies substantially by method, ranging from 5 (0.2\%) to 1712 (80.9\%) complexes, suggesting that some pipelines frequently produce strained or clash-prone raw geometries that are subsequently corrected.
Across methods, larger mean $|\Delta\mathrm{RMSD}|$ tends to coincide with more validity improvements (Pearson $r\approx0.75$), consistent with the need for larger coordinate adjustments when resolving steric or geometric violations.
Accordingly, relaxation can measurably shift RMSD-based outcomes for some methods, and results are therefore reported both without post-processing and with relaxation.

\begin{table*}[t]
    \centering
    \caption{Change in ligand RMSD-to-reference and PoseBusters validity after applying PoseX relaxation, evaluated on PoseX + Astex. $\Delta \mathrm{RMSD} = \mathrm{RMSD}_{\text{relax}} - \mathrm{RMSD}_{\text{raw}}$. “PB-validity” counts the number of complexes whose PoseBusters status improved (invalid $\rightarrow$ valid), worsened (valid $\rightarrow$ invalid), or remained unchanged.}
    \label{tab:paired_stats_relaxation}
        \begin{tabular}{lccccc}
        \toprule
        \textbf{Method} & \multicolumn{2}{c}{$|\Delta\mathrm{RMSD}|$} & \multicolumn{3}{c}{\textbf{PB-validity}}\\
        \cmidrule(lr){2-3}\cmidrule(lr){4-6}
        & Mean & IQR & $\mathrm{\#Improved}$ & $\mathrm{\#Worsened}$ & $\mathrm{\#Unchanged}$\\
        \midrule
        AutoDock Vina          & 0.218 & 0.124 & 200  & 14  & 1901 \\
        MOE                    & 0.075 & 0.068 & 6    & 30  & 2079 \\
        Glide (IFD)            & 0.073 & 0.070 & 5    & 28  & 2082 \\
        Schrödinger Glide      & 0.091 & 0.091 & 6    & 35  & 2074 \\
        Discovery Studio       & 0.075 & 0.078 & 16   & 27  & 2072 \\
        GNINA                  & 0.108 & 0.119 & 152  & 36  & 1927 \\
        Equibind               & 1.185 & 0.820 & 1705 & 0   & 410  \\
        FABind                 & 0.640 & 0.404 & 1223 & 1   & 891  \\
        DeepDock               & 0.299 & 0.314 & 1195 & 0   & 920  \\
        TankBind               & 0.685 & 0.396 & 1095 & 1   & 1019 \\
        DynamicBind            & 0.447 & 0.281 & 1712 & 1   & 402  \\
        DiffDock               & 0.329 & 0.192 & 1316 & 3   & 796  \\
        DiffDock-L             & 0.341 & 0.177 & 1142 & 10  & 963  \\
        DiffDock-Pocket        & 0.119 & 0.128 & 932  & 7   & 1176 \\
        Interformer            & 0.140 & 0.136 & 656  & 20  & 1439 \\
        Uni-Mol                & 0.123 & 0.120 & 97   & 25  & 1993 \\
        SurfDock               & 0.154 & 0.160 & 1069 & 8   & 1038 \\
        NeuralPLexer           & 0.655 & 0.421 & 1643 & 1   & 471  \\
        Boltz-1                & 0.215 & 0.134 & 440  & 18  & 1657 \\
        Chai-1                 & 0.262 & 0.121 & 243  & 21  & 1851 \\
        Protenix               & 0.177 & 0.110 & 211  & 22  & 1882 \\
        AlphaFold3             & 0.210 & 0.111 & 230  & 18  & 1867 \\
        \bottomrule
        \end{tabular}

\end{table*}

\subsection{Supplemental results}\label{app:verbose}

\subsubsection{Detailed performance over portfolio}
Detailed performances over algorithm portfolios and MolAS/VBS selection frequencies over \textit{MOAD-curated}, \textit{PoseX + Astex}, \textit{PoseX-SD}, \textit{PoseX-CD}, and \textit{PoseBusters} are shown in Figures~\ref{fig:detail_moad}, \ref{fig:detail_posex_astex}, \ref{fig:detail_posexsd}, \ref{fig:detail_posexcd}, and \ref{fig:detail_posebusters}.

\begin{figure*}
\centering
\hspace*{-2em}
% ==========================
% TOP LEFT
% ==========================
\scalebox{0.8}{
\begin{minipage}[t]{0.43\textwidth}
\raggedright
\textbf{\small(a) Averaged 5-fold results on MOAD-curated.}\\[4pt]
    \begin{tabular}{p{2.5cm}cc}
    \toprule
    \textbf{Method}  & 
    \multicolumn{2}{c}{$(\%)\ \text{RMSD}\leq x~\text{\AA}\ \&\ \text{PB-valid}$}\\
    \cmidrule{2-3} & $x=1$&$x=2$\\
    \midrule
    \multicolumn{3}{c}{\textbf{Physics-based methods}} \\
     Qvina-W & 11.20 & 19.28  \\
     Smina      & 11.23 & 19.50  \\
     GNINA (hybrid)      & 25.45 & 42.87  \\
    \midrule
    \multicolumn{3}{c}{\textbf{AI docking methods}} \\
     KarmaDock  &  3.08 & 8.50   \\
     DiffDock    & 8.60 & 11.40  \\
     DiffDock-L  & 10.22 & 14.72  \\
     SurfDock & 39.28 & 53.66  \\
     Uni-Mol  & \textbf{43.41} & \textbf{68.16}  \\
    \midrule
     \textbf{MolAS} & \textbf{50.01}& \textbf{74.68} \\
     \midrule
     VBS@1 & 81.31 & 97.23 \\
     VBS@2 & 69.20 & 95.56 \\
     VBS@3 & 50.96 & 84.11 \\
    \bottomrule
    \end{tabular}
\end{minipage}
}
\hfill
% ==========================
% TOP RIGHT
% ==========================
\begin{minipage}[t]{0.53\textwidth}
\raggedleft
\textbf{\small (b) (\%) Selection frequencies on MOAD-curated.}\\[4pt]
% \resizebox{\linewidth}{!}{%
    \begin{tikzpicture}
    \definecolor{vir1}{HTML}{fd9668}
    \definecolor{vir2}{HTML}{2a788e}
    
    \begin{axis}[
        xbar,
        y dir=reverse,
        legend reversed=true,
        bar width=5pt,
        width=7.5cm,
        height=8cm,
        xmin=0, xmax=70,
        symbolic y coords={
            MOADcur,
            PXAstex,
            PXSD,
            PXCD,
            PB,
            PXAstexNoRel,
            PXAstexRel,
            PXSDNoRel
        },
        yticklabels={
            {Smina},
            {Qvina},
            {KarmaDock},
            {DiffDock},
            {DiffDock-L},
            {Gnina},
            {SurfDock},
            {Uni-Mol}
        },
        ytick=data,
        yticklabel style={font=\small, align=center},
        enlarge y limits=0.075,
        % nodes near coords,
        point meta=x,
        nodes near coords align={horizontal},
        every node near coord/.append style={
            anchor=west, xshift=2pt, font=\scriptsize
        },
        xmajorgrids=true,
        major grid style={dashed, thin, black!30},
    ]
    
    % ---- MolAS Selection (vir2) ----
    % order bottom → top = karmadock, diffdock, autodock, smina,
    %                diffdock_l, gnina, unimol, surfdock
    \addplot+[
        fill=vir2!70, draw=vir2!90,
        every node near coord/.append style={text=vir2}
    ] coordinates {
        (0.377477,MOADcur)      % KarmaDock
        (0.44039,PXAstex)       % DiffDock
        (0.912237,PXSD)         % AutoDock (Qvina placeholder)
        (3.27147,PXCD)          % Smina
        (0.597672,PB)           % DiffDock-L
        (11.6389,PXAstexNoRel)  % Gnina
        (63.1331,PXAstexRel)    % Uni-Mol
        (19.6288,PXSDNoRel)     % SurfDock
    };
    
    % ---- Oracle (vir1) ----
    \addplot+[
        fill=vir1!70, draw=vir1!90,
        every node near coord/.append style={text=vir1}
    ] coordinates {
        (0.817867,MOADcur)      % KarmaDock
        (3.86914,PXAstex)       % DiffDock
        (3.99497,PXSD)          % AutoDock
        (4.02642,PXCD)          % Smina
        (5.1274,PB)             % DiffDock-L
        (15.854,PXAstexNoRel)   % Gnina
        (30.2296,PXAstexRel)    % Uni-Mol
        (36.0805,PXSDNoRel)     % SurfDock
    };
    
    \legend{MolAS, VBS}
    
    \end{axis}
    \end{tikzpicture}
% }
\end{minipage}

\caption{Table of performances over the algorithm portfolio (left) and histogram of MolAS/VBS selection frequencies (right) on \textbf{MOAD-curated}. In the table, the portfolio-wise best and MolAS performances are in \textbf{bold}. 
Performances are ordered by ascending percentage of poses below $2~\text{\AA}$ and PoseBusters-valid, 
while frequencies are ordered by ascending VBS frequency.}
\label{fig:detail_moad}
\end{figure*}

\begin{figure*}
\centering
\hspace*{-7em}
% ==========================
% BOTTOM LEFT
% ==========================
\scalebox{0.88}{
\begin{minipage}[t]{0.65\textwidth}
\raggedright
\textbf{Averaged 5-fold results on PoseX + Astex.}\\[6pt]

\begin{tabular}{p{3.9cm}cc}
\toprule
    \textbf{Method} &
    \multicolumn{2}{c}{$(\%)\ \text{RMSD}\leq x~\text{\AA}\ \&\ \text{PB-valid}$}\\
    \cmidrule(lr){2-3}
    & $x=1$ & $x=2$\\
    \midrule
    \multicolumn{3}{c}{\textbf{Physics-based methods}} \\
    AutoDock Vina             & 17.170 & 31.502 \\
    AutoDock Vina (relax)     & 18.496 & 33.112 \\
    MOE (relax)               & 18.258 & 33.350 \\
    MOE                       & 18.258 & 33.822 \\
    Schrödinger Glide (relax) & 22.186 & 39.262 \\
    Schrödinger Glide         & 21.098 & 39.498 \\
    Glide (IFD)               & 20.718 & 41.910 \\
    Glide (IFD) (relax)       & 21.284 & 41.912 \\
    Discovery Studio (relax)  & 27.722 & 45.788 \\
    Discovery Studio          & 27.672 & 46.356 \\
    GNINA (hybrid)            & 25.544 & 53.310 \\
    GNINA (hybrid) (relax)    & 27.340 & 56.054 \\
    \midrule
    \multicolumn{3}{c}{\textbf{AI docking methods}} \\
    EquiBind                  & 0.000 & 0.618 \\
    FABind                    & 1.938 & 4.778 \\
    TankBind                  & 2.268 & 5.346 \\
    DeepDock                  & 2.364 & 6.054 \\
    EquiBind (relax)          & 1.088 & 6.480 \\
    DynamicBind               & 5.060 & 9.510 \\
    DeepDock (relax)          & 5.346 & 15.656 \\
    TankBind (relax)          & 7.380 & 18.402 \\
    DiffDock                  & 13.860 & 23.180 \\
    FABind (relax)            & 9.462 & 23.274 \\
    DiffDock-L                & 21.054 & 30.038 \\
    DynamicBind (relax)       & 14.002 & 33.062 \\
    DiffDock-Pocket           & 21.806 & 35.144 \\
    SurfDock                  & 32.024 & 39.782 \\
    Interformer               & 22.326 & 42.382 \\
    DiffDock (relax)          & 25.214 & 44.796 \\
    DiffDock-L (relax)        & 33.348 & 50.946 \\
    DiffDock-Pocket (relax)   & 32.214 & 56.384 \\
    Uni-Mol                   & 37.654 & 57.900 \\
    Uni-Mol (relax)           & 37.656 & 57.948 \\
    Interformer (relax)       & 31.500 & 58.606 \\
    SurfDock (relax)          & \textbf{50.236} & \textbf{74.078} \\
    \midrule
    \multicolumn{3}{c}{\textbf{AI co-folding methods}} \\
    NeuralPLexer              & 0.522 & 1.986 \\
    RFAA                      & 4.350 & 7.806 \\
    NeuralPLexer (relax)      & 5.772 & 18.068 \\
    RFAA (relax)              & 10.358 & 24.218 \\
    Boltz-1                   & 30.842 & 43.710 \\
    Protenix                  & 35.006 & 48.532 \\
    Protenix (relax)          & 36.566 & 51.422 \\
    Chai-1                    & 39.450 & 51.988 \\
    Boltz-1 (relax)           & 36.142 & 53.124 \\
    AlphaFold 3               & 42.904 & 54.778 \\
    Chai-1 (relax)            & 42.384 & 57.428 \\
    AlphaFold 3 (relax)       & 44.416 & 59.462 \\
    Boltz-1x                  & 42.478 & 61.022 \\
    Boltz-1x (relax)          & 41.104 & 61.070 \\
    \midrule
     \textbf{MolAS} & \textbf{58.606} & \textbf{79.800} \\
     \midrule
     VBS@1 & 89.308 & 97.684 \\
     VBS@2 & 86.896 & 97.256 \\
     VBS@3 & 82.022 & 95.222 \\
\bottomrule
\end{tabular}

\end{minipage}
}
\hfill
% ==========================
% BOTTOM RIGHT
% ==========================
\begin{minipage}[t]{0.366\textwidth}
\raggedleft
\textbf{\scriptsize (\%) Selection frequencies on PoseX + Astex.}\\[4pt]
\hspace*{-4em}
\begin{tikzpicture}

% ---- Viridis picks ----
\definecolor{vir2}{HTML}{2a788e} % teal
\definecolor{vir1}{HTML}{fd9668} % orange
\begin{axis}[
    xbar,
    y dir=reverse,
    bar width=4pt,
    tick label style={/pgf/number format/fixed, /pgf/number format/precision=2},
    legend reversed=true,
    width=6cm,
    height=23.5cm,
    xmin=0, xmax=70,
    symbolic y coords={
        DiffDock,
        EquiBind,
        EquiBindRelax,
        FABind,
        NeuralPLexer,
        NeuralPLexerRelax,
        RFAA,
        TankBind,
        DeepDock,
        DynamicBind,
        TankBindRelax,
        AutoDockVina,
        FABindRelax,
        RFAARelax,
        DeepDockRelax,
        DynamicBindRelax,
        DiffDockL,
        IFD,
        Interformer,
        GNINA,
        IFDRelax,
        DiffDockRelax,
        AutoDockVinaRelax,
        Glide,
        DiffDockPocket,
        Boltz1xRelax,
        Boltz1Relax,
        GlideRelax,
        GNINARelax,
        MOE,
        MOERelax,
        Boltz1,
        UniMol,
        DiffDockLRelax,
        InterformerRelax,
        ProtenixRelax,
        DiscoveryStudio,
        Boltz1x,
        AlphaFold3Relax,
        Chai1Relax,
        UniMolRelax,
        Chai1,
        DiscoveryStudioRelax,
        DiffDockPocketRelax,
        Protenix,
        SurfDock,
        AlphaFold3,
        SurfDockRelax
    },
    yticklabels={
        {DiffDock},
        {EquiBind},
        {EquiBind (relax)},
        {FABind},
        {NeuralPLexer},
        {NeuralPLexer (relax)},
        {RFAA},
        {TankBind},
        {DeepDock},
        {DynamicBind},
        {TankBind (relax)},
        {AutoDock Vina},
        {FABind (relax)},
        {RFAA (relax)},
        {DeepDock (relax)},
        {DynamicBind (relax)},
        {DiffDock-L},
        {Glide (IFD)},
        {Interformer},
        {GNINA},
        {Glide (IFD) (relax)},
        {DiffDock (relax)},
        {AutoDock Vina (relax)},
        {Glide},
        {DiffDock-Pocket},
        {Boltz-1x (relax)},
        {Boltz-1 (relax)},
        {Glide (relax)},
        {GNINA (relax)},
        {MOE},
        {MOE (relax)},
        {Boltz-1},
        {Uni-Mol},
        {DiffDock-L (relax)},
        {Interformer (relax)},
        {Protenix (relax)},
        {Discovery Studio},
        {Boltz-1x},
        {AlphaFold 3 (relax)},
        {Chai-1 (relax)},
        {Uni-Mol (relax)},
        {Chai-1},
        {Discovery Studio (relax)},
        {DiffDock-Pocket (relax)},
        {Protenix},
        {SurfDock},
        {AlphaFold 3},
        {SurfDock (relax)}
    },
    ytick=data,
    yticklabel style={font=\scriptsize},
    enlarge y limits=0.015,
    % nodes near coords,
    point meta=x,
    nodes near coords align={horizontal},
    every node near coord/.append style={
        anchor=east,
        xshift=1pt,
        font=\tiny,
        /pgf/number format/fixed,
        /pgf/number format/precision=2
    },
    xmajorgrids=true,
    major grid style={dashed, thin, black!30}
]

% ---- MolAS (Selection_Count) ----
\addplot+[
    fill=vir2!70, draw=vir2!90,
    every node near coord/.append style={text=vir2}
] coordinates {
    (0,DiffDock)
    (0,EquiBind)
    (0,EquiBindRelax)
    (0,FABind)
    (0,NeuralPLexer)
    (0,NeuralPLexerRelax)
    (0,RFAA)
    (0,TankBind)
    (0,DeepDock)
    (0,DynamicBind)
    (0,TankBindRelax)
    (0.0473037,AutoDockVina)
    (0.189215,FABindRelax)
    (0.37843,RFAARelax)
    (0,DeepDockRelax)
    (0.0946074,DynamicBindRelax)
    (0,DiffDockL)
    (0.189215,IFD)
    (0.709555,Interformer)
    (0.283822,GNINA)
    (0.283822,IFDRelax)
    (0.567644,DiffDockRelax)
    (0.567644,AutoDockVinaRelax)
    (0.520341,Glide)
    (0.141911,DiffDockPocket)
    (4.11542,Boltz1xRelax)
    (3.7843,Boltz1Relax)
    (1.75024,GlideRelax)
    (2.79092,GNINARelax)
    (0.946074,MOE)
    (1.60833,MOERelax)
    (1.2299,Boltz1)
    (5.48723,UniMol)
    (2.74361,DiffDockLRelax)
    (2.22327,InterformerRelax)
    (3.21665,ProtenixRelax)
    (1.41911,DiscoveryStudio)
    (13.0085,Boltz1x)
    (6.00757,AlphaFold3Relax)
    (7.7105,Chai1Relax)
    (5.81835,UniMolRelax)
    (2.6017,Chai1)
    (1.18259,DiscoveryStudioRelax)
    (5.01419,DiffDockPocketRelax)
    (1.93945,Protenix)
    (0.189215,SurfDock)
    (6.52791,AlphaFold3)
    (14.7114,SurfDockRelax)
};

% ---- Oracle (Oracle_Count) ----
\addplot+[
    fill=vir1!70, draw=vir1!90,
    every node near coord/.append style={text=vir1}
] coordinates {
    (0,DiffDock)
    (0,EquiBind)
    (0,EquiBindRelax)
    (0,FABind)
    (0,NeuralPLexer)
    (0,NeuralPLexerRelax)
    (0,RFAA)
    (0,TankBind)
    (0.0473037,DeepDock)
    (0.0946074,DynamicBind)
    (0.283822,TankBindRelax)
    (0.331126,AutoDockVina)
    (0.331126,FABindRelax)
    (0.37843,RFAARelax)
    (0.425733,DeepDockRelax)
    (0.614948,DynamicBindRelax)
    (0.709555,DiffDockL)
    (0.756859,IFD)
    (0.804163,Interformer)
    (0.89877,GNINA)
    (1.18259,IFDRelax)
    (1.2299,DiffDockRelax)
    (1.2772,AutoDockVinaRelax)
    (1.41911,Glide)
    (1.46641,DiffDockPocket)
    (1.51372,Boltz1xRelax)
    (1.65563,Boltz1Relax)
    (1.75024,GlideRelax)
    (1.75024,GNINARelax)
    (1.79754,MOE)
    (1.79754,MOERelax)
    (1.89215,Boltz1)
    (1.98675,UniMol)
    (2.17597,DiffDockLRelax)
    (2.31788,InterformerRelax)
    (2.41249,ProtenixRelax)
    (2.5544,DiscoveryStudio)
    (2.79092,Boltz1x)
    (3.8789,AlphaFold3Relax)
    (3.92621,Chai1Relax)
    (3.92621,UniMolRelax)
    (4.21003,Chai1)
    (4.39924,DiscoveryStudioRelax)
    (5.1561,DiffDockPocketRelax)
    (5.58184,Protenix)
    (6.76443,SurfDock)
    (11.6367,AlphaFold3)
    (11.8732,SurfDockRelax)
};

\legend{MolAS, VBS}

\end{axis}

\end{tikzpicture}
\end{minipage}
\caption{Table of performances over the algorithm portofolio (left) and histogram of MolAS/VBS selection frequencies (right) on \textbf{PoseX + Astex}. In the table, the portfolio-wise best and MolAS performances are in \textbf{bold}. 
Performances are ordered by ascending percentage of poses below $2~\text{\AA}$ and PoseBusters-valid, 
while frequencies are ordered by ascending VBS frequency.}
\label{fig:detail_posex_astex}
\end{figure*}

\begin{figure*}
\centering
\hspace*{-7em}
% ==========================
% BOTTOM LEFT
% ==========================
\scalebox{0.88}{
\begin{minipage}[t]{0.65\textwidth}
\raggedright
\textbf{\small Averaged 5-fold results on PoseX-SD.}\\[6pt]

\begin{tabular}{p{3.9cm}cc}
\toprule
    \textbf{Method} &
    \multicolumn{2}{c}{$(\%)\ \text{RMSD}\leq x~\text{\AA}\ \&\ \text{PB-valid}$}\\
    \cmidrule(lr){2-3}
    & $x=1$ & $x=2$\\
    \midrule
    \multicolumn{3}{c}{\textbf{Physics-based methods}} \\
    AutoDock Vina             & 22.590 & 36.812 \\
    MOE (relax)               & 23.160 & 37.670 \\
    AutoDock Vina (relax)     & 23.708 & 38.348 \\
    MOE                       & 23.716 & 39.474 \\
    Glide (IFD) (relax)       & 24.822 & 45.460 \\
    Glide (IFD)               & 26.216 & 46.156 \\
    Schrödinger Glide (relax) & 31.380 & 46.292 \\
    Schrödinger Glide         & 30.826 & 47.966 \\
    Discovery Studio (relax)  & 35.714 & 52.584 \\
    Discovery Studio          & 36.554 & 53.976 \\
    GNINA (hybrid)            & 33.472 & 59.558 \\
    GNINA (hybrid) (relax)    & 34.030 & 61.788 \\
    \midrule
    \multicolumn{3}{c}{\textbf{AI docking methods}} \\
    EquiBind                  & 0.000  & 0.420  \\
    FABind                    & 1.254  & 2.092  \\
    TankBind                  & 1.394  & 4.186  \\
    EquiBind (relax)          & 0.558  & 4.326  \\
    DeepDock                  & 3.348  & 6.410  \\
    DynamicBind               & 3.624  & 7.810  \\
    FABind (relax)            & 4.464  & 12.690 \\
    TankBind (relax)          & 5.022  & 12.834 \\
    DeepDock (relax)          & 5.720  & 15.058 \\
    DiffDock                  & 9.342  & 16.034 \\
    DiffDock-L                & 17.428 & 25.102 \\
    DynamicBind (relax)       & 11.012 & 25.518 \\
    DiffDock-Pocket           & 18.684 & 29.006 \\
    DiffDock (relax)          & 17.986 & 34.588 \\
    SurfDock                  & 32.506 & 41.578 \\
    DiffDock-L (relax)        & 29.426 & 44.770 \\
    Interformer               & 24.402 & 48.108 \\
    DiffDock-Pocket (relax)   & 27.334 & 49.510 \\
    Uni-Mol (relax)           & 39.474 & 59.140 \\
    Uni-Mol                   & 40.730 & 60.530 \\
    Interformer (relax)       & 34.312 & 63.316 \\
    SurfDock (relax)          & \textbf{50.078} & \textbf{74.210} \\
    \midrule
    \multicolumn{3}{c}{\textbf{AI co-folding methods}} \\
    NeuralPLexer              & 0.698  & 1.396  \\
    RFAA                      & 4.180  & 7.106  \\
    NeuralPLexer (relax)      & 5.572  & 15.612 \\
    RFAA (relax)              & 12.126 & 27.052 \\
    Boltz-1                   & 25.660 & 34.306 \\
    Boltz-1 (relax)           & 29.142 & 42.392 \\
    Chai-1                    & 33.054 & 43.654 \\
    Protenix                  & 34.592 & 44.770 \\
    AlphaFold 3               & 34.730 & 45.468 \\
    Protenix (relax)          & 35.706 & 47.700 \\
    Chai-1 (relax)            & 35.566 & 49.512 \\
    Boltz-1x (relax)          & 34.866 & 51.880 \\
    AlphaFold 3 (relax)       & 38.354 & 51.886 \\
    Boltz-1x                  & 36.114 & 52.300 \\
    \midrule
     \textbf{MolAS} & \textbf{51.616} & \textbf{74.628} \\
     \midrule
     VBS@1 & 92.052 & 99.304 \\
     VBS@2 & 89.540 & 99.304 \\
     VBS@3 & 85.216 & 97.072 \\
\bottomrule
\end{tabular}

\end{minipage}
}
\hfill
% ==========================
% BOTTOM RIGHT
% ==========================
\begin{minipage}[t]{0.366\textwidth}
\raggedleft
\textbf{\scriptsize (\%) Selection frequencies on PoseX-SD.}\\[4pt]

\hspace*{-5em}
\begin{tikzpicture}

% ---- Viridis picks ----
\definecolor{vir2}{HTML}{2a788e} % teal
\definecolor{vir1}{HTML}{fd9668} % orange
\begin{axis}[
    xbar,
    y dir=reverse,
    bar width=3pt,
    tick label style={/pgf/number format/fixed, /pgf/number format/precision=2},
    legend reversed=true,
    width=6cm,
    height=23.5cm,
    xmin=0, xmax=70,
    symbolic y coords={
        DiffDock,
        EquiBind,
        EquiBindRelax,
        FABind,
        NeuralPLexer,
        NeuralPLexerRelax,
        RFAA,
        TankBind,
        TankBindRelax,
        DeepDock,
        Boltz1Relax,
        DynamicBind,
        FABindRelax,
        DeepDockRelax,
        DynamicBindRelax,
        AutoDockVina,
        Interformer,
        RFAARelax,
        Boltz1,
        DiffDockL,
        DiffDockPocket,
        Boltz1xRelax,
        DiffDockRelax,
        IFD,
        IFDRelax,
        MOERelax,
        GNINA,
        AlphaFold3Relax,
        AutoDockVinaRelax,
        ProtenixRelax,
        Boltz1x,
        Chai1Relax,
        DiffDockLRelax,
        UniMol,
        Glide,
        GNINARelax,
        DiffDockPocketRelax,
        GlideRelax,
        InterformerRelax,
        MOE,
        UniMolRelax,
        Chai1,
        Protenix,
        DiscoveryStudio,
        DiscoveryStudioRelax,
        SurfDock,
        SurfDockRelax,
        AlphaFold3
    },
    yticklabels={
        {DiffDock},
        {EquiBind},
        {EquiBind (relax)},
        {FABind},
        {NeuralPLexer},
        {NeuralPLexer (relax)},
        {RFAA},
        {TankBind},
        {TankBind (relax)},
        {DeepDock},
        {Boltz-1 (relax)},
        {DynamicBind},
        {FABind (relax)},
        {DeepDock (relax)},
        {DynamicBind (relax)},
        {AutoDock Vina},
        {Interformer},
        {RFAA (relax)},
        {Boltz-1},
        {DiffDock-L},
        {DiffDock-Pocket},
        {Boltz-1x (relax)},
        {DiffDock (relax)},
        {Glide (IFD)},
        {Glide (IFD) (relax)},
        {MOE (relax)},
        {GNINA},
        {AlphaFold 3 (relax)},
        {AutoDock Vina (relax)},
        {Protenix (relax)},
        {Boltz-1x},
        {Chai-1 (relax)},
        {DiffDock-L (relax)},
        {Uni-Mol},
        {Glide},
        {GNINA (relax)},
        {DiffDock-Pocket (relax)},
        {Glide (relax)},
        {Interformer (relax)},
        {MOE},
        {Uni-Mol (relax)},
        {Chai-1},
        {Protenix},
        {Discovery Studio},
        {Discovery Studio (relax)},
        {SurfDock},
        {SurfDock (relax)},
        {AlphaFold 3}
    },
    ytick=data,
    yticklabel style={font=\scriptsize},
    enlarge y limits=0.015,
    point meta=x,
    nodes near coords align={horizontal},
    every node near coord/.append style={
        anchor=east,
        xshift=1pt,
        font=\tiny,
        /pgf/number format/fixed,
        /pgf/number format/precision=2
    },
    xmajorgrids=true,
    major grid style={dashed, thin, black!30}
]

% ---- MolAS (Selection_Count) ----
\addplot+[
    fill=vir2!70, draw=vir2!90,
    every node near coord/.append style={text=vir2}
] coordinates {
    (0,DiffDock)
    (0,EquiBind)
    (0,EquiBindRelax)
    (0,FABind)
    (0,NeuralPLexer)
    (0,NeuralPLexerRelax)
    (0,RFAA)
    (0,TankBind)
    (0,TankBindRelax)
    (0,DeepDock)
    (0,Boltz1Relax)
    (0,DynamicBind)
    (0,FABindRelax)
    (0,DeepDockRelax)
    (0,DynamicBindRelax)
    (0,AutoDockVina)
    (0.41841,Interformer)
    (0,RFAARelax)
    (0,Boltz1)
    (0,DiffDockL)
    (0,DiffDockPocket)
    (0.55788,Boltz1xRelax)
    (0.13947,DiffDockRelax)
    (0,IFD)
    (0,IFDRelax)
    (0,MOERelax)
    (3.20781,GNINA)
    (0.83682,AlphaFold3Relax)
    (0,AutoDockVinaRelax)
    (0.69735,ProtenixRelax)
    (14.5049,Boltz1x)
    (0,Chai1Relax)
    (0.41841,DiffDockLRelax)
    (3.62622,UniMol)
    (0.55788,Glide)
    (4.1841,GNINARelax)
    (0,DiffDockPocketRelax)
    (0.13947,GlideRelax)
    (0.55788,InterformerRelax)
    (0.55788,MOE)
    (0.13947,UniMolRelax)
    (0,Chai1)
    (0.13947,Protenix)
    (1.67364,DiscoveryStudio)
    (1.67364,DiscoveryStudioRelax)
    (0.27894,SurfDock)
    (65.6904,SurfDockRelax)
    (0,AlphaFold3)
};

% ---- Oracle (Oracle_Count) ----
\addplot+[
    fill=vir1!70, draw=vir1!90,
    every node near coord/.append style={text=vir1}
] coordinates {
    (0,DiffDock)
    (0,EquiBind)
    (0,EquiBindRelax)
    (0,FABind)
    (0,NeuralPLexer)
    (0,NeuralPLexerRelax)
    (0,RFAA)
    (0,TankBind)
    (0,TankBindRelax)
    (0.13947,DeepDock)
    (0.27894,Boltz1Relax)
    (0.27894,DynamicBind)
    (0.27894,FABindRelax)
    (0.41841,DeepDockRelax)
    (0.41841,DynamicBindRelax)
    (0.55788,AutoDockVina)
    (0.55788,Interformer)
    (0.55788,RFAARelax)
    (0.69735,Boltz1)
    (0.83682,DiffDockL)
    (0.83682,DiffDockPocket)
    (0.97629,Boltz1xRelax)
    (0.97629,DiffDockRelax)
    (0.97629,IFD)
    (0.97629,IFDRelax)
    (1.53417,MOERelax)
    (1.81311,GNINA)
    (1.95258,AlphaFold3Relax)
    (1.95258,AutoDockVinaRelax)
    (1.95258,ProtenixRelax)
    (2.09205,Boltz1x)
    (2.09205,Chai1Relax)
    (2.09205,DiffDockLRelax)
    (2.37099,UniMol)
    (2.51046,Glide)
    (2.64993,GNINARelax)
    (2.7894,DiffDockPocketRelax)
    (2.7894,GlideRelax)
    (3.34728,InterformerRelax)
    (3.48675,MOE)
    (3.62622,UniMolRelax)
    (3.76569,Chai1)
    (5.43933,Protenix)
    (5.85774,DiscoveryStudio)
    (6.83403,DiscoveryStudioRelax)
    (8.50767,SurfDock)
    (10.3208,SurfDockRelax)
    (10.4603,AlphaFold3)
};

\legend{MolAS, VBS}

\end{axis}

\end{tikzpicture}
\end{minipage}
\caption{Table of performances over the algorithm portofolio (left) and histogram of MolAS/VBS selection frequencies (right) on \textbf{PoseX-SD}. In the table, the portfolio-wise best and MolAS performances are in \textbf{bold}. 
Performances are ordered by ascending percentage of poses below $2~\text{\AA}$ and PoseBusters-valid, 
while frequencies are ordered by ascending VBS frequency.}
\label{fig:detail_posexsd}
\end{figure*}

\begin{figure*}
\centering
\hspace*{-7em}
% ==========================
% BOTTOM LEFT
% ==========================
\scalebox{0.88}{
\begin{minipage}[t]{0.65\textwidth}
\raggedright
\textbf{\small Averaged 5-fold results on PoseX-CD.}\\[6pt]

\begin{tabular}{p{3.9cm}cc}
\toprule
    \textbf{Method} &
    \multicolumn{2}{c}{$(\%)\ \text{RMSD}\leq x~\text{\AA}\ \&\ \text{PB-valid}$}\\
    \cmidrule(lr){2-3}
    & $x=1$ & $x=2$\\
    \midrule
    \multicolumn{3}{c}{\textbf{Physics-based methods}} \\
    AutoDock Vina             & 13.338 & 27.288 \\
    AutoDock Vina (relax)     & 14.634 & 28.810 \\
    MOE                       & 13.870 & 29.270 \\
    MOE (relax)               & 14.178 & 29.422 \\
    Glide                     & 14.710 & 33.306 \\
    Glide (relax)             & 16.006 & 33.840 \\
    Glide (IFD)               & 17.230 & 37.958 \\
    Glide (IFD) (relax)       & 18.602 & 38.338 \\
    Discovery Studio (relax)  & 22.252 & 40.850 \\
    Discovery Studio          & 21.642 & 40.924 \\
    GNINA (hybrid)            & 19.740 & 48.088 \\
    GNINA (hybrid) (relax)    & 22.408 & 51.290 \\
    \midrule
    \multicolumn{3}{c}{\textbf{AI docking methods}} \\
    EquiBind                  & 0.000  & 0.608 \\
    DeepDock                  & 1.600  & 5.560 \\
    FABind                    & 1.906  & 5.942 \\
    TankBind                  & 2.744  & 5.944 \\
    EquiBind (relax)          & 1.450  & 7.470 \\
    DynamicBind               & 5.486  & 9.758 \\
    DeepDock (relax)          & 4.802  & 15.474 \\
    TankBind (relax)          & 8.078  & 20.502 \\
    DiffDock                  & 13.948 & 24.620 \\
    FABind (relax)            & 11.506 & 28.582 \\
    DiffDock-L                & 19.964 & 29.950 \\
    DynamicBind (relax)       & 14.788 & 35.522 \\
    DiffDock-Pocket           & 21.494 & 36.888 \\
    SurfDock                  & 30.414 & 37.350 \\
    Interformer               & 19.664 & 38.106 \\
    DiffDock (relax)          & 26.904 & 48.470 \\
    DiffDock-L (relax)        & 32.540 & 52.128 \\
    Interformer (relax)       & 28.046 & 54.572 \\
    Uni-Mol                   & 33.382 & 54.648 \\
    Uni-Mol (relax)           & 33.916 & 55.562 \\
    DiffDock-Pocket (relax)   & 32.238 & 58.382 \\
    SurfDock (relax)          & \textbf{49.080} & \textbf{73.016} \\
    \midrule
    \multicolumn{3}{c}{\textbf{AI co-folding methods}} \\
    NeuralPLexer              & 0.456  & 2.436 \\
    RFAA                      & 4.264  & 8.080 \\
    NeuralPLexer (relax)      & 5.412  & 18.598 \\
    RFAA (relax)              & 9.298  & 22.102 \\
    Boltz-1                   & 32.700 & 48.244 \\
    Protenix                  & 33.456 & 49.384 \\
    Protenix (relax)          & 35.208 & 52.130 \\
    Chai-1                    & 41.312 & 55.338 \\
    AlphaFold 3               & 45.808 & 58.534 \\
    Boltz-1 (relax)           & 39.248 & 58.686 \\
    Chai-1 (relax)            & 44.742 & 60.670 \\
    AlphaFold 3 (relax)       & 46.264 & 62.500 \\
    Boltz-1x                  & 44.816 & 65.162 \\
    Boltz-1x (relax)          & 43.520 & 65.390 \\
    \midrule
     \textbf{MolAS} & \textbf{66.386} & \textbf{87.496} \\
     \midrule
     VBS@1 & 87.114 & 96.646 \\
     VBS@2 & 84.598 & 95.960 \\
     VBS@3 & 79.264 & 93.900 \\
\bottomrule
\end{tabular}

\end{minipage}
}
\hfill
% ==========================
% BOTTOM RIGHT
% ==========================
\begin{minipage}[t]{0.366\textwidth}
\raggedleft
\textbf{\scriptsize (\%) Selection frequencies on PoseX-CD.}\\[4pt]

\hspace*{-5em}
\begin{tikzpicture}

% ---- Viridis picks ----
\definecolor{vir2}{HTML}{2a788e} % teal
\definecolor{vir1}{HTML}{fd9668} % orange
\begin{axis}[
    xbar,
    y dir=reverse,
    bar width=3pt,
    tick label style={/pgf/number format/fixed, /pgf/number format/precision=2},
    legend reversed=true,
    width=6cm,
    height=23.5cm,
    xmin=0, xmax=70,
    symbolic y coords={
        DeepDock,
        DiffDock,
        DynamicBind,
        EquiBind,
        EquiBindRelax,
        FABind,
        NeuralPLexer,
        NeuralPLexerRelax,
        RFAA,
        TankBind,
        AutoDockVina,
        RFAARelax,
        DiffDockL,
        FABindRelax,
        DeepDockRelax,
        GNINA,
        TankBindRelax,
        DynamicBindRelax,
        IFD,
        DiscoveryStudio,
        Interformer,
        Glide,
        MOE,
        AutoDockVinaRelax,
        GlideRelax,
        GNINARelax,
        IFDRelax,
        DiffDockPocket,
        DiffDockRelax,
        Boltz1xRelax,
        UniMol,
        InterformerRelax,
        MOERelax,
        DiffDockLRelax,
        BoltzRelax,
        Boltz1,
        ProtenixRelax,
        DiscoveryStudioRelax,
        Boltz1x,
        UniMolRelax,
        Chai1,
        Protenix,
        AlphaFold3Relax,
        Chai1Relax,
        SurfDock,
        DiffDockPocketRelax,
        AlphaFold3,
        SurfDockRelax
    },
    yticklabels={
        {DeepDock},
        {DiffDock},
        {DynamicBind},
        {EquiBind},
        {EquiBind (relax)},
        {FABind},
        {NeuralPLexer},
        {NeuralPLexer (relax)},
        {RFAA},
        {TankBind},
        {AutoDock Vina},
        {RFAA (relax)},
        {DiffDock-L},
        {FABind (relax)},
        {DeepDock (relax)},
        {GNINA},
        {TankBind (relax)},
        {DynamicBind (relax)},
        {Glide (IFD)},
        {Discovery Studio},
        {Interformer},
        {Glide},
        {MOE},
        {AutoDock Vina (relax)},
        {Glide (relax)},
        {GNINA (relax)},
        {Glide (IFD) (relax)},
        {DiffDock-Pocket},
        {DiffDock (relax)},
        {Boltz-1x (relax)},
        {Uni-Mol},
        {Interformer (relax)},
        {MOE (relax)},
        {DiffDock-L (relax)},
        {Boltz-1 (relax)},
        {Boltz-1},
        {Protenix (relax)},
        {Discovery Studio (relax)},
        {Boltz-1x},
        {Uni-Mol (relax)},
        {Chai-1},
        {Protenix},
        {AlphaFold 3 (relax)},
        {Chai-1 (relax)},
        {SurfDock},
        {DiffDock-Pocket (relax)},
        {AlphaFold 3},
        {SurfDock (relax)}
    },
    ytick=data,
    yticklabel style={font=\scriptsize},
    enlarge y limits=0.015,
    point meta=x,
    nodes near coords align={horizontal},
    every node near coord/.append style={
        anchor=east,
        xshift=1pt,
        font=\tiny,
        /pgf/number format/fixed,
        /pgf/number format/precision=2
    },
    xmajorgrids=true,
    major grid style={dashed, thin, black!30}
]

% ---- MolAS (Selection_Count) ----
\addplot+[
    fill=vir2!70, draw=vir2!90,
    every node near coord/.append style={text=vir2}
] coordinates {
    (0,DeepDock)
    (0,DiffDock)
    (0,DynamicBind)
    (0,EquiBind)
    (0,EquiBindRelax)
    (0,FABind)
    (0,NeuralPLexer)
    (0,NeuralPLexerRelax)
    (0,RFAA)
    (0,TankBind)
    (0,AutoDockVina)
    (0.381098,RFAARelax)
    (0,DiffDockL)
    (0.0762195,FABindRelax)
    (0,DeepDockRelax)
    (0,GNINA)
    (0,TankBindRelax)
    (0.0762195,DynamicBindRelax)
    (0.304878,IFD)
    (1.14329,DiscoveryStudio)
    (0.914634,Interformer)
    (0.685976,Glide)
    (0.304878,MOE)
    (0.0762195,AutoDockVinaRelax)
    (1.75305,GlideRelax)
    (0.990854,GNINARelax)
    (0.304878,IFDRelax)
    (0.609756,DiffDockPocket)
    (1.06707,DiffDockRelax)
    (3.96341,Boltz1xRelax)
    (2.89634,UniMol)
    (3.04878,InterformerRelax)
    (0.914634,MOERelax)
    (1.06707,DiffDockLRelax)
    (7.24085,BoltzRelax)
    (1.37195,Boltz1)
    (3.27744,ProtenixRelax)
    (0.990854,DiscoveryStudioRelax)
    (7.0122,Boltz1x)
    (6.6311,UniMolRelax)
    (1.75305,Chai1)
    (4.26829,Protenix)
    (9.7561,AlphaFold3Relax)
    (10.8232,Chai1Relax)
    (0.0762195,SurfDock)
    (5.25915,DiffDockPocketRelax)
    (12.0427,AlphaFold3)
    (8.84146,SurfDockRelax)
};

% ---- Oracle (Oracle_Count) ----
\addplot+[
    fill=vir1!70, draw=vir1!90,
    every node near coord/.append style={text=vir1}
] coordinates {
    (0,DeepDock)
    (0,DiffDock)
    (0,DynamicBind)
    (0,EquiBind)
    (0,EquiBindRelax)
    (0,FABind)
    (0,NeuralPLexer)
    (0,NeuralPLexerRelax)
    (0,RFAA)
    (0,TankBind)
    (0.228659,AutoDockVina)
    (0.304878,RFAARelax)
    (0.381098,DiffDockL)
    (0.381098,FABindRelax)
    (0.457317,DeepDockRelax)
    (0.457317,GNINA)
    (0.457317,TankBindRelax)
    (0.685976,DynamicBindRelax)
    (0.685976,IFD)
    (0.762195,DiscoveryStudio)
    (0.838415,Interformer)
    (0.914634,Glide)
    (0.914634,MOE)
    (0.990854,AutoDockVinaRelax)
    (1.29573,GlideRelax)
    (1.29573,GNINARelax)
    (1.37195,IFDRelax)
    (1.44817,DiffDockPocket)
    (1.44817,DiffDockRelax)
    (1.67683,Boltz1xRelax)
    (1.67683,UniMol)
    (1.75305,InterformerRelax)
    (2.05793,MOERelax)
    (2.28659,DiffDockLRelax)
    (2.43902,BoltzRelax)
    (2.51524,Boltz1)
    (2.82012,ProtenixRelax)
    (3.04878,DiscoveryStudioRelax)
    (3.35366,Boltz1x)
    (3.73476,UniMolRelax)
    (4.34451,Chai1)
    (4.80183,Protenix)
    (4.95427,AlphaFold3Relax)
    (5.10671,Chai1Relax)
    (6.17378,SurfDock)
    (6.32622,DiffDockPocketRelax)
    (12.4238,AlphaFold3)
    (13.186,SurfDockRelax)
};

\legend{MolAS, VBS}

\end{axis}

\end{tikzpicture}
\end{minipage}
\caption{Table of performances over the algorithm portofolio (left) and histogram of MolAS/VBS selection frequencies (right) on \textbf{PoseX-CD}. In the table, the portfolio-wise best and MolAS performances are in \textbf{bold}. 
Performances are ordered by ascending percentage of poses below $2~\text{\AA}$ and PoseBusters-valid, 
while frequencies are ordered by ascending VBS frequency.}
\label{fig:detail_posexcd}
\end{figure*}

\begin{figure*}
\centering
\hspace*{-7em}
% ==========================
% BOTTOM LEFT
% ==========================
\begin{minipage}[t]{0.595\textwidth}
\raggedright
\textbf{\small(c) Averaged 5-fold results on PoseBusters.}\\[6pt]

\scalebox{0.9}{
\begin{tabular}{p{4.1cm}cc}
\toprule
    \textbf{Method} &
    \multicolumn{2}{c}{$(\%)\ \text{RMSD}\leq x~\text{\AA}\ \&\ \text{PB-valid}$}\\
    \cmidrule(lr){2-3}
    & $x=1$ & $x=2$\\
    \midrule
    \multicolumn{3}{c}{\textbf{Physics-based methods}} \\
    AutoDock Vina (MM-min)     & 28.724 & 44.150 \\
    GOLD (MM-min)              & 27.582 & 45.102 \\
    GOLD                      & 32.712 & 48.364 \\
    AutoDock Vina             & \textbf{34.338} & \textbf{51.166} \\
    \midrule
    \multicolumn{3}{c}{\textbf{AI docking methods}} \\
    EquiBind                  & 0.000  & 0.704  \\
    Uni-Mol                   & 1.400  & 2.096  \\
    TankBind                  & 0.936  & 2.570  \\
    DeepDock                  & 2.572  & 4.910  \\
    EquiBind (MM-min)          & 2.342  & 7.488  \\
    DiffDock                  & 8.888  & 14.026 \\
    TankBind (MM-min)          & 5.842  & 14.252 \\
    DeepDock (MM-min)          & 6.544  & 15.188 \\
    Uni-Mol (MM-min)           & 7.944  & 19.848 \\
    DiffDock (MM-min)          & 17.052 & 35.982 \\
    \midrule
    \textbf{MolAS}            & \textbf{36.688} & \textbf{54.910} \\
    \midrule
    VBS@1                     & 60.750 & 82.714 \\
    VBS@2                     & 48.600 & 78.746 \\
    VBS@3                     & 38.312 & 68.004 \\
\bottomrule
\end{tabular}
}

\end{minipage}
\hfill
% ==========================
% BOTTOM RIGHT
% ==========================
\begin{minipage}[t]{0.366\textwidth}
\raggedleft
\textbf{\small(d) (\%) Selection frequencies on PoseBusters.}\\[4pt]

\hspace*{-7em}
\begin{tikzpicture}

% ---- Viridis picks ----
\definecolor{vir2}{HTML}{2a788e} % teal
\definecolor{vir1}{HTML}{fd9668} % orange
\begin{axis}[
    xbar,
    y dir=reverse,
    bar width=4.5pt,
    tick label style={/pgf/number format/fixed, /pgf/number format/precision=2},
    legend reversed=true,
    width=7.7cm,
    height=10.5cm,
    xmin=0, xmax=70,
    symbolic y coords={
        TankBind,
        EquiBind,
        UniMol,
        EquiBindRelax,
        TankBindRelax,
        DiffDock,
        DeepDockRelax,
        UniMolRelax,
        DeepDock,
        AutoDockVinaRelax,
        GOLDRelax,
        DiffDockRelax,
        GOLD,
        AutoDockVina
    },
    yticklabels={
        {TankBind},
        {EquiBind},
        {Uni-Mol},
        {EquiBind (MM-min)},
        {TankBind (MM-min)},
        {DiffDock},
        {DeepDock (MM-min)},
        {Uni-Mol (MM-min)},
        {DeepDock},
        {AutoDock Vina (MM-min)},
        {GOLD (MM-min)},
        {DiffDock (MM-min)},
        {GOLD},
        {AutoDock Vina}
    },
    ytick=data,
    yticklabel style={font=\small},
    enlarge y limits=0.05,
    point meta=x,
    nodes near coords align={horizontal},
    every node near coord/.append style={
        anchor=east,
        xshift=1pt,
        font=\tiny,
        /pgf/number format/fixed,
        /pgf/number format/precision=2
    },
    xmajorgrids=true,
    major grid style={dashed, thin, black!30}
]

% ---- MolAS (Selection_Count) ----
\addplot+[
    fill=vir2!70, draw=vir2!90,
    every node near coord/.append style={text=vir2}
] coordinates {
    (0,TankBind)
    (0,EquiBind)
    (0.233645,UniMol)
    (0,EquiBindRelax)
    (0,TankBindRelax)
    (0,DiffDock)
    (0.46729,DeepDockRelax)
    (0.46729,UniMolRelax)
    (0,DeepDock)
    (2.1028,AutoDockVinaRelax)
    (14.2523,GOLDRelax)
    (20.5607,DiffDockRelax)
    (24.0654,GOLD)
    (37.8505,AutoDockVina)
};

% ---- Oracle (Oracle_Count) ----
\addplot+[
    fill=vir1!70, draw=vir1!90,
    every node near coord/.append style={text=vir1}
] coordinates {
    (0,TankBind)
    (0.233645,EquiBind)
    (0.700935,UniMol)
    (2.33645,EquiBindRelax)
    (2.80374,TankBindRelax)
    (3.50467,DiffDock)
    (3.97196,DeepDockRelax)
    (5.60748,UniMolRelax)
    (7.24299,DeepDock)
    (9.34579,AutoDockVinaRelax)
    (14.9533,GOLDRelax)
    (15.4206,DiffDockRelax)
    (16.3551,GOLD)
    (17.5234,AutoDockVina)
};

\legend{MolAS, VBS}

\end{axis}

\end{tikzpicture}
\end{minipage}
\caption{Table of performances over the algorithm portofolio (left) and histogram of MolAS/VBS selection frequencies (right) on \textbf{PoseBusters}. In the table, the portfolio-wise best and MolAS performances are in \textbf{bold}. 
Performances are ordered by ascending percentage of poses below $2~\text{\AA}$ and PoseBusters-valid, 
while frequencies are ordered by ascending VBS frequency.}
\label{fig:detail_posebusters}
\end{figure*}

\subsubsection{Embedding-space diagnostics (PCA/t-SNE/centroids).}\label{app:failure_supplements}
\begin{figure*}[h!]
\centering

% ---------- Row 1 ----------
\begin{minipage}[t]{0.325\textwidth}
    \centering
    {\small \textbf{(a)} PCA spectral decay}
    \hspace*{-8em}
    \includegraphics[width=1.5\textwidth]{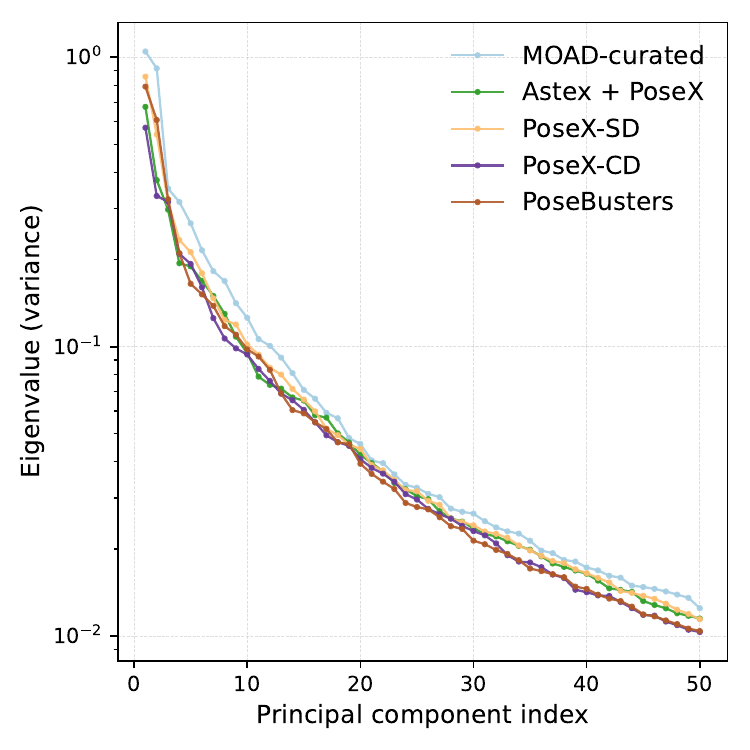}
\end{minipage}
\hfill
\begin{minipage}[t]{0.325\textwidth}
    \centering
    {\small \textbf{(b)} t-SNE of \textit{MOAD-curated}}
    \hspace*{-2em}
    \includegraphics[width=1.4\textwidth]{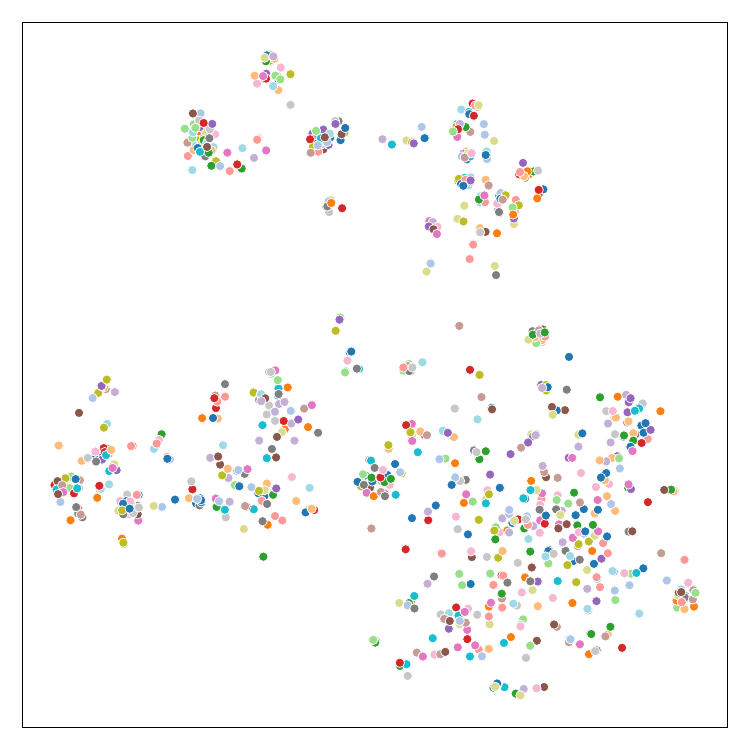}
\end{minipage}
\hfill
\begin{minipage}[t]{0.325\textwidth}
    \centering
    {\small \textbf{(c)} t-SNE of \textit{PoseX + Astex}}
    \hspace*{2em}
    \includegraphics[width=1.4\textwidth]{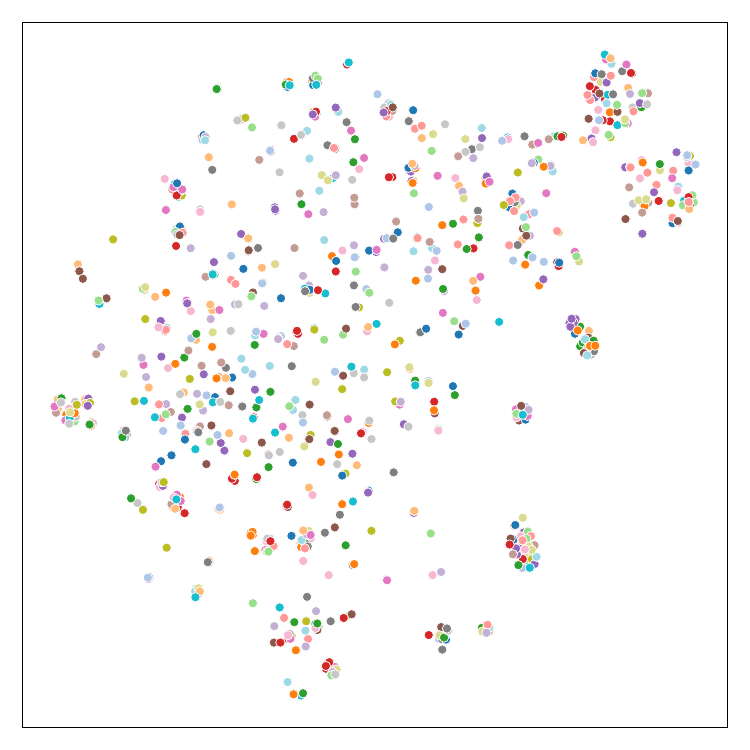}
\end{minipage}

% \vspace{0.5em} % spacing between rows

% ---------- Row 2 ----------
\begin{minipage}[t]{0.325\textwidth}
    \centering
    {\small \textbf{(d)} t-SNE of \textit{PoseX-SD}}
    \hspace*{-7em}
    \includegraphics[width=1.4\textwidth]{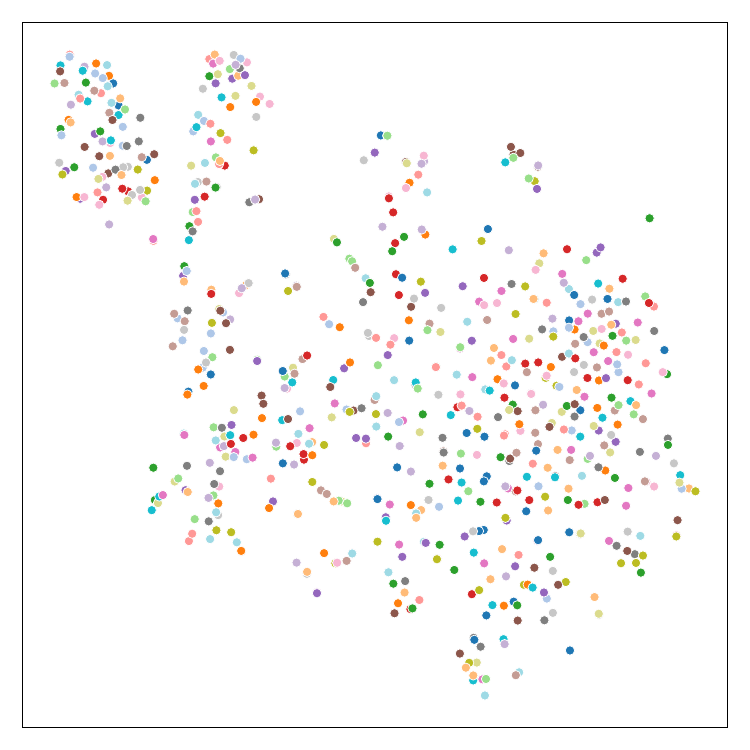}
\end{minipage}
\hfill
\begin{minipage}[t]{0.325\textwidth}
    \centering
    {\small \textbf{(e)} t-SNE of \textit{PoseX-CD}}
    \hspace*{-2em}
    \includegraphics[width=1.4\textwidth]{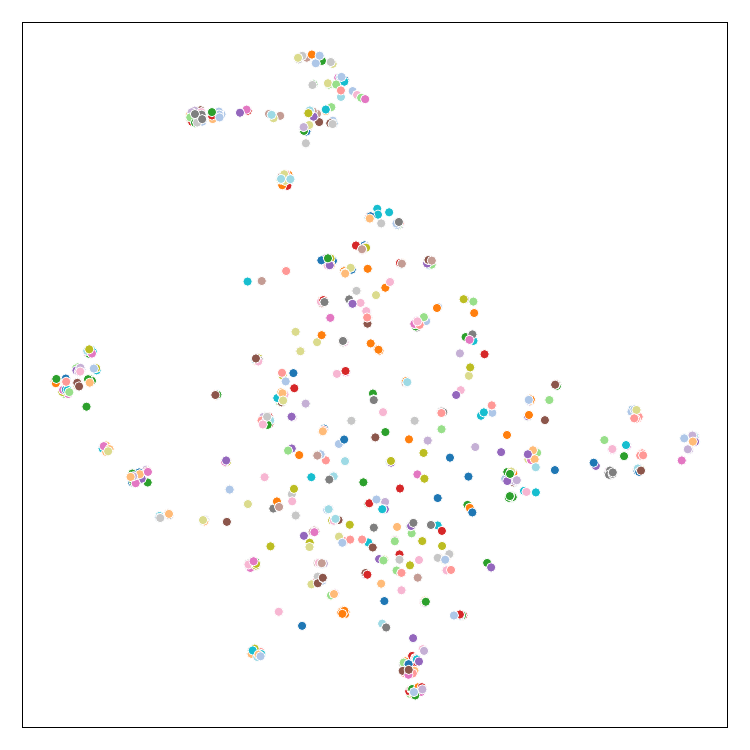}
\end{minipage}
\hfill
\begin{minipage}[t]{0.325\textwidth}
    \centering
    {\small \textbf{(f)} t-SNE of \textit{PoseBusters}}
    \hspace*{2em}
    \includegraphics[width=1.4\textwidth]{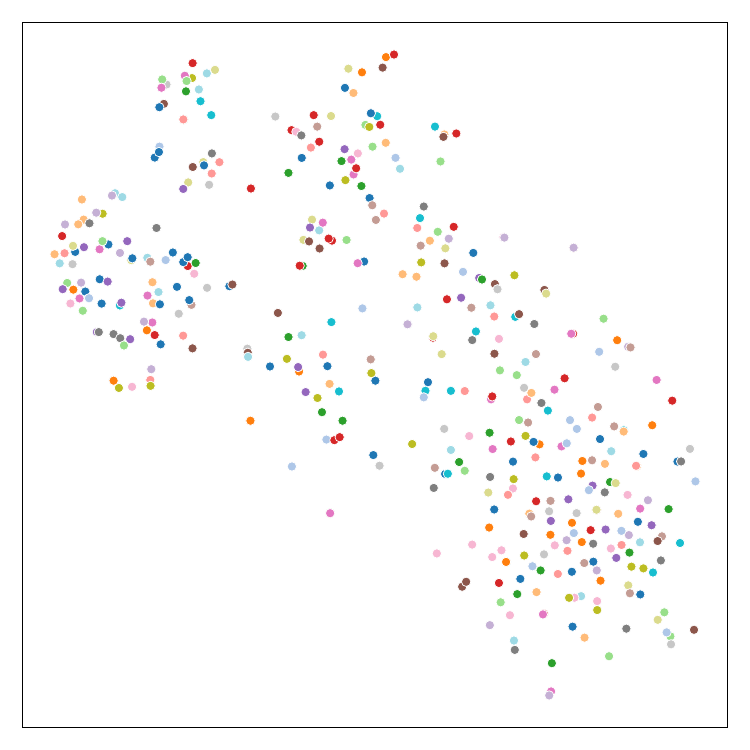}
\end{minipage}

\caption{(a) Eigenvalue spectra via PCA. (b - f) t-SNE visualisations of embedding spaces across benchmarks. Each panel shows a 2D projection of embedding clusters coloured by their oracle-selected algorithm.  }
\label{fig:tsne}
\end{figure*}

\begin{figure*}
    \centering
    \hspace*{-10em}
    \includegraphics[width=1.5\textwidth]{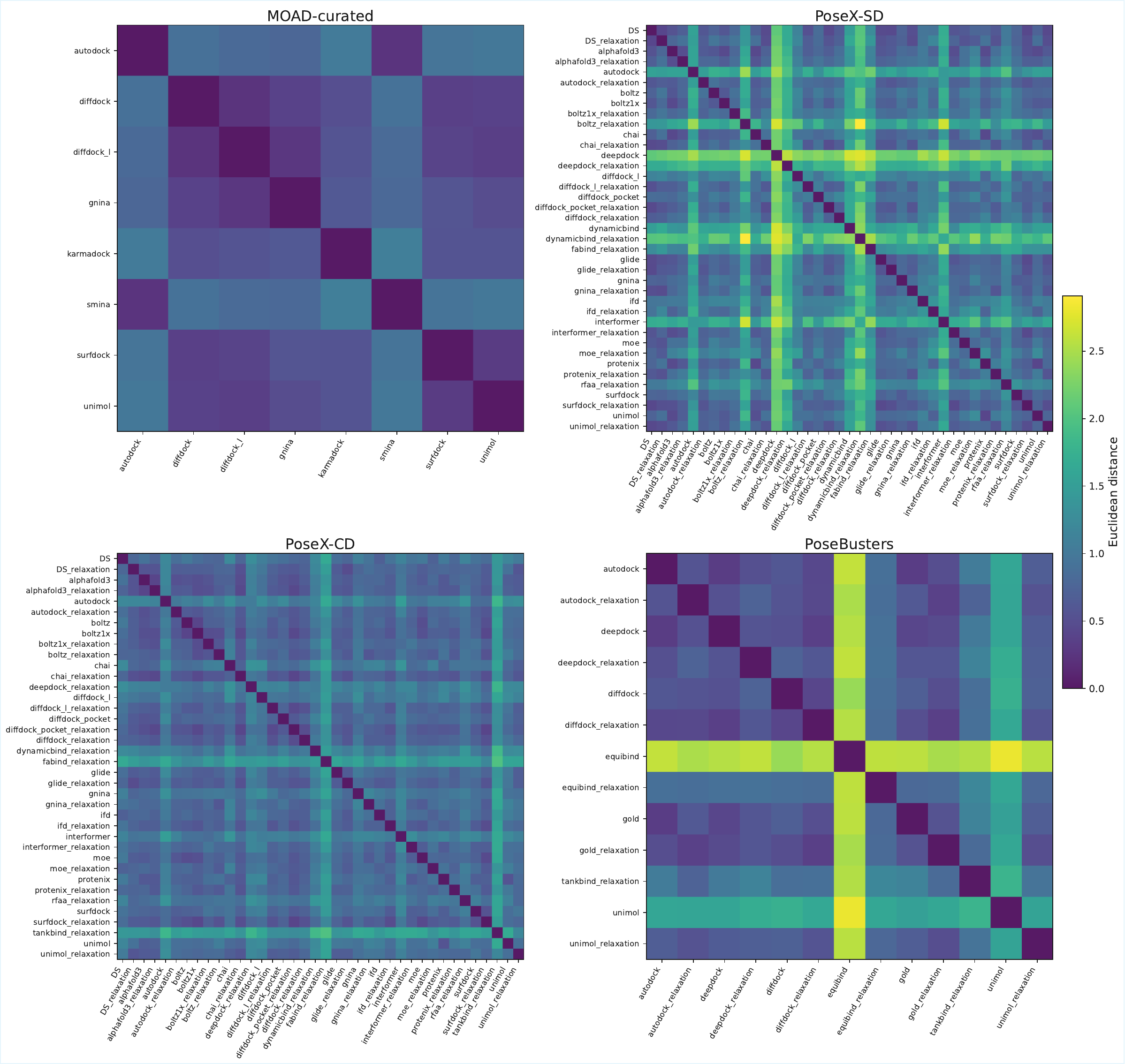}
    \caption{Pairwise Euclidean distances between embedding-space centroids of docking algorithms for each benchmark. Darker values indicate smaller distances; diagonal entries are zero by definition. Separate panels show MOAD-curated, PoseX-SD, PoseX-CD, and PoseBusters.}
    \label{fig:inter-algo-distance}
\end{figure*}

We examined whether MolAS failures correlate with gross differences in the pooled protein+ligand embedding space using PCA spectra, t-SNE projections, and solver-wise centroid statistics (Fig.~\ref{fig:tsne}, Fig.~\ref{fig:inter-algo-distance}, Table~\ref{tab:embed_stats}). 
Across benchmarks, the PCA eigenvalue spectra are broadly similar, with comparable decay and no evidence of a low-rank collapse. This suggests that the difficult regimes (e.g., PoseX-SD and PoseBusters) are unlikely to be explained by a trivial loss of embedding variability alone.

At the same time, both the t-SNE projections and the cluster statistics indicate weak alignment between the pretrained embedding geometry and oracle-defined solver regions. The projections form multi-lobed clouds rather than clean solver-separated clusters, which is expected because the embeddings are not trained to discriminate docking pipelines. Consistently, silhouette scores are negative across benchmarks (Table~\ref{tab:embed_stats}), indicating substantial overlap between solver-associated regions under this oracle colouring, with the poorest separability observed on PoseX-SD and PoseX-CD.

Centroid-distance heatmaps provide a complementary solver-level view. On MOAD-curated, centroid separations are relatively homogeneous, consistent with a compact embedding geometry in which solver centroids are not strongly isolated. On PoseX benchmarks, the matrices become denser with repeated block structure, suggesting that many solvers occupy overlapping regions of the embedding space, which is compatible with the high oracle diversity observed there. PoseBusters exhibits a different pattern in which a small number of solvers (notably EquiBind, and to a lesser extent Uni-Mol) form more distinct centroid blocks relative to the remaining methods.

Overall, these diagnostics do not support a collapse of the embedding feature spectrum, but they do suggest that pretrained embeddings provide limited solver-specific separability under several workflows. In regimes where oracle diversity is high and embedding separability is weak (Table~\ref{tab:embed_stats}), MolAS tends to default towards an SBS-like strategy, consistent with the selection-collapse patterns in the main text.

\begin{table*}
    \centering
    \caption{Summary of embedding-space statistics and oracle-landscape properties across benchmarks. Cluster tightness and Silhouette score quantify per-algorithm embedding dispersion and separability. VBS score and VBS entropy summarise the oracle success rate and the distribution of oracle-best algorithms.}
    \begin{tabular}{l|cccc}
    \toprule
         Dataset&   Cluster tightness&Silhouette score &VBS score &VBS entropy\\ \midrule
         MOAD-curated& 
     3.305& -0.036& 0.867&2.395\\
 PoseX + Astex& 2.709& -0.143& 0.956&4.699\\
 PoseX-SD& 2.960& -0.173& 0.972&4.675\\
 PoseX-CD& 2.431& -0.183& 0.944&4.590\\
 PoseBusters& 2.982& -0.047& 0.763&3.205\\
 \bottomrule
 \end{tabular}
    \label{tab:embed_stats}
\end{table*}

\subsubsection{Top-k overlap across post-processing settings}\label{app:topk-overlap}
\begin{figure}[t]
    \centering
    \includegraphics[width=0.75\linewidth]{figures/JoC/topk_jaccard_mixed.pdf}\\[0.8em]
    \includegraphics[width=0.75\linewidth]{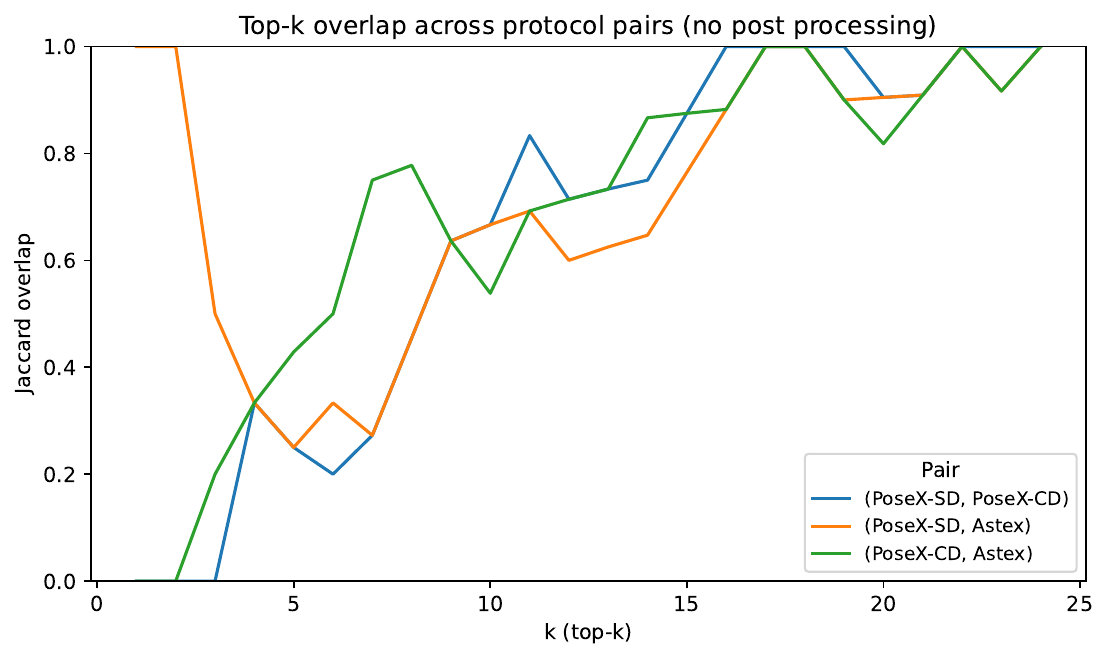}\\[0.8em]
    \includegraphics[width=0.75\linewidth]{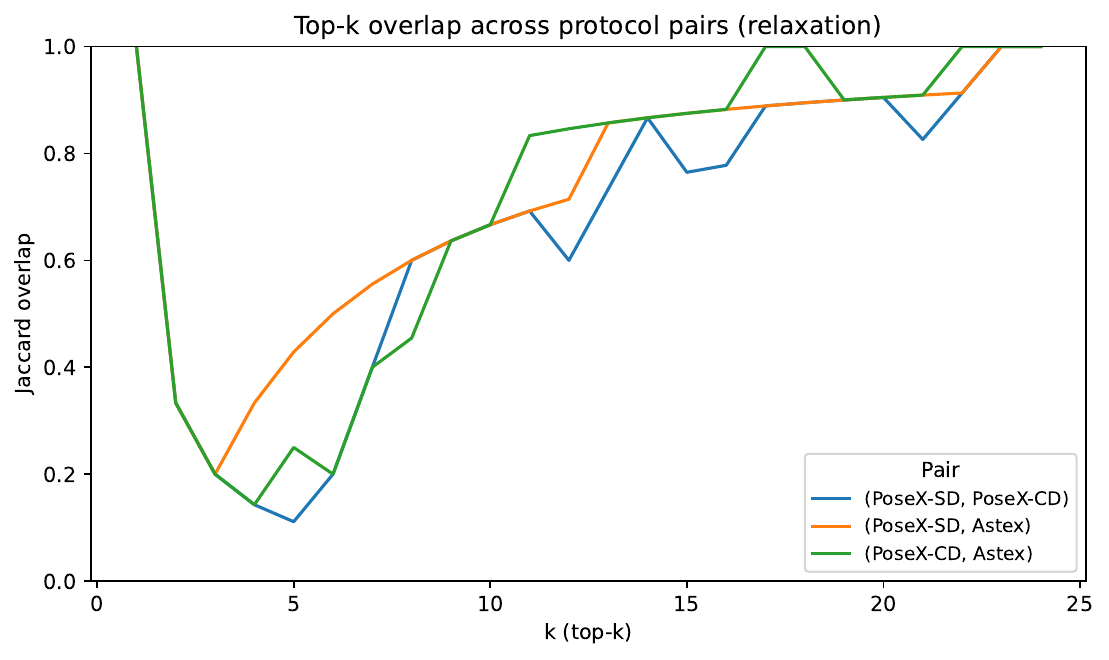}
    \caption{Top-$k$ overlap of solver rankings across docking protocols under three post-processing settings (mixed, no post-processing, relaxation). Each curve reports the Jaccard similarity between the sets of the top-$k$ solvers under a benchmark pair $(p,q)$, for $k=1,\dots,m$, where $m$ is the number of solvers. Higher overlap indicates a more stable solver hierarchy across protocols, while low overlap at small $k$ highlights instability among the highest-ranked solvers.}
    \label{fig:topk_jaccard_appendix}
\end{figure}

Figure~\ref{fig:topk_jaccard_appendix} reports the top-$k$ overlap curves for all three post-processing settings.
Across settings, the qualitative pattern is consistent: overlap is typically low for small-to-moderate $k$ (apart from some $k=1$ when the top solver coincides) and increases towards 1 as $k$ approaches $m$, reflecting instability among near-top solvers.
Accordingly, these curves primarily diagnose upper-tail hierarchy shifts across protocols, which are most relevant for oracle labels and algorithm-selection transfer.

\end{appendices}


\bibliographystyle{plainnat}
\bibliography{JoC/sn-bibliography}

@article{su2018comparative,
  title={Comparative assessment of scoring functions: the CASF-2016 update},
  author={Su, Minyi and Yang, Qifan and Du, Yu and Feng, Guoqin and Liu, Zhihai and Li, Yan and Wang, Renxiao},
  journal={Journal of chemical information and modeling},
  volume={59},
  number={2},
  pages={895--913},
  year={2018},
  publisher={ACS Publications},
  doi={10.1021/acs.jcim.8b00545}
}

@article{francoeur2020three,
  title={Three-dimensional convolutional neural networks and a cross-docked data set for structure-based drug design},
  author={Francoeur, Paul G and Masuda, Tomohide and Sunseri, Jocelyn and Jia, Andrew and Iovanisci, Richard B and Snyder, Ian and Koes, David R},
  journal={Journal of chemical information and modeling},
  volume={60},
  number={9},
  pages={4200--4215},
  year={2020},
  publisher={ACS Publications},
  doi={10.1021/acs.jcim.0c00411}
}

@article{cheng2010knowledge,
  title={A knowledge-guided strategy for improving the accuracy of scoring functions in binding affinity prediction},
  author={Cheng, Tiejun and Liu, Zhihai and Wang, Renxiao},
  journal={BMC bioinformatics},
  volume={11},
  number={1},
  pages={193},
  year={2010},
  publisher={Springer},
  doi={10.1186/1471-2105-11-193}
}

@article{meng2011molecular,
  title={Molecular docking: a powerful approach for structure-based drug discovery},
  author={Meng, Xuan-Yu and Zhang, Hong-Xing and Mezei, Mihaly and Cui, Meng},
  journal={Current computer-aided drug design},
  volume={7},
  number={2},
  pages={146--157},
  year={2011},
  publisher={Bentham Science Publishers direct},
  doi={10.2174/157340911795677602}
}

@article{ballester2010machine,
  title={A machine learning approach to predicting protein--ligand binding affinity with applications to molecular docking},
  author={Ballester, Pedro J and Mitchell, John BO},
  journal={Bioinformatics},
  volume={26},
  number={9},
  pages={1169--1175},
  year={2010},
  publisher={Oxford University Press},
  doi = {10.1093/bioinformatics/btq112}
}

@article{feurer2015efficient,
  title={Efficient and robust automated machine learning},
  author={Feurer, Matthias and Klein, Aaron and Eggensperger, Katharina and Springenberg, Jost and Blum, Manuel and Hutter, Frank},
  journal={Advances in neural information processing systems},
  volume={28},
  year={2015}, 
  url = {https://proceedings.neurips.cc/paper_files/paper/2015/file/11d0e6287202fced83f79975ec59a3a6-Paper.pdf}
}

@inproceedings{thornton2013auto,
  title={Auto-WEKA: Combined selection and hyperparameter optimization of classification algorithms},
  author={Thornton, Chris and Hutter, Frank and Hoos, Holger H and Leyton-Brown, Kevin},
  booktitle={Proceedings of the 19th ACM SIGKDD international conference on Knowledge discovery and data mining},
  pages={847--855},
  year={2013}, 
  doi = {10.1145/2487575.2487629}
}

@article{trott2010autodock,
  title={AutoDock Vina: improving the speed and accuracy of docking with a new scoring function, efficient optimization, and multithreading},
  author={Trott, Oleg and Olson, Arthur J},
  journal={Journal of computational chemistry},
  volume={31},
  number={2},
  pages={455--461},
  year={2010},
  publisher={Wiley Online Library},
  doi={10.1002/jcc.21334}
}

@article{ahmad2022chemberta,
  title={Chemberta-2: Towards chemical foundation models},
  author={Ahmad, Walid and Simon, Elana and Chithrananda, Seyone and Grand, Gabriel and Ramsundar, Bharath},
  journal={arXiv preprint arXiv:2209.01712},
  year={2022}
}

@software{evolutionaryscale_2024,
  author       = {Zeming Lin and
                  Chetan Mishra and
                  santiag0m and
                  Jun Gong and
                  Neil Thomas and
                  Ishaan Mathur and
                  tina-z-jia and
                  Tom Sercu and
                  Steve Chan and
                  Salvatore Candido and
                  Jenna MacCarley and
                  Imran Qureshi and
                  Jordan Safer and
                  robert-verkuil and
                  Ventura Rivera and
                  blenderwang and
                  carolynkim and
                  halilakin},
  title        = {evolutionaryscale/esm: v3.2.3},
  month        = oct,
  year         = 2025,
  publisher    = {Zenodo},
  version      = {v3.2.3},
  doi          = {10.5281/zenodo.17353381},
  url          = {https://doi.org/10.5281/zenodo.17353381},
  swhid        = {swh:1:dir:3974fe98e61bd75f6a8d44d0d7f5d5ddaa55b0ea
                   ;origin=https://doi.org/10.5281/zenodo.14219303;vi
                   sit=swh:1:snp:414464ba93a3745ba5471fdd54e1f91b78f9
                   3b89;anchor=swh:1:rel:a97f7b2f7b2b400392ee0ce7a600
                   add2a2021653;path=evolutionaryscale-esm-23b084b
                  },
}

@article{hu2005binding,
  title={Binding MOAD (mother of all databases)},
  author={Hu, Liegi and Benson, Mark L and Smith, Richard D and Lerner, Michael G and Carlson, Heather A},
  journal={Proteins: Structure, Function, and Bioinformatics},
  volume={60},
  number={3},
  pages={333--340},
  year={2005},
  publisher={Wiley Online Library},
  doi={10.1002/prot.20512}
}

@article{zayed2025optimizing,
  title={Optimizing protein-ligand docking through machine learning: algorithm selection with AutoDock Vina},
  author={Zayed, Ala’Omar Hasan},
  journal={Discover Chemistry},
  volume={2},
  number={1},
  pages={164},
  year={2025},
  publisher={Springer},
  doi={10.1007/s44371-025-00246-4}
}

@inproceedings{yuan2024algorithm,
  title={Algorithm Selection on Molecular Docking for State-of-the-Art Performance},
  author={Yuan, Yiliang and Misir, Mustafa},
  booktitle={Proceedings of the 2024 8th International Conference on Computer Science and Artificial Intelligence},
  pages={295--302},
  year={2024}, 
  doi = {10.1145/3709026.3709046}
}

@inproceedings{yuan2024enhancing,
  title={Enhancing Molecular Docking Performance with a GNN-Based Algorithm Selection Model},
  author={Yuan, Yiliang and Misir, Mustafa},
  booktitle={Proceedings of the 2024 8th International Conference on Computational Biology and Bioinformatics},
  pages={14--19},
  year={2024},
  doi = {10.1145/3715020.3715040}
}

@incollection{rice1976algorithm,
  title={The algorithm selection problem},
  author={Rice, John R},
  booktitle={Advances in computers},
  volume={15},
  pages={65--118},
  year={1976},
  publisher={Elsevier},
  doi={10.1016/S0065-2458(08)60520-3}
}

@article{sieg2019need,
  title={In need of bias control: evaluating chemical data for machine learning in structure-based virtual screening},
  author={Sieg, Jochen and Flachsenberg, Florian and Rarey, Matthias},
  journal={Journal of chemical information and modeling},
  volume={59},
  number={3},
  pages={947--961},
  year={2019},
  publisher={ACS Publications}
}

@article{gorantla2023proteins,
  title={From proteins to ligands: decoding deep learning methods for binding affinity prediction},
  author={Gorantla, Rohan and Kubincov{\'a}, Alzbeta and Wei{\ss}e, Andrea Y and Mey, Antonia SJS},
  journal={Journal of Chemical Information and Modeling},
  volume={64},
  number={7},
  pages={2496--2507},
  year={2023},
  publisher={ACS Publications}
}

@misc{esm2024cambrian,
  author = {{ESM Team}},
  title = {ESM Cambrian: Revealing the mysteries of proteins with unsupervised learning},
  year = {2024},
  publisher = {EvolutionaryScale Website},
  url = {https://evolutionaryscale.ai/blog/esm-cambrian},
  urldate = {2024-12-04}
}

@article{mqawass2024graphlambda,
  title={graphLambda: fusion graph neural networks for binding affinity prediction},
  author={Mqawass, Ghaith and Popov, Petr},
  journal={Journal of Chemical Information and Modeling},
  volume={64},
  number={7},
  pages={2323--2330},
  year={2024},
  publisher={ACS Publications},
  doi={10.1021/acs.jcim.3c00771}
}

@article{olier2018meta,
  title={Meta-QSAR: a large-scale application of meta-learning to drug design and discovery},
  author={Olier, Ivan and Sadawi, Noureddin and Bickerton, G Richard and Vanschoren, Joaquin and Grosan, Crina and Soldatova, Larisa and King, Ross D},
  journal={Machine Learning},
  volume={107},
  number={1},
  pages={285--311},
  year={2018},
  publisher={Springer},
  doi={10.1007/s10994-017-5685-x}
}

@article{liu2017forging,
  title={Forging the basis for developing protein--ligand interaction scoring functions},
  author={Liu, Zhihai and Su, Minyi and Han, Li and Liu, Jie and Yang, Qifan and Li, Yan and Wang, Renxiao},
  journal={Accounts of chemical research},
  volume={50},
  number={2},
  pages={302--309},
  year={2017},
  publisher={ACS Publications},
  doi={10.1021/acs.accounts.6b00491}
}

@article{hartshorn2007diverse,
  title={Diverse, high-quality test set for the validation of protein- ligand docking performance},
  author={Hartshorn, Michael J and Verdonk, Marcel L and Chessari, Gianni and Brewerton, Suzanne C and Mooij, Wijnand TM and Mortenson, Paul N and Murray, Christopher W},
  journal={Journal of medicinal chemistry},
  volume={50},
  number={4},
  pages={726--741},
  year={2007},
  publisher={ACS Publications}, 
  doi={10.1021/jm061277y}
}

@article{cao2025mc,
  title={MC-GNNAS-Dock: Multi-criteria GNN-based Algorithm Selection for Molecular Docking},
  author={Cao, Siyuan and Wu, Hongxuan and Wang, Jiabao Brad and Yuan, Yiliang and Misir, Mustafa},
  journal={arXiv preprint arXiv:2509.26377},
  year={2025}
}

@inproceedings{alcaide2025uni,
  title={Uni-mol docking v2: Towards realistic and accurate binding pose prediction},
  author={Alcaide, Eric and Gao, Zhifeng and Ke, Guolin and Li, Yaqi and Zhang, Linfeng and Zheng, Hang and Zhou, Gengmo},
  booktitle={International Conference on Artificial Neural Networks},
  pages={34--41},
  year={2025},
  organization={Springer},
  doi={10.1007/978-3-032-04552-2_5}
}

@article{buttenschoen2024posebusters,
  title={PoseBusters: AI-based docking methods fail to generate physically valid poses or generalise to novel sequences},
  author={Buttenschoen, Martin and Morris, Garrett M and Deane, Charlotte M},
  journal={Chemical Science},
  volume={15},
  number={9},
  pages={3130--3139},
  year={2024},
  publisher={Royal Society of Chemistry},
  doi={10.1039/D3SC04185A}
}

@article{cao2025surfdock,
  title={SurfDock is a surface-informed diffusion generative model for reliable and accurate protein--ligand complex prediction},
  author={Cao, Duanhua and Chen, Mingan and Zhang, Runze and Wang, Zhaokun and Huang, Manlin and Yu, Jie and Jiang, Xinyu and Fan, Zhehuan and Zhang, Wei and Zhou, Hao and others},
  journal={Nature Methods},
  volume={22},
  number={2},
  pages={310--322},
  year={2025},
  publisher={Nature Publishing Group US New York},
  doi = {10.1038/s41592-024-02516-y}
}

@article{chen2023algorithm,
  title={Algorithm selection for protein--ligand docking: strategies and analysis on ACE},
  author={Chen, Tianlai and Shu, Xiwen and Zhou, Huiyuan and Beckford, Floyd A and Misir, Mustafa},
  journal={Scientific Reports},
  volume={13},
  number={1},
  pages={8219},
  year={2023},
  publisher={Nature Publishing Group UK London},
  doi = {10.1038/s41598-023-35132-5}
}

@article{corso2024deep,
  title={Deep Confident Steps to New Pockets: Strategies for Docking Generalization},
  author={Corso, Gabriele and Deng, Arthur and Fry, Benjamin and Polizzi, Nicholas and Barzilay, Regina and Jaakkola, Tommi},
  journal={arXiv preprint arXiv:2402.18396},
  year={2024}
}

@article{corso2022diffdock,
  title={Diffdock: Diffusion steps, twists, and turns for molecular docking},
  author={Corso, Gabriele and St{\"a}rk, Hannes and Jing, Bowen and Barzilay, Regina and Jaakkola, Tommi},
  journal={arXiv preprint arXiv:2210.01776},
  year={2022}
}

@article{fan2019progress,
  title={Progress in molecular docking},
  author={Fan, Jiyu and Fu, Ailing and Zhang, Le},
  journal={Quantitative Biology},
  volume={7},
  number={2},
  pages={83--89},
  year={2019},
  publisher={Springer},
  doi = {10.1007/s40484-019-0172-y}
}

@article{hassan2017protein,
  title={Protein-ligand blind docking using QuickVina-W with inter-process spatio-temporal integration},
  author={Hassan, Nafisa M and Alhossary, Amr A and Mu, Yuguang and Kwoh, Chee-Keong},
  journal={Scientific reports},
  volume={7},
  number={1},
  pages={15451},
  year={2017},
  publisher={Nature Publishing Group UK London},
  doi = {10.1038/s41598-017-15571-7}
}

@article{he2020resnet,
  title={Why resnet works? residuals generalize},
  author={He, Fengxiang and Liu, Tongliang and Tao, Dacheng},
  journal={IEEE transactions on neural networks and learning systems},
  volume={31},
  number={12},
  pages={5349--5362},
  year={2020},
  publisher={IEEE},
  doi = {10.1109/TNNLS.2020.2966319}
}

@misc{jiang2025posex,
    title={PoseX: {AI} Defeats Physics-based Methods on Protein Ligand Cross-Docking},
    author={Yize Jiang and Xinze Li and Yuanyuan Zhang and Jin Han and Youjun Xu and Ayush Pandit and ZAIXI ZHANG and Mengdi Wang and Mengyang Wang and Chong Liu and Guang Yang and Yejin Choi and Wu-Jun Li and Tianfan Fu and Fang Wu and Junhong Liu},
    year={2025},
    url={https://openreview.net/forum?id=qHOU1aAyXA}
}

@article{kerschke2019automated,
  title={Automated algorithm selection: Survey and perspectives},
  author={Kerschke, Pascal and Hoos, Holger H and Neumann, Frank and Trautmann, Heike},
  journal={Evolutionary computation},
  volume={27},
  number={1},
  pages={3--45},
  year={2019},
  publisher={mIT Press},
  doi = {10.1162/evco_a_00242}
}

@article{koes2013lessons,
  title={Lessons learned in empirical scoring with smina from the CSAR 2011 benchmarking exercise},
  author={Koes, David Ryan and Baumgartner, Matthew P and Camacho, Carlos J},
  journal={Journal of chemical information and modeling},
  volume={53},
  number={8},
  pages={1893--1904},
  year={2013},
  publisher={ACS Publications},
  doi = {10.1021/ci300604z}
}

@article{mcnutt2021gnina,
  title={GNINA 1.0: molecular docking with deep learning},
  author={McNutt, Andrew T and Francoeur, Paul and Aggarwal, Rishal and Masuda, Tomohide and Meli, Rocco and Ragoza, Matthew and Sunseri, Jocelyn and Koes, David Ryan},
  journal={Journal of cheminformatics},
  volume={13},
  number={1},
  pages={43},
  year={2021},
  publisher={Springer},
  doi = {10.1186/s13321-021-00522-2}
}

@article{misir2017alors,
  title={Alors: An algorithm recommender system},
  author={M{\i}s{\i}r, Mustafa and Sebag, Mich{\`e}le},
  journal={Artificial Intelligence},
  volume={244},
  pages={291--314},
  year={2017},
  publisher={Elsevier},
  doi = {10.1016/j.artint.2016.12.001}
}

@article{wolpert2002no,
  title={No free lunch theorems for optimization},
  author={Wolpert, David H and Macready, William G},
  journal={IEEE transactions on evolutionary computation},
  volume={1},
  number={1},
  pages={67--82},
  year={2002},
  publisher={IEEE},
  doi={10.1109/4235.585893}
}

@inproceedings{yuan2024gnnas,
  title={GNNAS-Dock: Budget Aware Algorithm Selection with Graph Neural Networks for Molecular Docking},
  author={Yuan, Yiliang and Misir, Mustafa},
  booktitle={NeurIPS 2024 Workshop on AI for New Drug Modalities}, 
  year={2024},
  url={https://openreview.net/forum?id=W4d1a4odSc}
}

@article{zhang2023efficient,
  title={Efficient and accurate large library ligand docking with KarmaDock},
  author={Zhang, Xujun and Zhang, Odin and Shen, Chao and Qu, Wanglin and Chen, Shicheng and Cao, Hanqun and Kang, Yu and Wang, Zhe and Wang, Ercheng and Zhang, Jintu and others},
  journal={Nature Computational Science},
  volume={3},
  number={9},
  pages={789--804},
  year={2023},
  publisher={Nature Publishing Group US New York},
  doi = {10.1038/s43588-023-00511-5}
}

@article{cenikj2025landscape,
  title={Landscape features in single-objective continuous optimization: Have we hit a wall in algorithm selection generalization?},
  author={Cenikj, Gjorgjina and Petelin, Ga{\v{s}}per and Seiler, Moritz and Cenikj, Nikola and Eftimov, Tome},
  journal={Swarm and Evolutionary Computation},
  volume={94},
  pages={101894},
  year={2025},
  publisher={Elsevier}
}

\end{document}